\documentclass{aa} 
\usepackage{threeparttable,framed,rotating,xcolor,caption}
\usepackage{amsmath}
\usepackage{lineno}
\usepackage[utf8]{inputenc}
\usepackage[title]{appendix}%
\usepackage{graphicx}
\usepackage{txfonts}
\begin{document}

   \title{The chemical footprint of AGN feedback
in the outflowing circumnuclear disk of NGC 1068}
    \titlerunning{Chemical footprint in the outflowing CND of NGC 1068}
    \authorrunning{Huang et al.}


   \author{K.-Y. Huang
          \inst{1}\fnmsep\thanks{kyhuang@strw.leidenuniv.nl}, 
          S. Viti
          \inst{1,2}, 
          J. Holdship
          \inst{1,2}, 
          S. Garc{\'{\i}}a-Burillo
          \inst{3}, 
          K. Kohno
          \inst{4}, 
          A. Taniguchi
          \inst{5}, 
          S. Mart\'{{\i}}n
          \inst{6,7}, 
          R. Aladro
          \inst{8},
          A. Fuente
          \inst{3}, 
          \and
          M. S{\'a}nchez-Garc{\'{\i}}a
          \inst{9}
          }

   \institute{Leiden Observatory, Leiden University, PO Box 9513, 2300 RA Leiden, The Netherlands
         \and
              Department of Physics and Astronomy, University College London, Gower Street, London WC1E 6BT, UK
         \and
              Observatorio Astron{\'{o}}mico Nacional (OAN-IGN)-Observatorio de Madrid, Alfonso XII, 3, 28014-Madrid, Spain
         \and
              Institute of Astronomy, The University of Tokyo, Osawa, Mitaka, Tokyo 181-0015, Japan
         \and
              Division of Particle and Astrophysical Science, Graduate School of Science, Nagoya University, Furocho, Chikusa-ku, Nagoya, Aichi 464-8602, Japan
         \and
              European Southern Observatory, Alonso de C\'{o}rdova, 3107, Vitacura, Santiago 763-0355, Chile
         \and
              Joint ALMA Observatory, Alonso de C\'{o}rdova, 3107, Vitacura, Santiago 763-0355, Chile
         \and
              Max-Planck-Institut f\"{u}r Radioastronomie, Auf dem H\"{u}gel 69, D-53121 Bonn, Germany
         \and
              Centro de Astrobiolog{\'{\i}}a (CSIC/INTA), Ctra de Torrej{\'o}n a Ajalvir, km 4, 28850 Torrej{\'o}n de Ardoz, Madrid, Spain
             }

   \date{Submitted Dec. 2021; accepted Feb. 2022}

 
  \abstract
   {In the nearby (D=14 Mpc) AGN-starburst composite galaxy NGC 1068, it has been found that the molecular gas in the Circum-nuclear Disk (CND) is outflowing, which is a manifestation of ongoing AGN feedback. The outflowing gas has a large spread of velocities, which likely drive different shock chemistry signatures at different locations in the CND. }
   {We perform a multi-line molecular study using two shock tracers, SiO and HNCO, with the aim to determine the gas properties traced by these two species, and explore the possibility of reconstructing the shock history in the CND. }
   {Five SiO transitions and three HNCO transitions were imaged at high resolution $0''.5-0''.8$ with the Atacama Large Millimeter/submillimeter Array (ALMA). We performed both LTE and non-LTE radiative transfer analysis coupled with Bayesian inference process in order to characterize the gas properties, such as molecular gas density and gas temperature. }
   {We found clear evidence of chemical differentiation between SiO and HNCO, with the SiO/HNCO ratio ranging from greater than one on the east of CND to lower than one on the west side. The non-LTE radiative transfer analysis coupled with Bayesian inference confirms that the gas traced by SiO has different densities - and possibly temperatures - than that traced by HNCO. We find that SiO  traces gas affected by fast shocks while the gas traced by HNCO is either just affected by slow shocks or not shocked at all. }
   {A distinct differentiation between SiO and HNCO has been revealed in our observations and the further analysis of the gas properties traced by both species, which confirms the results from previous chemical modelings. }


   \maketitle
%
\section{Introduction}
Multi-line molecular observations are an ideal tool to trace the physical and chemical processes in external galaxies, given the wide range of critical densities of different molecular species and the associated transitions, and the dependencies of chemical reactions on the energy available to the system. 
Observationally, there are several molecules found to trace different regions within a galaxy, such as HCO and HOC\textsuperscript{+} in photon-dominated regions (PDRs) \citep[e.g.][]{Savage_Ziurys_2004,GB+2002,Gerin+2009_HCO_PDR,Martin+2009_PDR_ngc253}
, and HCN and CS in dense gas clumps \citep[e.g.][]{Gao_Solomon_2004,Bayet+2008,Aladro+2011_CS}. 
In reality, it is seldom the case that one can identify a single gas component using one particular molecular species \citep{Kauffmann+2017_hcn,Pety+2017_densegas,Viti_2017}, for often the same species can be found in diverse environments, and the different transitions from the same species might trace different gas components due to the shaping of the energy distribution of the molecular ladders by the energetics present in the field. 
Therefore, molecular tracers that are uniquely sensitive to certain environments are particularly valuable in characterizing the gas conditions both physically and chemically. 

The molecules silicon monoxide, SiO, and isocyanic acid, HNCO, are both well known tracers of shocks \citep{sio_MP+1997,Huttemeister+1998,Zinchenko+2000,J-S+2008_shocktracers,Martin+2008_HNCO_galactic,hnco+RF+2010} and have been used observationally as shock tracers in nearby galaxies \citep[e.g.][]{GB+2000_sio_253,Meier_Turner_2005,Usero+2006,Martin+2009,GB+2010,Meier_Truner_2012_maffei2,Martin+2015_shocktracer_ngc1097,Meier+2015_hncosio_253,Kelly+2017}. 
HNCO may form mainly on dust grain mantles \citep{Fedoseev+2015}, or possibly form in the gas phase and then freezes out onto the dust grain \citep{LS+2015}. 
In either scenarios its presence on the icy mantles of the dust grain means that HNCO can be easily sublimated even in weakly shocked regions; hence HNCO may be a useful tracer of low-velocity shocks ( $\varv_{s}\sim20$ km s\textsuperscript{-1}). 
On the other hand, silicon is significantly sputtered from the core of the dust grains and released into the gas phase by higher-velocity shocks ($\varv_{s}\geq50$ km s\textsuperscript{-1}). 
Once silicon is in the gas phase, it can quickly react with molecular oxygen or a hydroxyl radical to form SiO \citep{Schilke+1997}. 
Therefore the enhanced abundance of SiO may be an indication to the presence of more heavily shocked regions. 

The simultaneous detection of HNCO and SiO  in a galaxy where shocks are believed to take place may provide us a more comprehensive picture of the shock history of the gas. Indeed, these two species have  already 
been  proposed  to distinguish and characterize different types of shocks (fast vs. slow) in the AGN-host galaxies NGC 1068 \citep{Kelly+2017} and NGC 1097 \citep{Martin+2015_shocktracer_ngc1097}, and in the nearby starburst galaxy NGC 253 \citep{Meier+2015_hncosio_253}. 
For example, in NGC 253, HNCO was found distinctively prominent in the outer part of the nuclear disk, and the varying HNCO/SiO ratio, which drops dramatically in the inner disk, has been suggested to signal both the decreasing shock strength and the erased shock chemistry of HNCO in the presence of dominating central radiation fields \citep{Meier+2015_hncosio_253}. In AGN-dominated galaxies, determining the origin and nature of the shocked gas may reveal its connection (or lack of) with the AGN feedback.

NGC1068 is a nearby (D = 14 Mpc \citealp{Bland-Hawthorn+1997}, $1'' \sim 70$ pc) Seyfert 2 galaxy and is considered to be the archetype of a composite AGN-starburst system. 
The proximity of this composite galaxy makes it an ideal laboratory to resolve the feedback from the starburst regions that are spatially distinct from the AGN activity. NGC 1068 has been extensively investigated by many single-dish and interferometric campaigns focused on the study of the fuelling of its central region and related feedback activity using molecular line observations \citep[e.g.][]{Usero+2004,Israel_2009,Kamenetzky+2011,Hailey-Dunsheath+2012,Aladro+2013,GB+2014,Viti+2014,GB+2017,GB+2019,Impellizzeri+2019,Imanishi+2020}. 
CO Observations of NGC 1068 by \citet{Schinnerer+2000} reveal the molecular gas distributing over three regions, also confirmed by e.g. \citet{GB+2014,GB+2019} and \citet{MSG+2022}: a starburst ring (SB ring) with a radius $\sim 1.5$ kpc, a circumnuclear disk (CND) of radius $\sim 200$ pc, and a $\sim 2$ kpc stellar bar running north east, along PA $\sim48^{\circ}$ \citep{Scoville+1988}, from the CND. 
In \citet{GB+2014} and \citet{Viti+2014}, five chemically distinct regions were found to be present within the CND: the AGN, the East Knot, West Knot and regions to the north and south of the AGN (CND-N and CND-S) using data from the Atacama Large Millimeter/submillimeter Array (ALMA). 
\citet{Viti+2014} combined these ALMA data with Plateau de Bure Interferometer (PdBI) data and determined the physical and chemical properties of each region. 
It was found that a pronounced chemical differentiation is present across the CND and that each sub-region could be characterised by a three-phase component interstellar medium, where one of the component is comprised of shocked gas. 
In fact, \citet{GB+2010} used the PdBI to map NGC 1068 and found strong emission of SiO(2-1) in the east and west of CND.  
The SiO kinematics of the CND point to an overall rotating structure, and is distorted by non-circular and/or non-coplanar motions. 
The authors concluded that this could be due to large scale shocks through cloud-cloud collisions, or through a jet-ISM interaction. 
Such shock-related non-circular kinematics of gas was also identified later by \citet{Krips+2011} using several molecular ratios of CO, \textsuperscript{13}CO, HCN, and HCO\textsuperscript{+}. 
However, due to strong CN emission not easily explained by shock models nor photon-dominated region (PDR) chemistry, they also suggest that the CND could actually be one large X-ray dominated region (XDR). 

In a more recent study  \citet{Kelly+2017} analyse PdBI observations at spatial resolution $\sim 1''.1$ of both SiO and HNCO and found that the SiO (3-2) emission was  stronger in the East Knot than in the West Knot, while 
HNCO (6-5) was found to be strongest in the West Knot, with a less prominent local peak in the East Knot. 
Furthermore, the local peaks of HNCO and SiO on both sides of CND were found spatially displaced from each other, hinting at the possibility that these two species were tracing distinct gas components. 
To verify this, \citet{Kelly+2017} performed a chemical modeling for the SiO and HNCO emission by considering a plane-parallel C-type shock propagating with the velocity $\varv_{s}$ through the ambient medium \citep{J-S+2008_shocktracers,Viti+2014}, and confirmed that fast shocks ($\varv_{s}=60$ km s\textsuperscript{-1}) are likely to be producing SiO; while weak shocks ($\varv_{s}=20$ km s\textsuperscript{-1}) are likely responsible for the abundance enhancement in the observed HNCO. 
The shocks, especially the high-velocity ones ($\varv_{s}=60$ km s\textsuperscript{-1}), are likely set by the molecular outflow with velocity at $\sim 100$ km s\textsuperscript{-1} scale in the CND\citep{GB+2019}, which is possibly a manifestation of AGN feedback onto the CND molecular gas. 
With the limited spatial resolution and limited number of transitions per species of their data, however, they were not able to firmly conclude whether HNCO is indeed associated with slower shocks, or with the gas that is simply warm, dense and non-shocked. 

In the current work we present higher resolution ($0''.5-0''.8$) ALMA observations of the CND of NGC 1068 for five SiO and three HNCO transitions. 
The main goal is to spatially resolve  the gas properties of potentially shocked gas in the CND by the use of multiple  transitions of these two shock tracers at better spatial resolution compared to the previous work. 
The paper is structured as follows. 
In Section 2 we describe the observations and the data reduction process. 
In Section 3 we present the molecular line intensity maps, and the comparison of intensity across transitions using overlay and ratio maps. 
In Section 4 we perform an LTE and a non-LTE radiative transfer analysis in order to constrain the physical conditions of the gas. 
We briefly summarise our findings in Section 5. 
\section{Observations and Data Reduction}
\label{sec:obs}
\begin{table*}[t!]
  \centering
  \caption{Observational details and the spatial resolution of the data used in this paper. A distance of 14 Mpc is assumed. }
  \label{tab:table_obsinfo}
  \begin{tabular}{c|cccccc}
  \hline
    Transition & Rest Frequency & E\textsubscript{u} & ALMA project ID & Band & Spatial resolution & mJy/beam to K  \\
    {} & [GHz] & [K] & {} & {} & {}\\
    \hline
    HNCO(4\textsubscript{0,4}-3\textsubscript{0,4}) & 87.925 & 10.55 & 2018.1.01506.S & 3 & $0.8" \times 0.7"$ ($53$ pc $\times 48$ pc) & 0.29\\
    HNCO(5\textsubscript{0,5}-4\textsubscript{0,4}) & 109.906 & 15.82 & 2018.1.01506.S & 3 & $0.7" \times 0.4"$ ($48$ pc $\times 27$ pc) & 0.27\\
    HNCO(6\textsubscript{0,6}-5\textsubscript{0,5}) & 131.886 & 22.15 & 2018.1.01506.S & 4 & $0.6" \times 0.5"$ ($ 41$ pc $\times 34$ pc) & 0.26\\
    SiO(2-1) & 86.847 & 6.25 & 2013.1.00221.S & 3 & $0.8" \times 0.5"$ ($53$ pc $\times 34$ pc) & 0.38\\
    SiO(3-2) & 130.269 & 12.50 & 2013.1.00221.S & 4 & $0.4" \times 0.4"$ ($27$ pc $\times 27$ pc) & 0.45\\
    SiO(5-4) & 217.105 & 31.26 & 2013.1.00221.S & 6 & $0.5" \times 0.5"$ ($34$ pc $\times 34$ pc) & 0.10\\
    SiO(6-5) & 260.518 & 43.76 & 2013.1.00221.S & 6 & $0.5" \times 0.4"$ ($34$ pc $\times 27$ pc) & 0.089\\
    SiO(7-6) & 303.927 & 58.35 & 2015.1.01144.S & 7 & $0.5" \times 0.4"$ ($34$ pc $\times 27$ pc) & 0.068\\
    \hline
  \end{tabular}
\end{table*}
\begin{table*}[t!]
  \begin{minipage}{0.5\linewidth}
  \centering
  \caption{Coordinates (RA and Dec) of the five selected regions within the CND. }
  \label{tab:table_7regs}
  \begin{tabular}{c|cc}
  \hline
    Name & RA & DEC   \\
    \hline
    AGN & 02\textsuperscript{h}42\textsuperscript{m}40\textsuperscript{s}.710 & -00$^{\circ}$00${'}$47${''}$.94  \\
    CND R1  & 02\textsuperscript{h}42\textsuperscript{m}40\textsuperscript{s}.7617 & -00$^{\circ}$00${'}$48${''}$.1200 \\
    CND R2  & 02\textsuperscript{h}42\textsuperscript{m}40\textsuperscript{s}.7243 & -00$^{\circ}$00${'}$49${''}$.2400 \\
    CND R3  & 02\textsuperscript{h}42\textsuperscript{m}40\textsuperscript{s}.6030 & -00$^{\circ}$00${'}$48${''}$.9600 \\
    CND R4  & 02\textsuperscript{h}42\textsuperscript{m}40\textsuperscript{s}.6590 & -00$^{\circ}$00${'}$47${''}$.7000 \\
    \hline
  \end{tabular}
  \end{minipage}\hfill
  \begin{minipage}{0.4\linewidth}
	\centering
	\includegraphics[width=40mm]{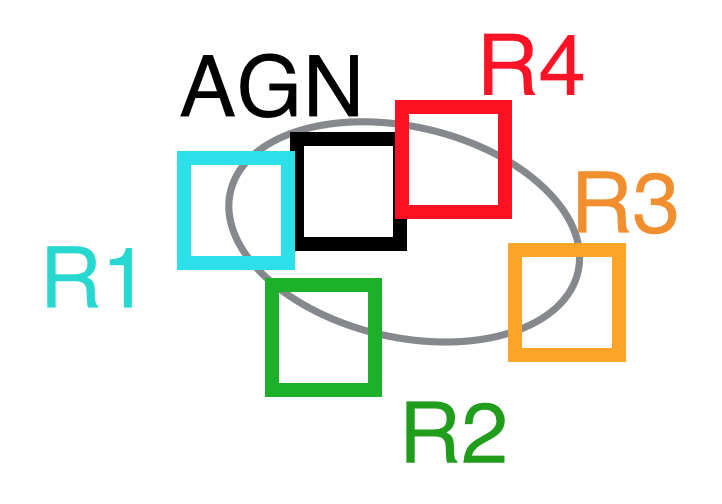}
	\captionof{figure}{The selected $0''.8\times0''.8$ regions in the CND (the grey ellipse in the background),  as listed in Table \ref{tab:table_7regs}, and with the color coding that is consistent with the spectra presented in Figures \ref{fig:allHs_spec}-\ref{fig:allSs_spec3}. These are the same regions used for the analyses from Section \ref{sec:gas_properties} onward. }
	\label{fig:Regs_schematic}
  \end{minipage}
\end{table*}

The HNCO and SiO transitions of NGC 1068 used in this paper were observed using ALMA. 
The HNCO data were obtained during cycle 6 (project-ID: 2018.1.01506.S) with HNCO(4-3) and HNCO(5-4) using band 3 receivers, and HNCO(6-5) using band 4 receivers. 
The SiO(7-6) data was obtained during cycle 3 (project-ID: 2015.1.01144.S) using band 7 receivers. 
The above mentioned data  were calibrated and imaged using the ALMA reduction package CASA\footnote{http://casa.nrao.edu} \citep{CASA_2007}. 
The rest of the SiO observations were obtained during cycle 2 (project ID: 2013.1.00221.S). 

The rest frequencies were defined using the systemic velocity determined by \citet{GB+2019}, as $v_{sys}$(LSR) = 1120 km s\textsuperscript{-1} (radio convention). 
The relative velocities throughout the paper refer to this $v_{sys}$. 
The phase tracking center was set to $\alpha_{2000}$ = (02\textsuperscript{h}42\textsuperscript{m}40.771\textsuperscript{s}), $\delta_{2000}$ = (–00$^{\circ}$00$'$47. 84$''$). 
The relevant information of each observation is listed in Table \ref{tab:table_obsinfo}. 
This Table includes the target molecular transition, the observation project ID, and the synthesized beam size for each observation. 
The beam sizes of our observations range between $0''.5-0''.8$, or 35-56 pc in physical scales. 
This is comparable to the typical scale of Giant Molecular Clouds (GMC). 

\begin{figure*}
        \centering
    \includegraphics[scale=0.48]{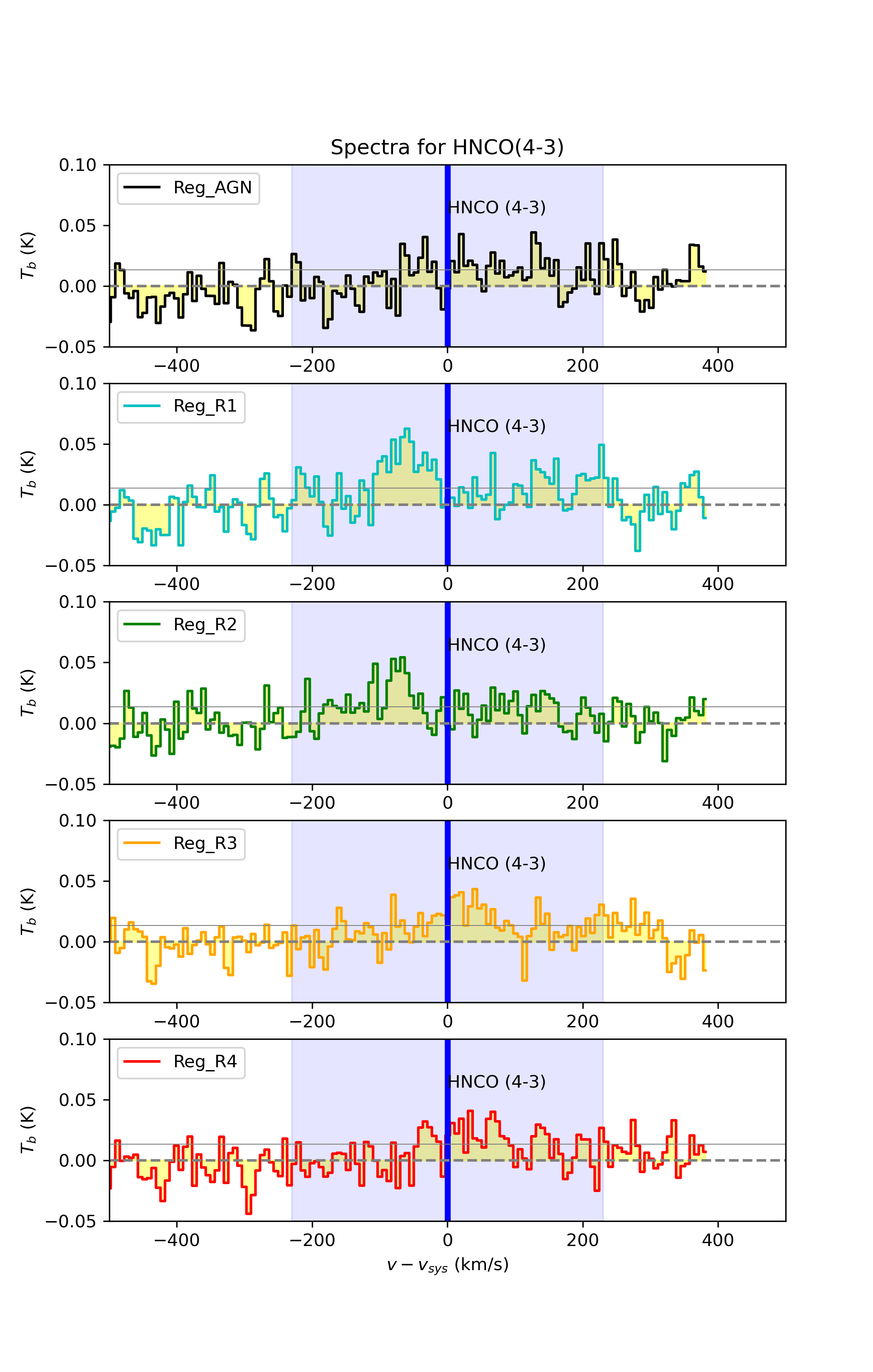}
    \includegraphics[scale=0.48]{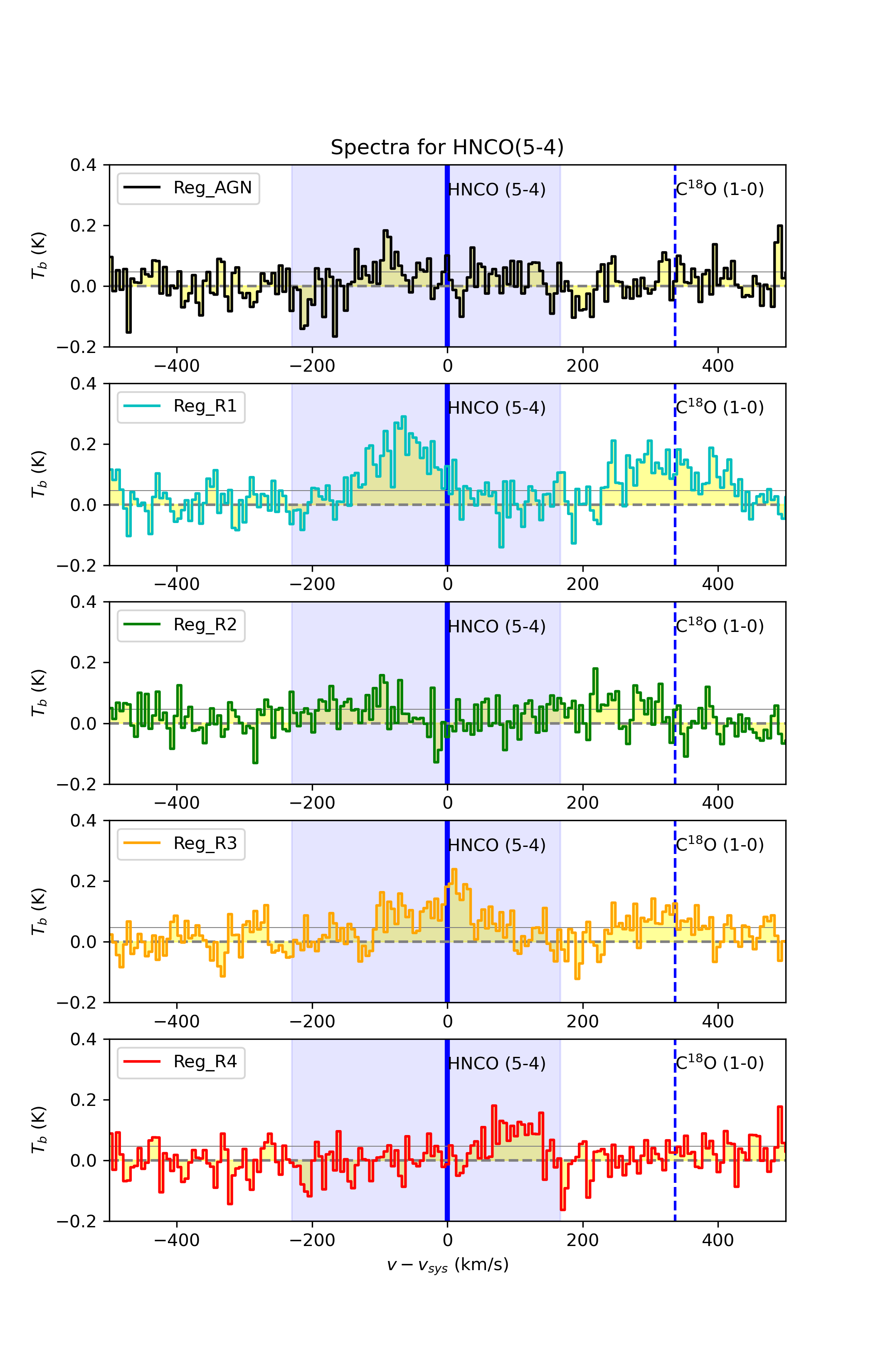}
    \caption{Spectra of the HNCO (4-3) and HNCO (5-4) transitions. Each color-coded solid curve plots the spectral data from each selected $0''.8\times0''.8$ region listed in Table \ref{tab:table_7regs} from the data cube at their original spectral resolution. The spectral resolution of these lines are: 6.6 km s\textsuperscript{-1} for HNCO (4-3), and 5.4 km s\textsuperscript{-1} for HNCO (5-4). The solid blue vertical lines are the HNCO lines including all the splittings; the rest dashed blue vertical lines are bonus lines potentially covered by our spectral setup. Velocities refer to $v_{sys, LSRK}=1120$ km s\textsuperscript{-1}. The blue shaded area indicates the velocity coverage we use to derive the velocity-integrated line intensities in our analysis at later stage. The grey solid horizontal line refers to the $1\sigma$ level for each transition. }
    \label{fig:allHs_spec}
\end{figure*}
\begin{figure*}
        \centering
    \includegraphics[scale=0.48]{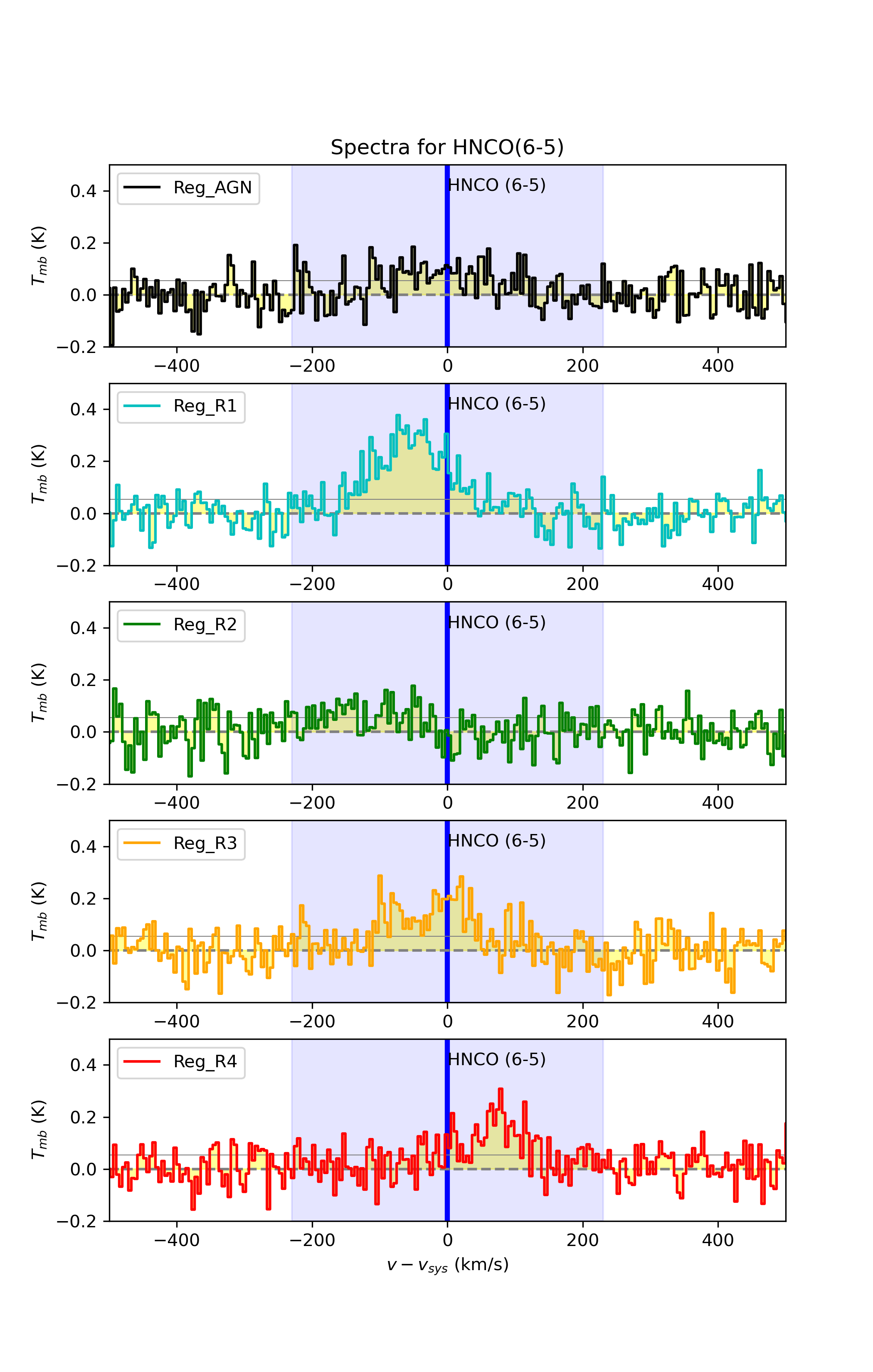}
    \includegraphics[scale=0.48]{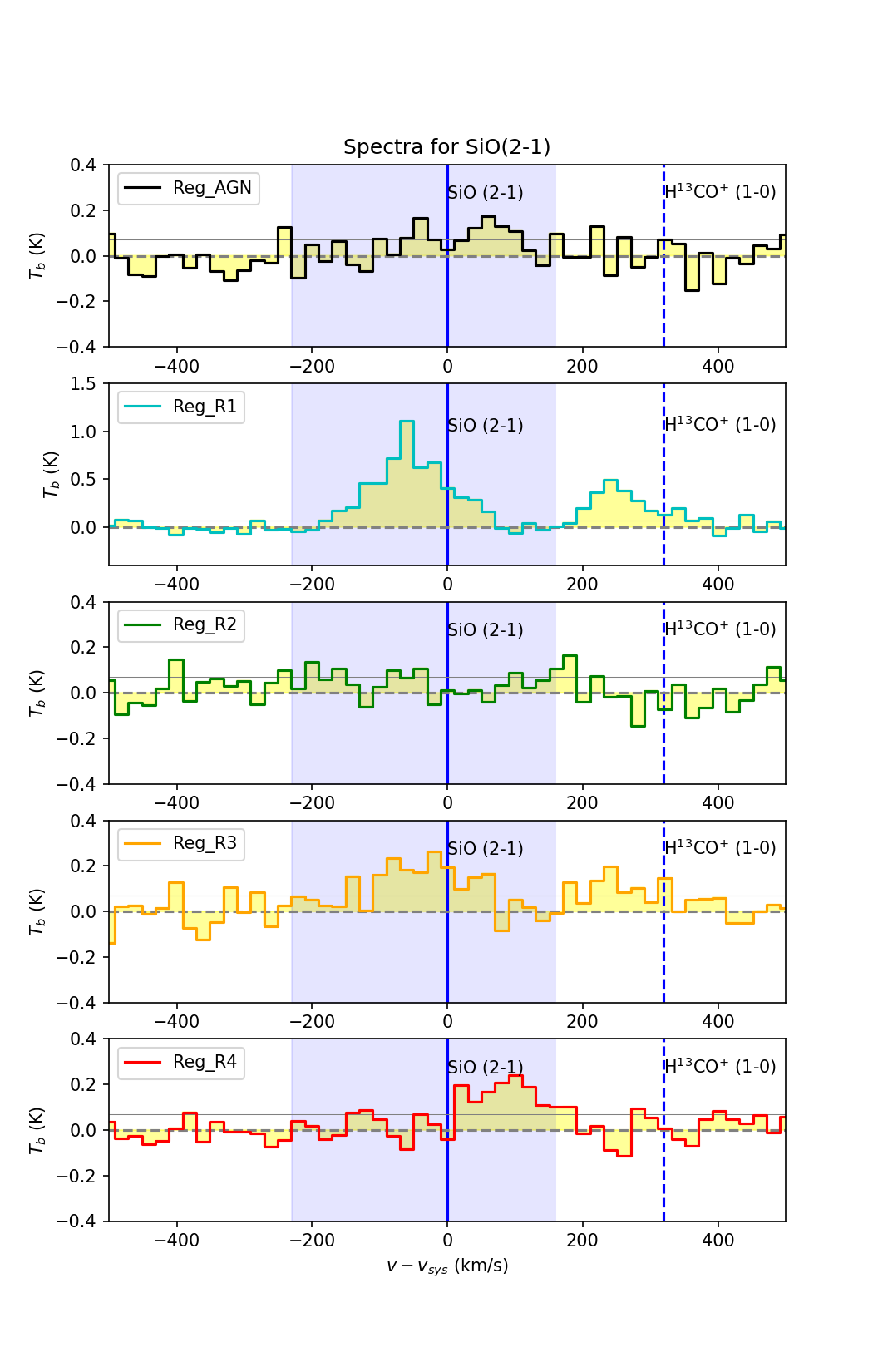}
    \caption{Spectra of the SiO (2-1) and HNCO (6-5) transitions. Each color-coded solid curve plots the spectral data from each selected $0''.8\times0''.8$ region listed in Table \ref{tab:table_7regs} from the data cube at their original spectral resolution. The spectral resolution of these lines are: 4.5 km s\textsuperscript{-1} for HNCO (6-5), and 20.1 km s\textsuperscript{-1} for SiO (2-1). The solid blue vertical lines are the HNCO line (including all the splittings) and SiO line, and the rest dashed blue vertical lines are bonus lines potentially covered by our spectral setup. Velocities refer to $v_{sys, LSRK}=1120$ km s\textsuperscript{-1}. The blue shaded area indicates the velocity coverage we use to derive the velocity-integrated line intensities in our analysis at later stage. The grey solid horizontal line refers to the $1\sigma$ level for each transition. }
    \label{fig:allSs_spec1}
\end{figure*}
\begin{figure*}
        \centering
    \includegraphics[scale=0.48]{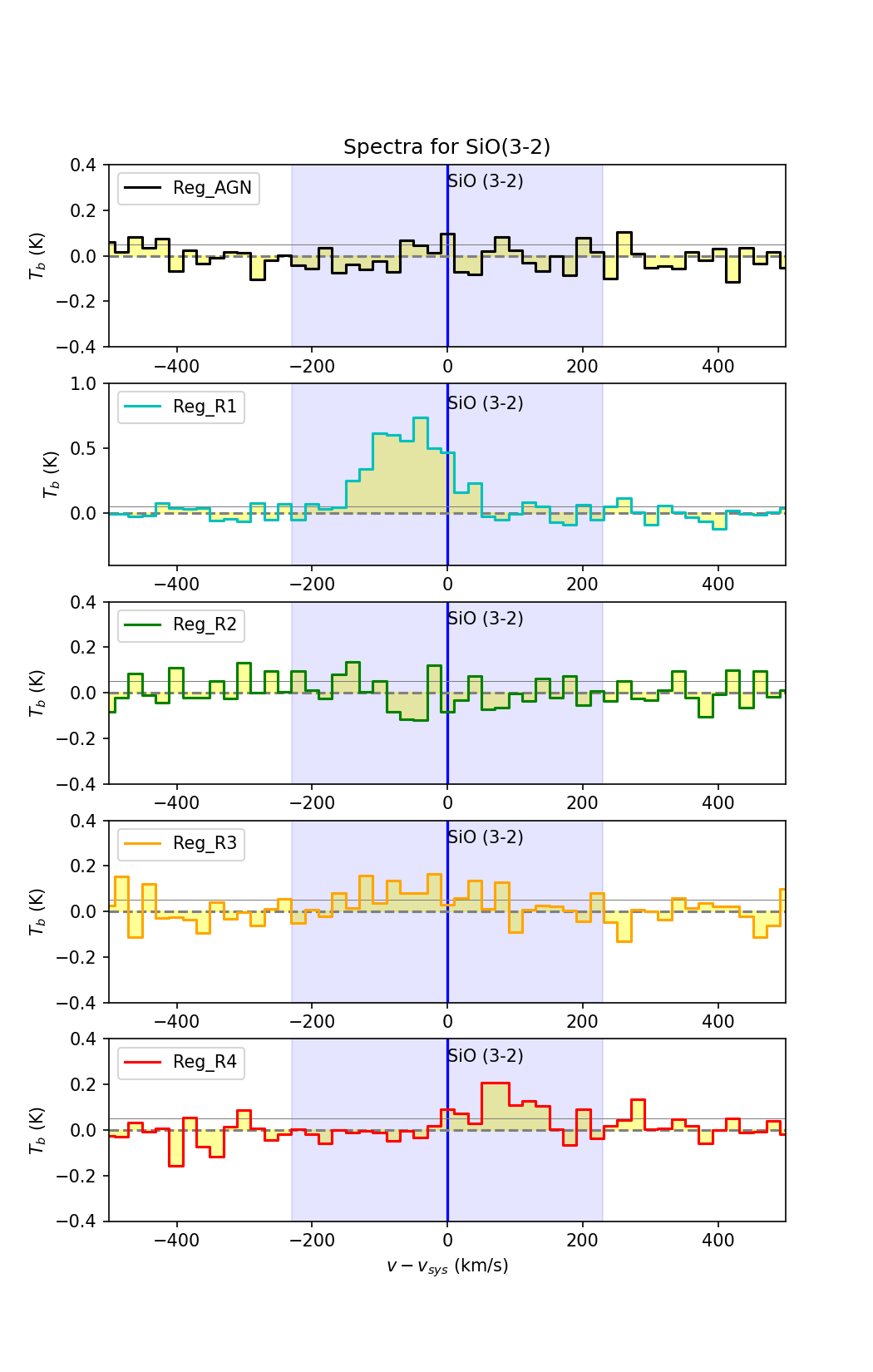}
    \includegraphics[scale=0.48]{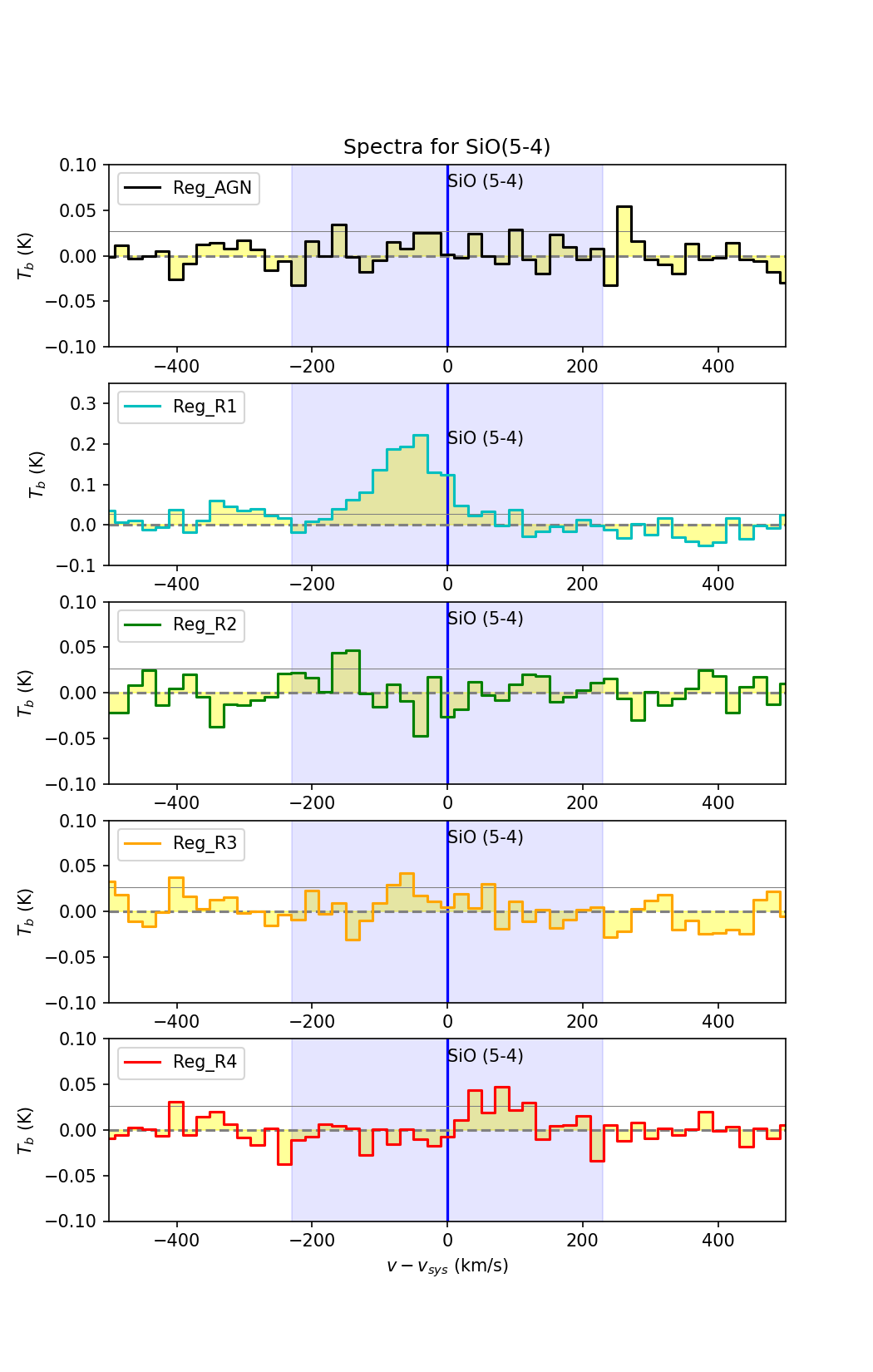}
    \caption{Spectra of the SiO (3-2) and SiO(5-4) transitions. Each color-coded solid curve plots the spectral data from each selected $0''.8\times0''.8$ region listed in Table \ref{tab:table_7regs} from the data cube at their original spectral resolution. The spectral resolution of these lines are: 20.1 km s\textsuperscript{-1} for SiO (3-2) and (5-4). The solid blue vertical lines are the SiO lines, the rest dashed blue vertical lines are bonus lines potentially covered by our spectral setup. Velocities refer to $v_{sys, LSRK}=1120$ km s\textsuperscript{-1}. The blue shaded area indicates the velocity coverage we use to derive the velocity-integrated line intensities in our analysis at later stage. The grey solid horizontal line refers to the $1\sigma$ level for each transition. }
    \label{fig:allSs_spec2}
\end{figure*}
\begin{figure*}
        \centering
    \includegraphics[scale=0.48]{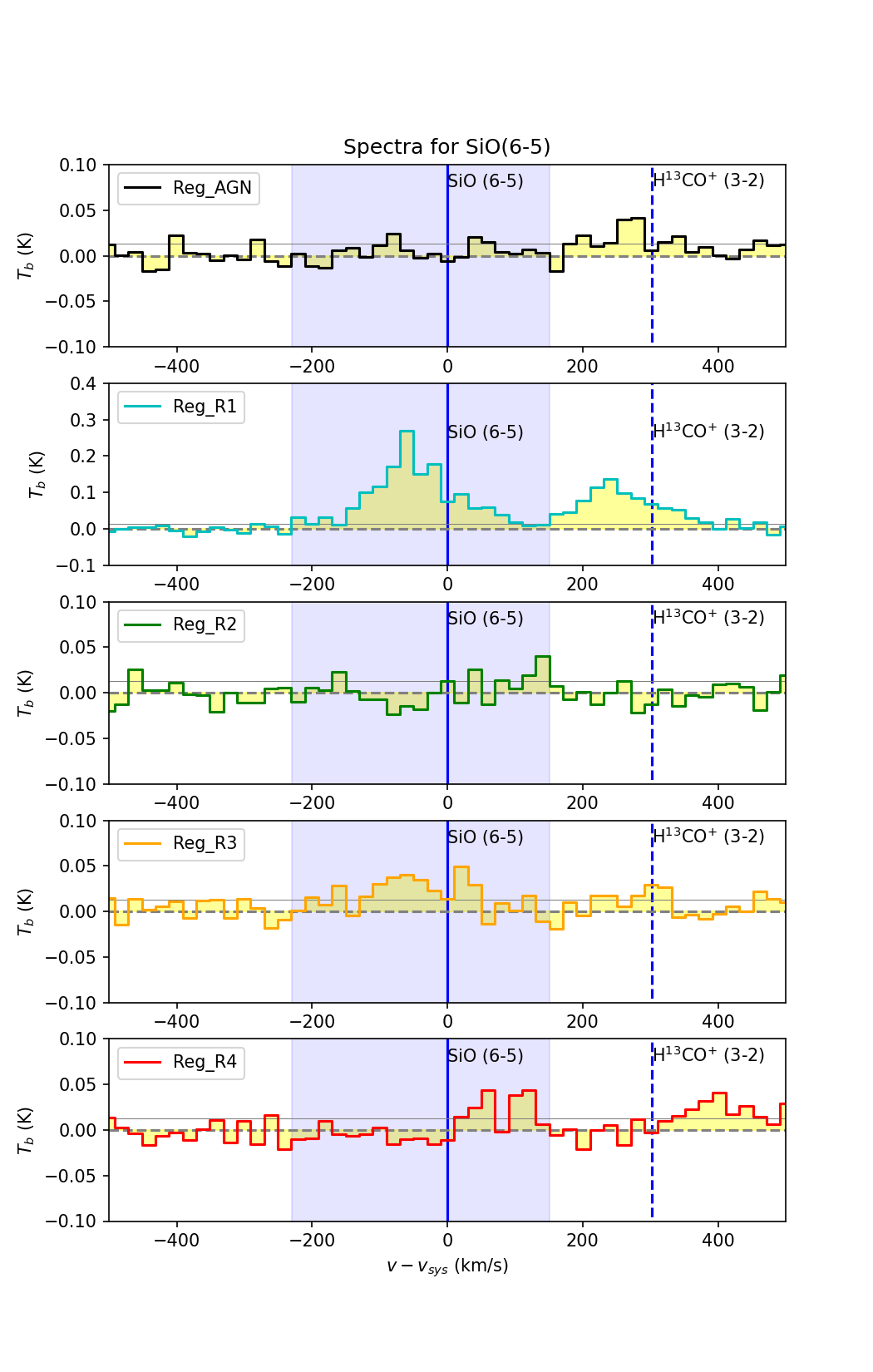}
    \includegraphics[scale=0.48]{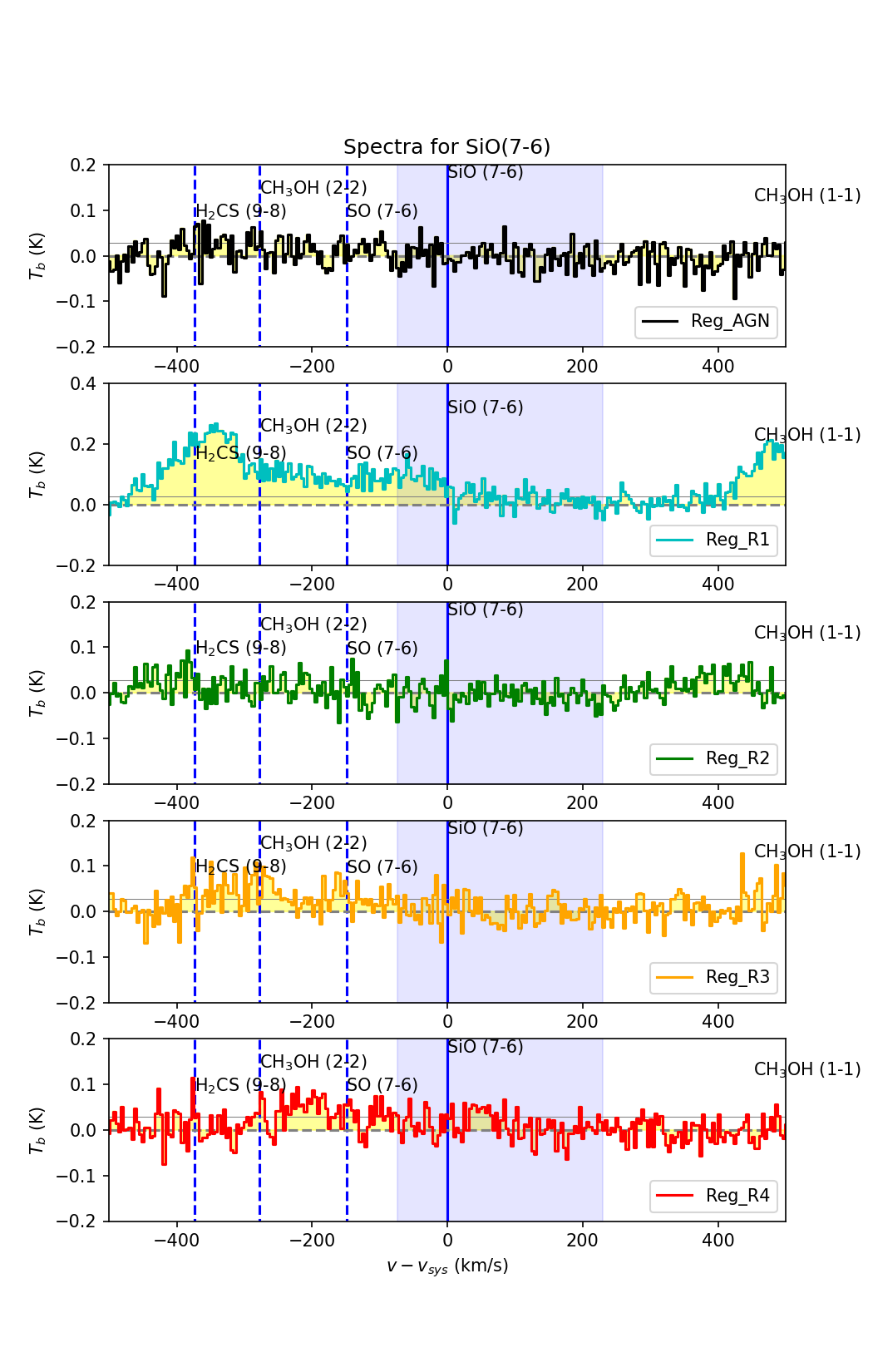}
    \caption{Spectra of the SiO (6-5) and SiO (7-6) transitions. Each color-coded solid curve plots the spectral data from each selected $0''.8\times0''.8$ region listed in Table \ref{tab:table_7regs} from the data cube at their original spectral resolution. The spectral resolution of these lines are: 20.1 km s\textsuperscript{-1} for SiO (6-5), and 3.9 km s\textsuperscript{-1} for SiO(7-6). The solid blue vertical lines are the SiO lines, the rest dashed blue vertical lines are bonus lines potentially covered by our spectral setup. Velocities refer to $v_{sys, LSRK}=1120$ km s\textsuperscript{-1}. The blue shaded area indicates the velocity coverage we use to derive the velocity-integrated line intensities in our analysis at later stage. The grey solid horizontal line refers to the $1\sigma$ level for each transition. }
    \label{fig:allSs_spec3}
\end{figure*}
\section{Molecular line emission}
\subsection{Molecular-line spectra}
In order to investigate the physical structure traces by HNCO and SiO, we selected two regions (R1, R2) in the east of CND, two regions (R3, R4) in the west of CND, and one region centered at the AGN, each of $0''.8\times0''.8$ size to match the lowest angular resolution among our data. 
The selection criterion for these five regions (AGN and CND R1-R4) is based on the emission peaks in our data, locally and above the {$3.0 \sigma$} threshold for HNCO lines, and lower-J (3-2 and 2-1) SiO lines (see Section \ref{sec:mom0}). 
A {$3.0 \sigma$} threshold is chosen to be reasonably inclusive of weak signals, especially for the HNCO lines. 
Table \ref{tab:table_7regs} lists these five selected positions within NGC 1068 with their coordinates, which are the center of the individual $0''.8\times0''.8$ apertures. 
In Figure \ref{fig:Regs_schematic} the schematic of the relative layout of these regions on the CND is shown. 
Compared to the chemically distinct regions identified by \citet{Viti+2014}, the spots on the east (CND-R1 and East Knot) and the west (CND-R3 and West Knot) are quite close, although there is a minor vertical offset due to the selection of local peaks, especially in the HNCO transitions, which causes further offsets for the CND-R2 and CND-R4 from the previously highlighted north and south regions by \citet{Viti+2014}. 

Figures \ref{fig:allHs_spec}-\ref{fig:allSs_spec3} show the spectra for all transitions used in this study in unit of [K], with all the [mJy beam\textsuperscript{-1}] to [K] conversion factors listed in Table \ref{tab:table_obsinfo}, from the five selected positions listed in Table \ref{tab:table_7regs}.  
The observed HNCO(4-3) and (5-4) spectra are in Figure \ref{fig:allHs_spec}, HNCO(6-5) and SiO(2-1) spectra in Figure \ref{fig:allSs_spec1}, SiO(3-2) and SiO(5-4) in Figure \ref{fig:allSs_spec2}, and SiO(6-5) and SiO(7-6) in Figure \ref{fig:allSs_spec3}
These spectral data are extracted from the data cube in their own original spatial and spectral resolution, and then averaged over a common-size aperture of $0''.8\times0''.8$ box centering at each selected position. 
All the adjacent bonus lines are also displayed, labelled by the blue dashed vertical lines in the spectra as opposed to the targeted HNCO and SiO lines in blue, solid vertical lines (including the splittings). 
Based on the strongest spectral line features in all transitions, the estimated line width is about $100$ km/s, which is slightly smaller but consistent with the estimate of the line width in the CND region of NGC 1068 from previous studies \citep{Kelly+2017}. 

The spot CND-R1 which is at the east-side of CND shows a very strong signal across all SiO transitions, and this is in  contrast to the rest of CND where the SiO emission is much weaker. 
On the other hand, HNCO is much more evenly distributed throughout the CND. 
This will be more obvious in the intensity maps presented later in Section \ref{sec:mom0}, and is in broad agreement  with the findings from the lower-resolution observations towards the AGN-host galaxy NGC 1097 \citep{Martin+2015_shocktracer_ngc1097}. 
For all three HNCO transitions there is prominent line emission on both sides (east and west) of the CND. 

\subsection{Moment-0 maps}
\label{sec:mom0}
\begin{figure*}
  \centering
  \begin{tabular}[b]{@{}p{0.45\textwidth}@{}}
    \centering\includegraphics[width=1.0\linewidth]{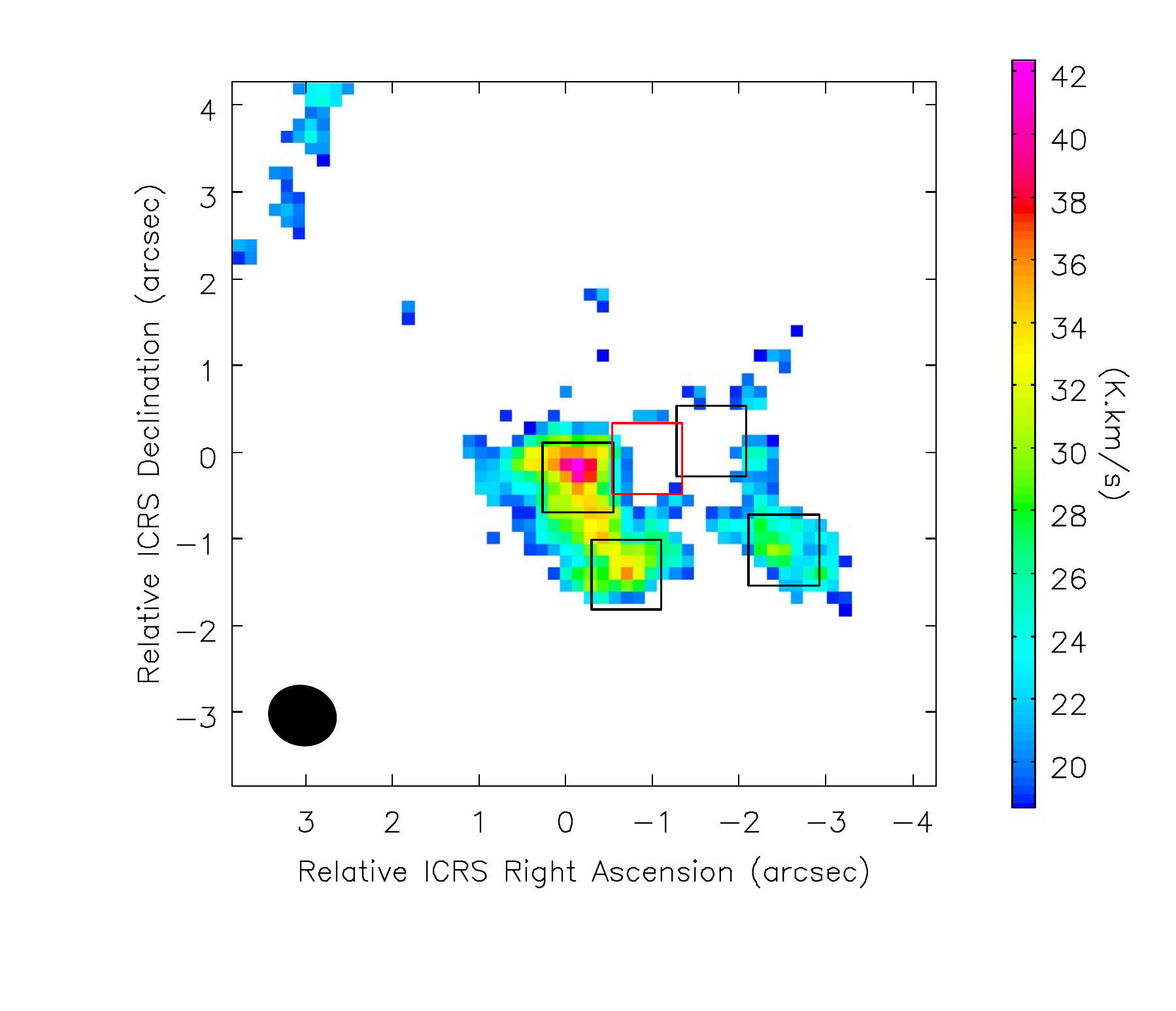} \\
    \centering\small (a) 
  \end{tabular}%
  \quad
  \begin{tabular}[b]{@{}p{0.45\textwidth}@{}}
    \centering\includegraphics[width=1.0\linewidth]{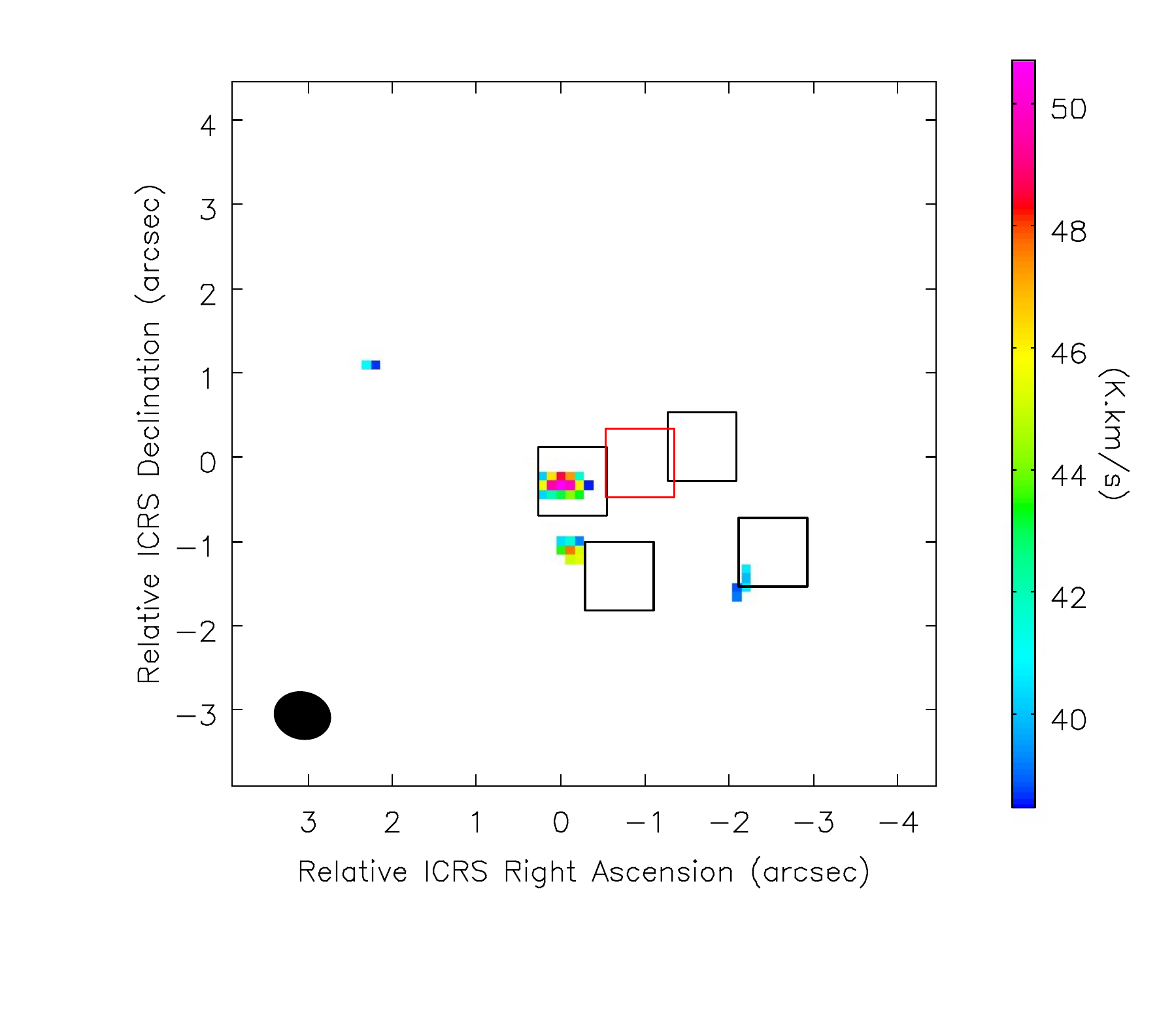} \\
    \centering\small (b) 
  \end{tabular}
  \begin{tabular}[b]{@{}p{0.45\textwidth}@{}}
    \centering\includegraphics[width=1.0\linewidth]{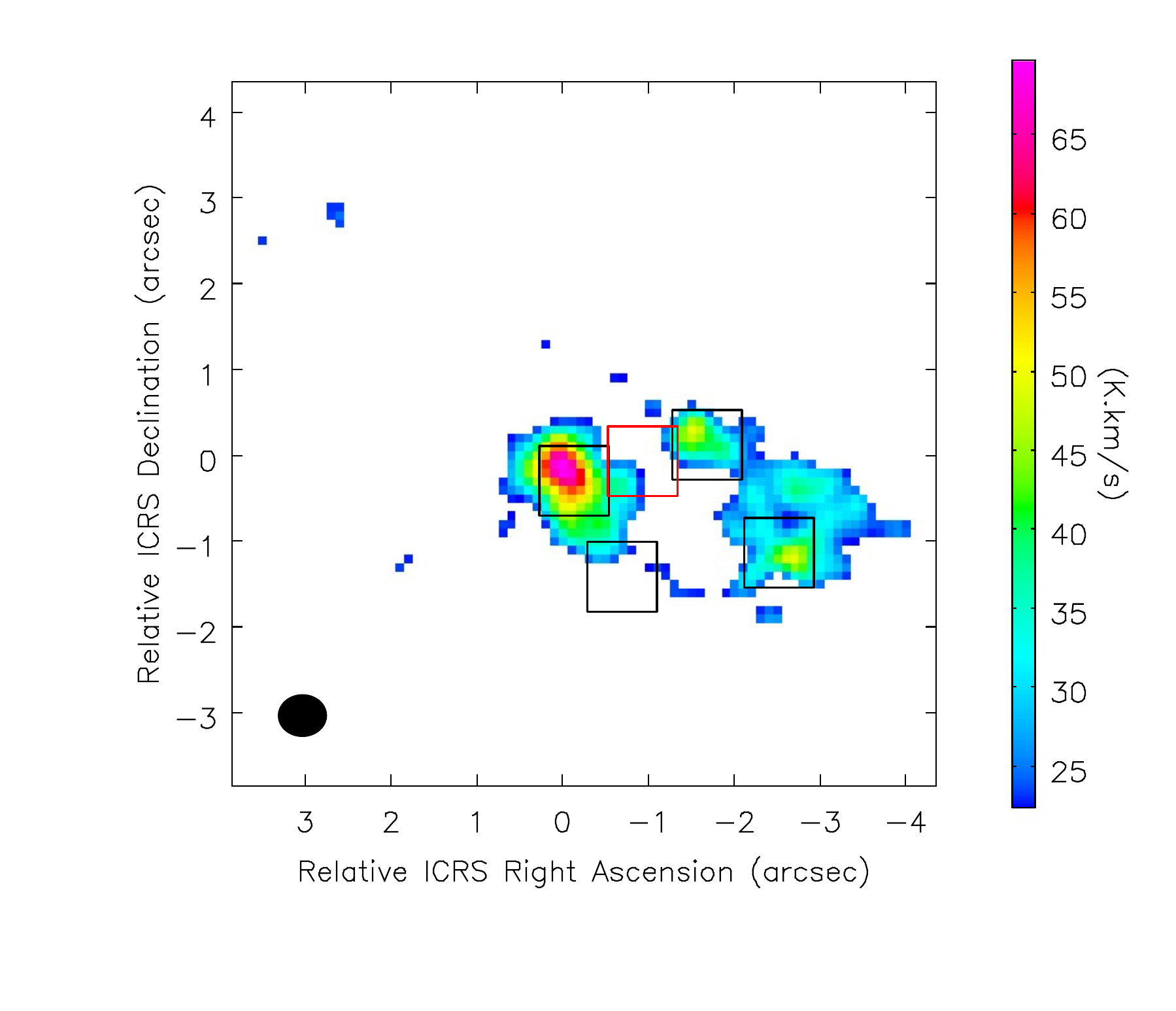} \\
    \centering\small (c) 
  \end{tabular}
  \caption{The velocity-integrated intensity maps of (a) HNCO(4-3), (b) HNCO(5-4), and (c) HNCO(6-5) at their  original spatial resolution. The black and red boxes mark the regions listed in Table \ref{tab:table_7regs}, where AGN is marked with the red box. These maps are masked with a {$3.0 \sigma$} threshold after the integration over velocity. The integrated velocity range for HNCO(4-3) and HNCO(6-5) are [-230km s\textsuperscript{-1}, 230km s\textsuperscript{-1}], and HNCO(5-4) with [-230km s\textsuperscript{-1}, 167km s\textsuperscript{-1}] so that to exclude the nearby line. }
  \label{fig:mom0_hnco_zoomin}
\end{figure*}

\begin{figure*}
  \centering
  \begin{tabular}[b]{@{}p{0.45\textwidth}@{}}
    \centering\includegraphics[width=1.0\linewidth]{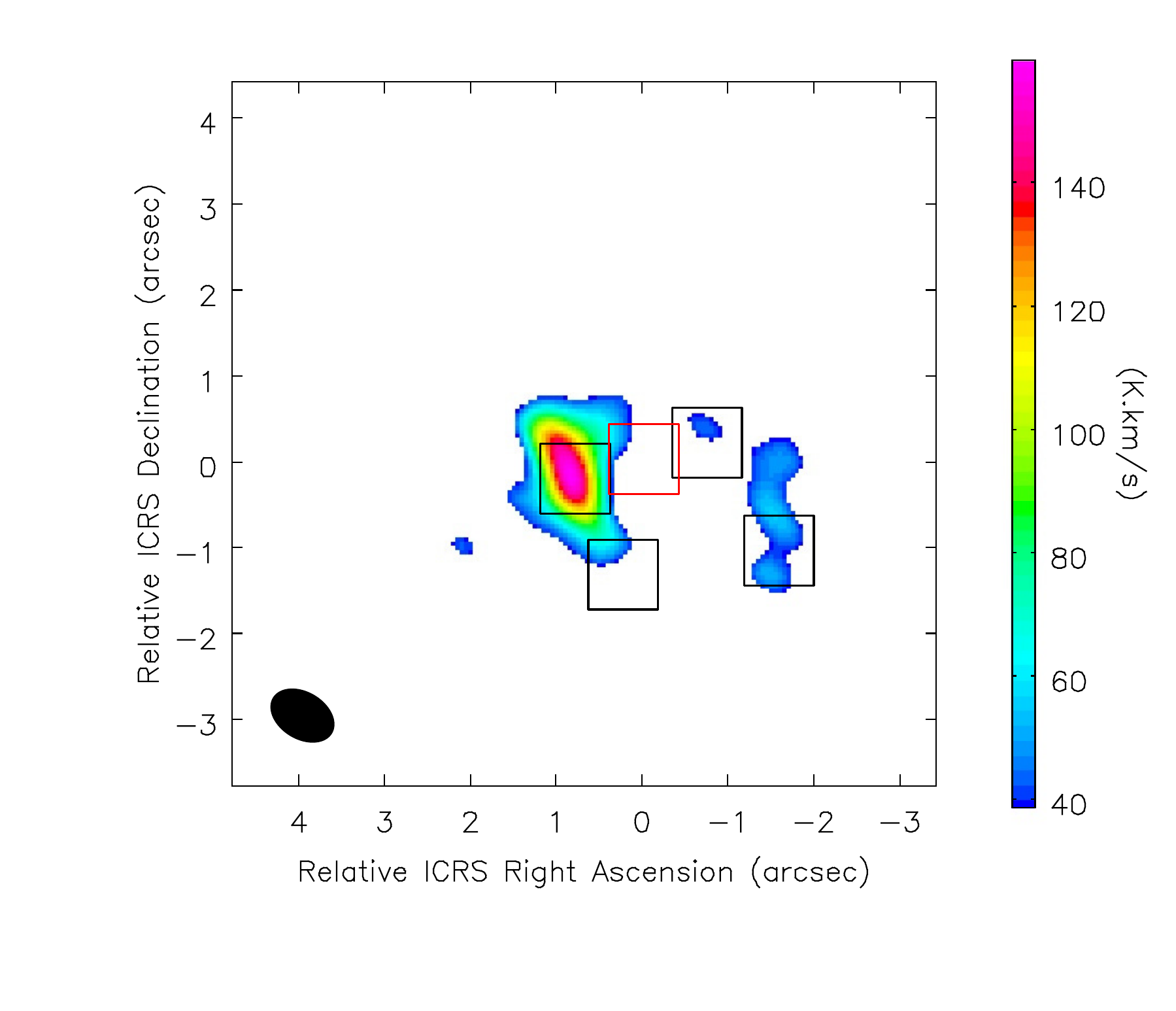} \\
    \centering\small (a) 
  \end{tabular}%
  \quad
  \begin{tabular}[b]{@{}p{0.45\textwidth}@{}}
    \centering\includegraphics[width=1.0\linewidth]{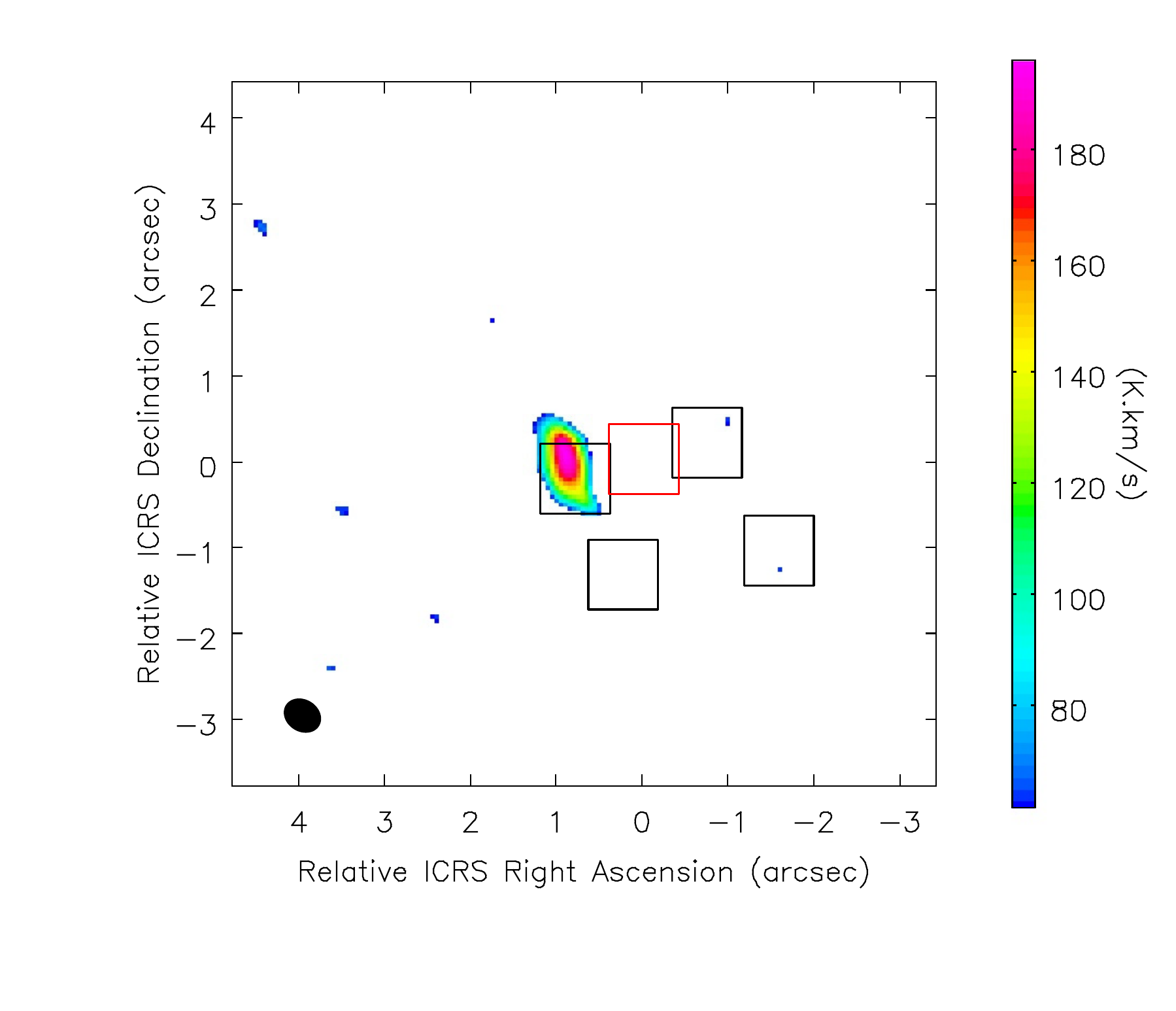} \\
    \centering\small (b) 
  \end{tabular}
  \begin{tabular}[b]{@{}p{0.45\textwidth}@{}}
    \centering\includegraphics[width=1.0\linewidth]{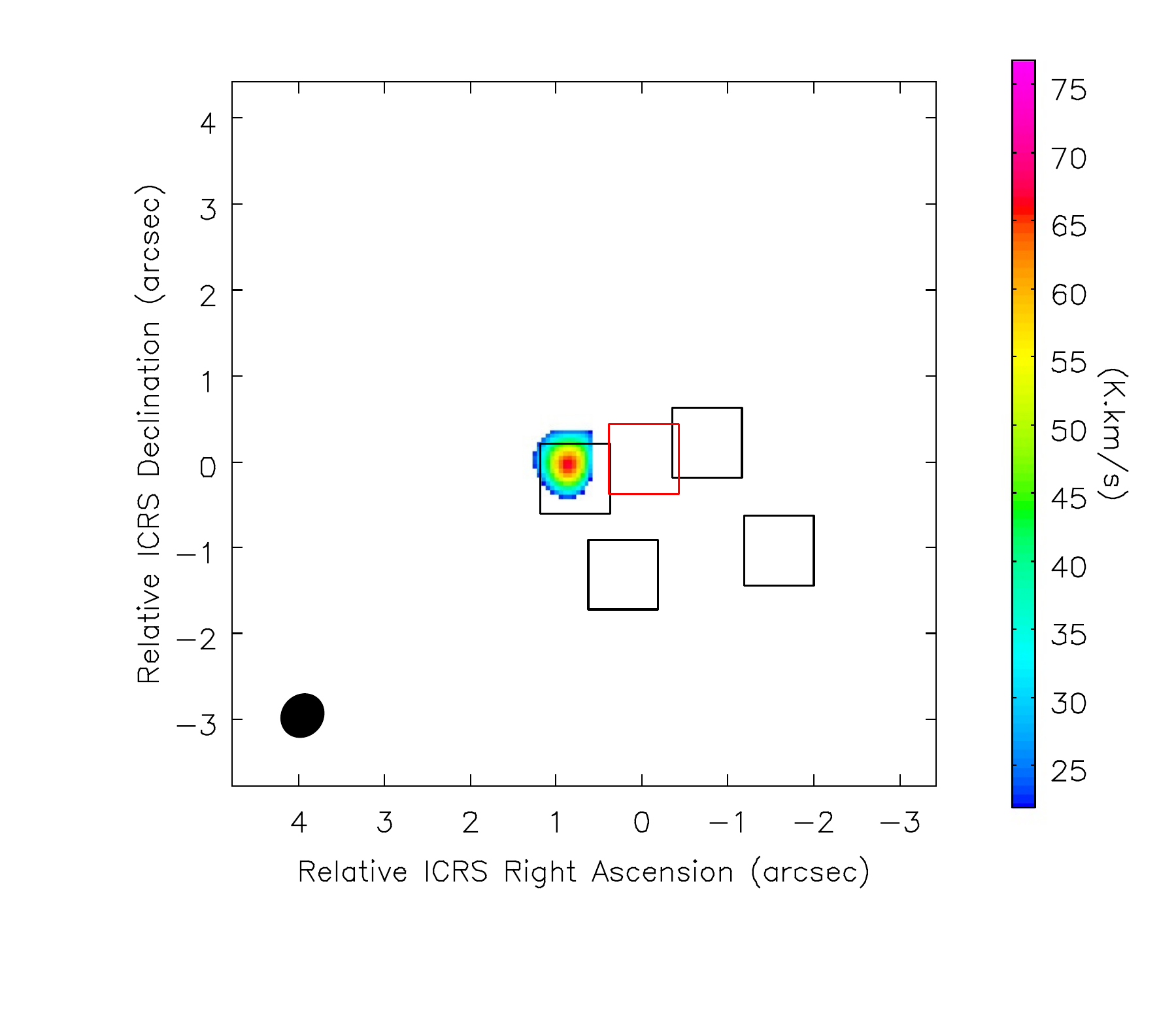} \\
    \centering\small (c) 
  \end{tabular}
  \begin{tabular}[b]{@{}p{0.45\textwidth}@{}}
    \centering\includegraphics[width=1.0\linewidth]{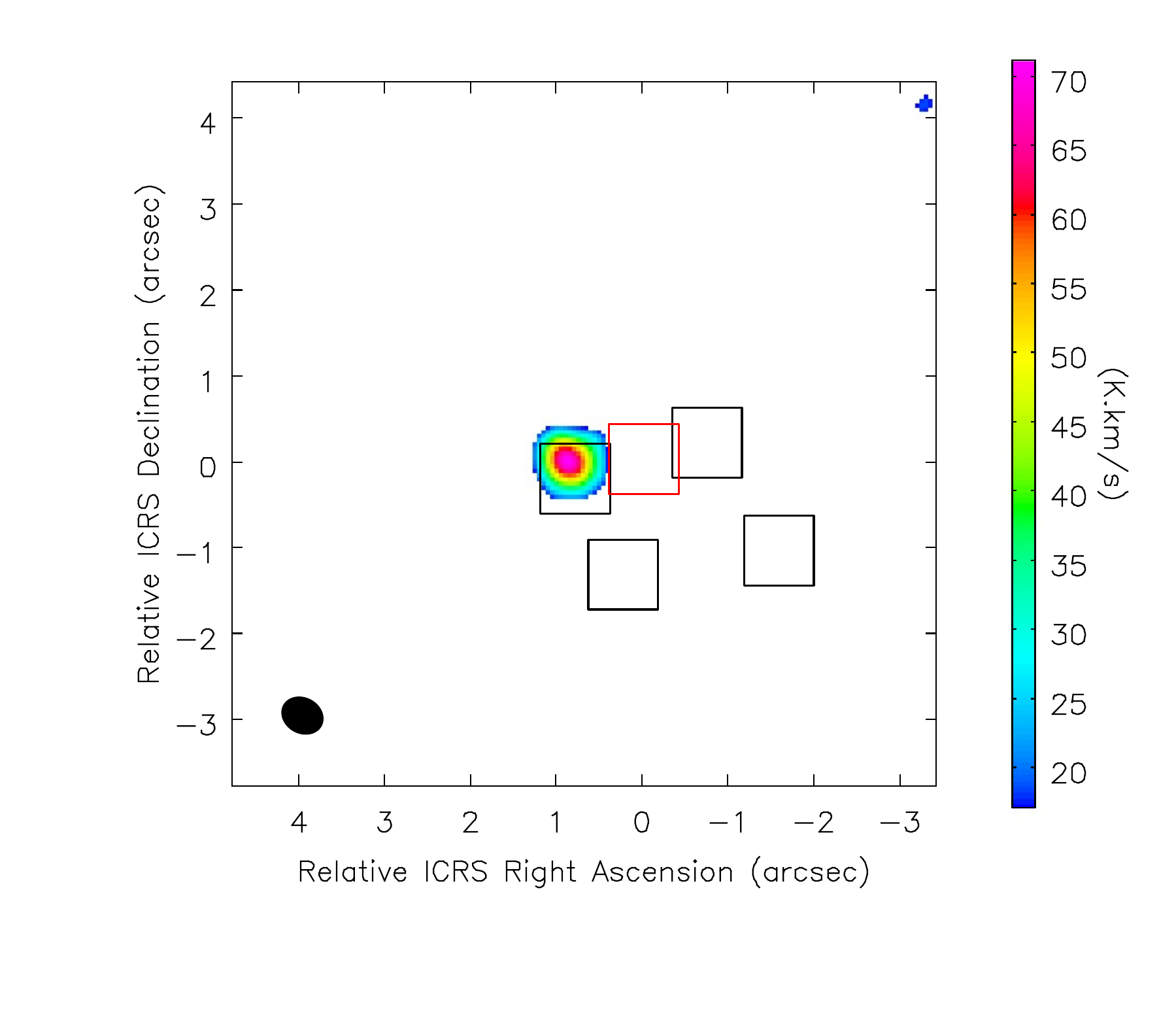} \\
    \centering\small (d) 
  \end{tabular}
  \begin{tabular}[b]{@{}p{0.45\textwidth}@{}}
    \centering\includegraphics[width=1.0\linewidth]{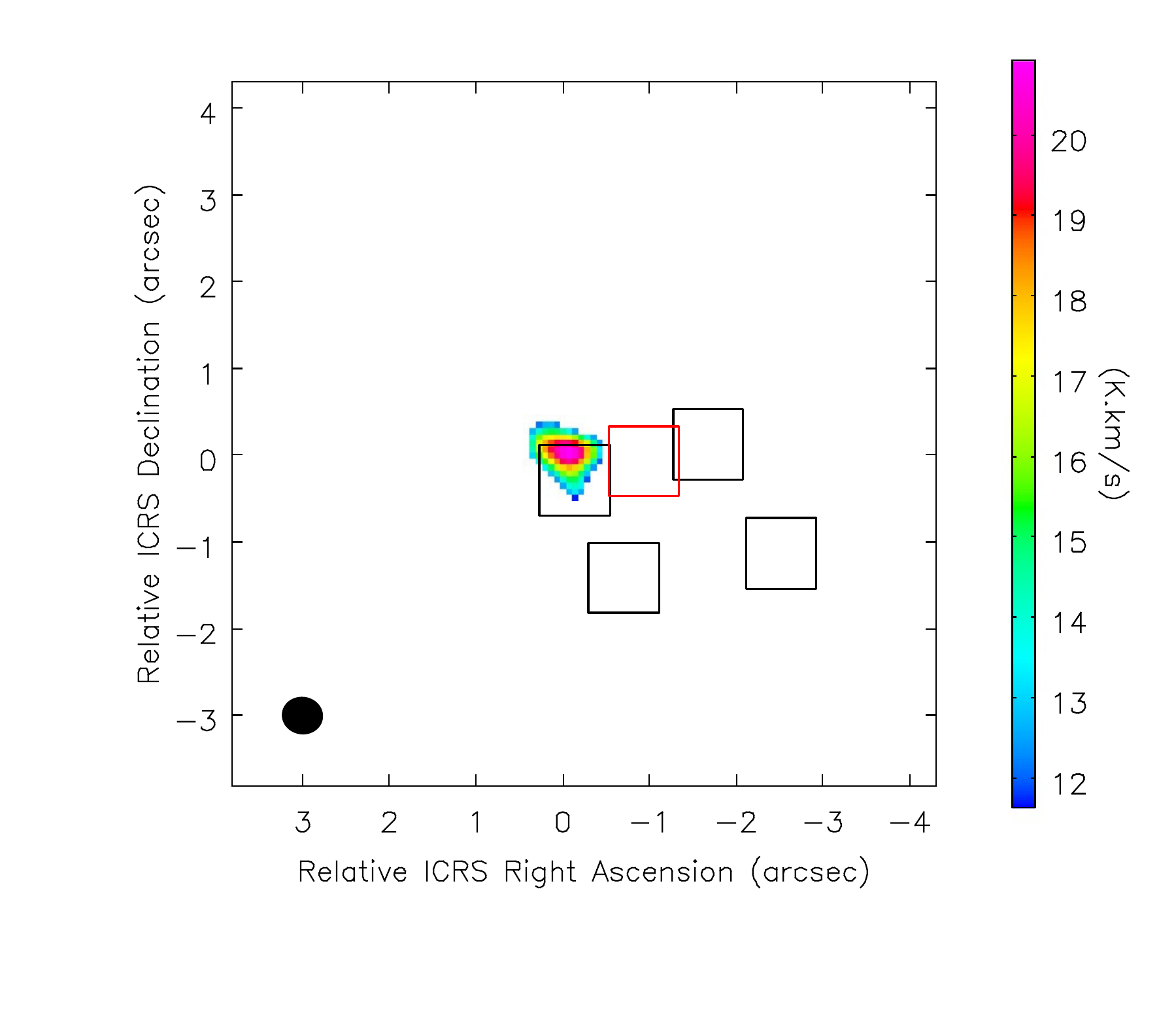} \\
    \centering\small (e) 
  \end{tabular}
  \caption{The velocity-integrated intensity maps of (a) SiO(2-1), (b) SiO(3-2), (c) SiO(5-4), (d) SiO(6-5), (e) SiO(7-6) at their  original spatial resolution. This is zoomed-in at the CND scale. The black and red boxes mark the regions listed in Table \ref{tab:table_7regs}, where the AGN is marked with the red box. These maps are masked with a {$3.0 \sigma$} threshold after the integration over velocity. The velocity ranges for integration are: SiO(2-1) with [-230km s\textsuperscript{-1}, 160km s\textsuperscript{-1}], SiO(6-5) with [-230km s\textsuperscript{-1}, 151km s\textsuperscript{-1}], and SiO(7-6) with [-74km s\textsuperscript{-1}, 230km s\textsuperscript{-1}], and for the rest is the default [-230km s\textsuperscript{-1}, 230km s\textsuperscript{-1}]}
  \label{fig:mom0_sio_zoomin}
\end{figure*}

Figure \ref{fig:mom0_hnco_zoomin} shows the velocity-integrated intensity maps at each original spatial resolution of HNCO(4-3), HNCO(5-4), and HNCO(6-5) at the CND scale. 
Figure \ref{fig:mom0_sio_zoomin} shows the velocity-integrated intensity maps at the original spatial resolution for SiO(2-1), SiO(3-2), SiO(5-4), SiO(6-5), and SiO(7-6) at the CND scale. 
These line-intensity maps were integrated over velocity, and a {$3.0 \sigma$} threshold clipping was applied to the moment-0 maps. 
The line fluxes were integrated to include any significant emission that arises over the velocity span due to rotation and outflow motions in NGC 1068. 
Here we used $|v-v_{sys}^{LSRK} |\leq$ 230 km s\textsuperscript{-1} to cover such range,  except for lines that have adjacent transitions that step in to such velocity span. 
In the latter case, which involves potential line contamination, the velocity integration was performed in a narrower range to the midpoint between the target line and the adjacent line, instead of performing further spectral fitting in an attempt to disentangle the multiple molecular transitions, for two reasons: 
(1) HNCO transitions actually involve multiple splitting components per transition (e.g. HNCO (4-3) line), which makes it much more complicated for the spectral fitting; and 
(2) there is no serious overlapping/contamination over different molecular transitions, thus narrowing the velocity range  was acceptable for our purpose. 
For the HNCO transitions that involve multiple fine and/or hyperfine splittings, the velocity is referring to the frequency associated with the expected strongest component. 
These lines with narrowed velocity coverage are: HNCO(5-4) with [-230km s\textsuperscript{-1}, 167km s\textsuperscript{-1}], SiO(2-1) with [-230km s\textsuperscript{-1}, 160km s\textsuperscript{-1}], SiO(6-5) with [-230km s\textsuperscript{-1}, 151km s\textsuperscript{-1}], and SiO(7-6) with [-74km s\textsuperscript{-1}, 230km s\textsuperscript{-1}]. 
The velocity covered to derive the velocity-integrated line intensities are also displayed in shaded blue in the individual spectra shown in Figure \ref{fig:allHs_spec}-\ref{fig:allSs_spec3}. 
{We note that the chosen velocity range may cover multiple gas  components (e.g. the double-peak feature in CND-R3 of both HNCO(5-4) (middle panel) and HNCO(6-5) (right panel) in Figure \ref{fig:allHs_spec}-\ref{fig:allSs_spec1}). For the purpose of this study, we simply aim at investigating the properties of the averaged gas within the beam.  }

{The CND ring structure traced by HNCO and SiO in the velocity-integrated intensity maps is noticeably off-centered from the inferred location of the AGN in the literature \citep{Roy+1998,GB+2014}, but coincides with the CO observations at similar and higher spatial resolution \citep{GB+2014,Viti+2014,GB+2019}, 
dust continuum \citep{GB+2014}, and observations of the dense gas tracers such as HCN and HCO\textsuperscript{+} \citep{Krips+2011,GB+2014,Imanishi+2016}. 
The contribution from atomic hydrogen to the total neutral gas content is negligible in the central $\sim2$ kpc of the disk \citep{Brinks+1997,GB+2014,GB+2017} compared to the molecular counterpart. 
This off-centered ring morphology could be an indication that the ring is possibly being shaped by the feedback of nuclear activity \citep{GB+2019}. 
Moreover, the CND ring revealed by our HNCO and SiO emissions shows substructure with several knots along the ring morphology. }
At the CND, the emission from all the transitions peaks near the R1 region. 
In particular for HNCO(6-5), such trend is in contrast to what was found in the previous work by \citet{Kelly+2017} where the strongest HNCO(6-5)  peaked at the CND west side. 
The strongest HNCO(6-5) emission ($69.79$ K km s\textsuperscript{-1}) is in the east of the CND, and this intensity is comparable to the peak intensity of the HNCO(6-5) emission of $60$ K km s\textsuperscript{-1} reported by \citet{Kelly+2017}, although the latter occurred in the west of the CND. 
The peak emission of HNCO(4-3) and HNCO(5-4) are also at the east side of CND, at $42.41$ K km s\textsuperscript{-1} and $50.73$ K km s\textsuperscript{-1} respectively. 

Compared to the SiO transitions, which are mostly centered around CND-R1, the HNCO(4-3) and HNCO(6-5) emission appears to be more evenly extended  to the west side throughout the entire CND. 
This is indeed similar to the findings of \citet{Martin+2015_shocktracer_ngc1097} obtained  from lower-resolution observations towards NGC 1097, where HNCO was found to be more extended on its CND, and SiO was found more likely as an unresolved source. 

The two emission knots (east and west) are connected by a "bridge" structure of weaker emission  that can be seen on both the north and south side of the AGN. 
In both HNCO(4-3) and HNCO(6-5) there are several local knots in this "bridge", part of the CND ring (e.g. CND-R2 and CND-R4). 
Intriguingly HNCO(4-3) shows more prominent "bridge" emission in the south bridge, while HNCO(6-5) appears more prominent in the north bridge. 
HNCO(5-4) also shows comparable emission on both sides of the CND; however, probably due to the higher noise level for this data set, most of the regions are below the threshold. 
\subsection{Molecular line ratios and overlay maps}
\label{sec:common0.8}
In this section we analyse selected line ratios at our observed locations in the CND of NGC 1068. 
Prior to deriving the ratio and overlay maps, all the {unmasked line intensity maps} were smoothed\footnote{Using CASA task "imsmooth" with parameter "targetres=True" specified to used the indicated beam geometry as the final aimed, the lower "common" resolution shared in any given transition pair. } to the lowest common resolution across the data, $0''.8\times0''.8$, re-projected to the same spatial grids 
\footnote{Using CASA task "imregrid"} in order to have a consistent map coordinate, and then we masked out pixels below the {$3.0 \sigma$} threshold for further image analysis. 
{In general, the noise will also be suppressed when performing the image smoothing and some weak signals may reach  the cutoff threshold (e.g. {$3.0 \sigma$}). Hence, we first smooth and re-project the unmasked intensity maps, followed by the re-sampling of the the noise and the mask application on the new images. }
From this section onward, all the data products have been spatially degraded to this common resolution ($0''.8$) and processed according to the description above, for all transitions. 

Differences in line intensity ratios  can arise from several physical processes such as differences in systematic gas density and temperature, different radiation fields (e.g UV, X-rays), and the presence and type of shocks \citep[see detailed discussion in][]{Krips+2008}. 
Therefore molecular ratios can be powerful diagnostics in determining the physical characteristics as well as the energetic processes of a galaxy. 
In section \ref{sec:mom0}, we highlighted that the HNCO transitions appear to be more evenly distributed over all CND regions, in  contrast to the SiO emission which is mostly in the East-CND region. 
By inspecting both molecular-line ratio maps and overlay maps, we aim to systematically compare such differences between the two molecular tracers in a morphological and quantitative way. 
{Quantitatively, the uncertainty of the intensity ratio can be estimated through standard error propagation from the uncertainty measured in the line intensity in the ideal case. With a threshold $\geq 3.0 \sigma$ for the quantities involved, the uncertainty of the ratio is  $\leq 47\%$ of the ratio itself. 
We note that for ratios near $1.0$, such uncertainty may reverse the ratio from below to above 1 (or vice versa). }

\begin{figure}
  \centering
  \begin{tabular}[b]{@{}p{0.5\textwidth}@{}}
    \centering\includegraphics[width=1.0\linewidth]{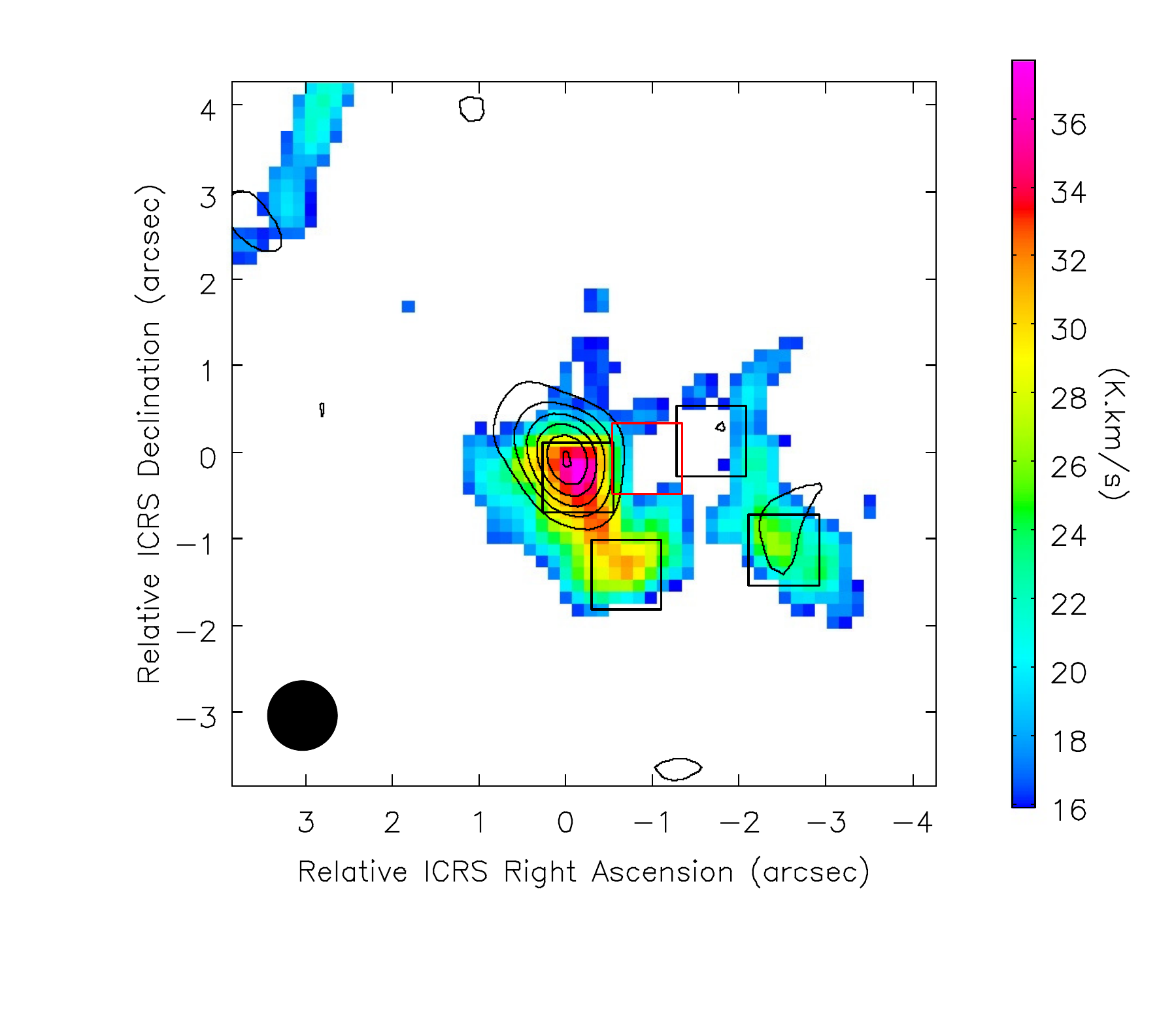} \\
    \centering\small (a) Overlay map of HNCO(4-3) and SiO(3-2) at CND scale. 
  \end{tabular}
  \begin{tabular}[b]{@{}p{0.5\textwidth}@{}}
    \centering\includegraphics[width=1.0\linewidth]{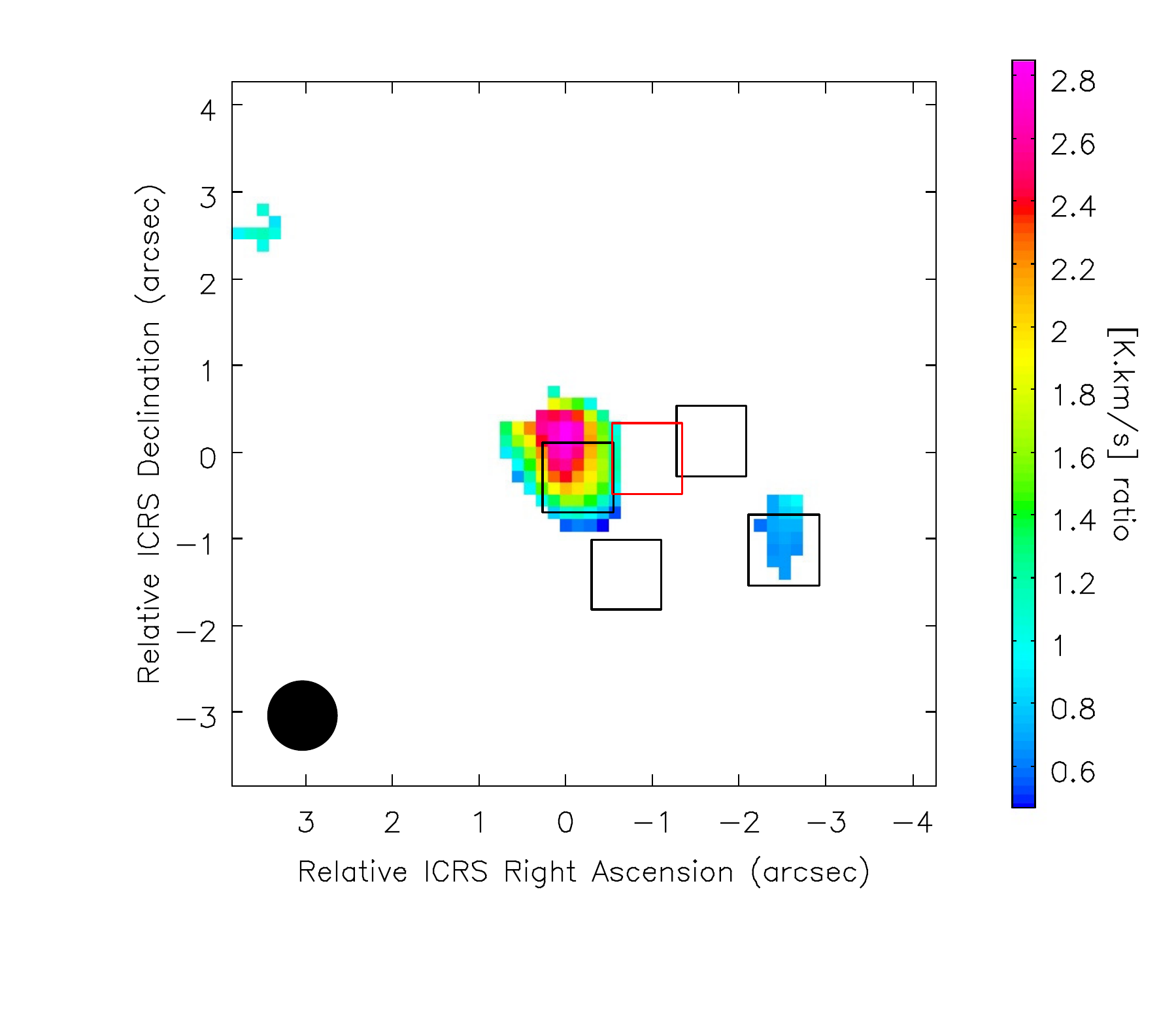} \\
    \centering\small (b) Ratio map of  SiO(3-2)/HNCO(4-3) at CND scale. 
  \end{tabular}
  \caption{The comparison maps between  SiO(3-2) and HNCO(4-3)  in overlay and ratio maps. In the overlay(a) map the HNCO(4-3) is shown in color map, and SiO(3-2) in contours. The contour starts from {$3.0 \sigma$}, with stpdf of {$3.0 \sigma$}. In (b) the SiO(3-2)/HNCO(4-3) ratio map is displayed.  The black and red square boxes mark the $0''.8\times0''.8$ regions listed in Table \ref{tab:table_7regs}, where the AGN is marked in red. The contour starts from {$3.0 \sigma$}, with stpdf of {$3.0 \sigma$}. These maps were pre-smoothed to the common resolution of $0''.8$ and masked by a {$3.0 \sigma$} threshold. }
  \label{fig:overlay_n_ratio_h43s32}
\end{figure}

\begin{figure}
  \centering
  \begin{tabular}[b]{@{}p{0.5\textwidth}@{}}
    \centering\includegraphics[width=1.0\linewidth]{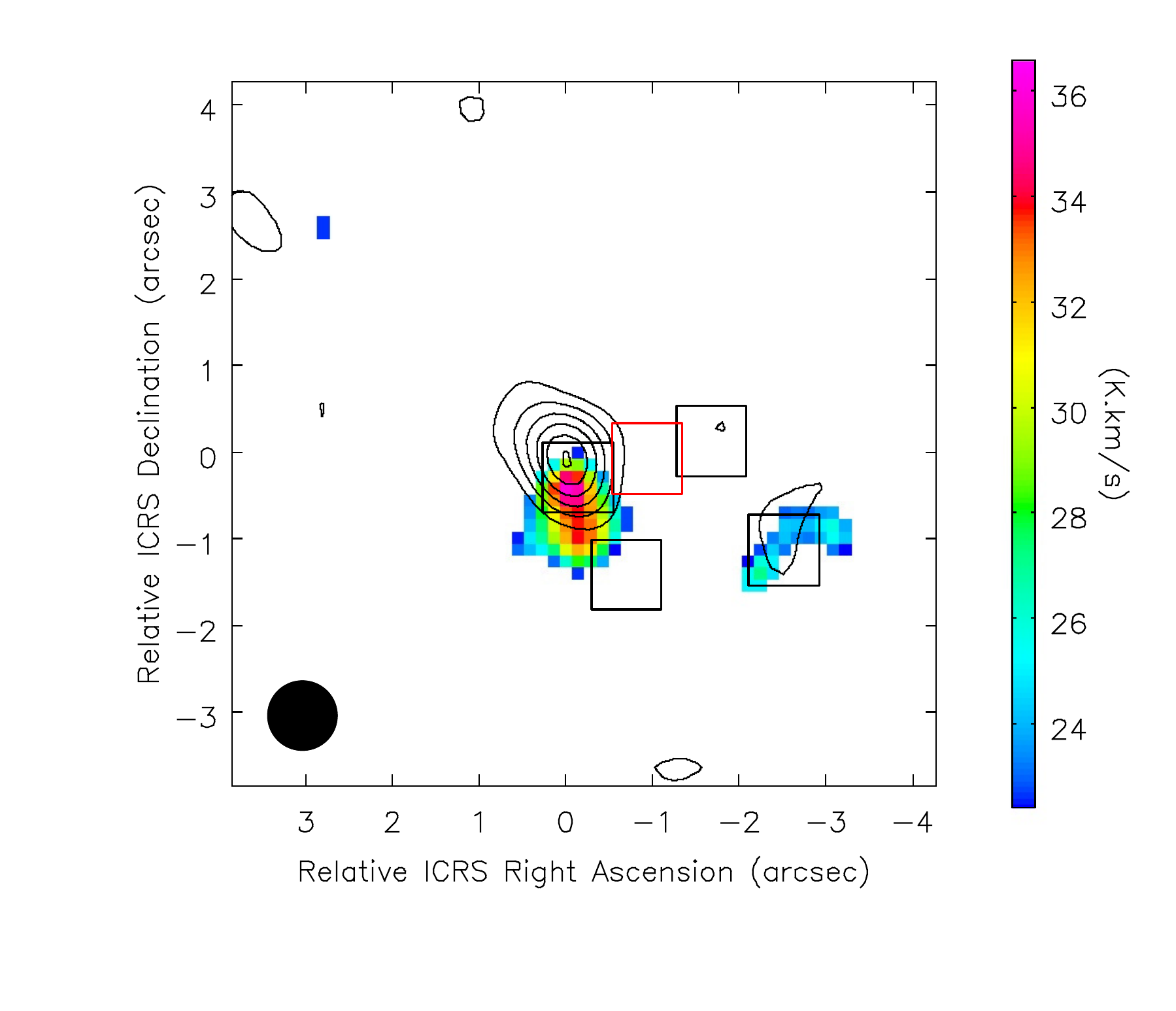} \\
    \centering\small (a) Overlay map of HNCO(5-4) and SiO(3-2) at CND scale.  
  \end{tabular}
  \begin{tabular}[b]{@{}p{0.5\textwidth}@{}}
    \centering\includegraphics[width=1.0\linewidth]{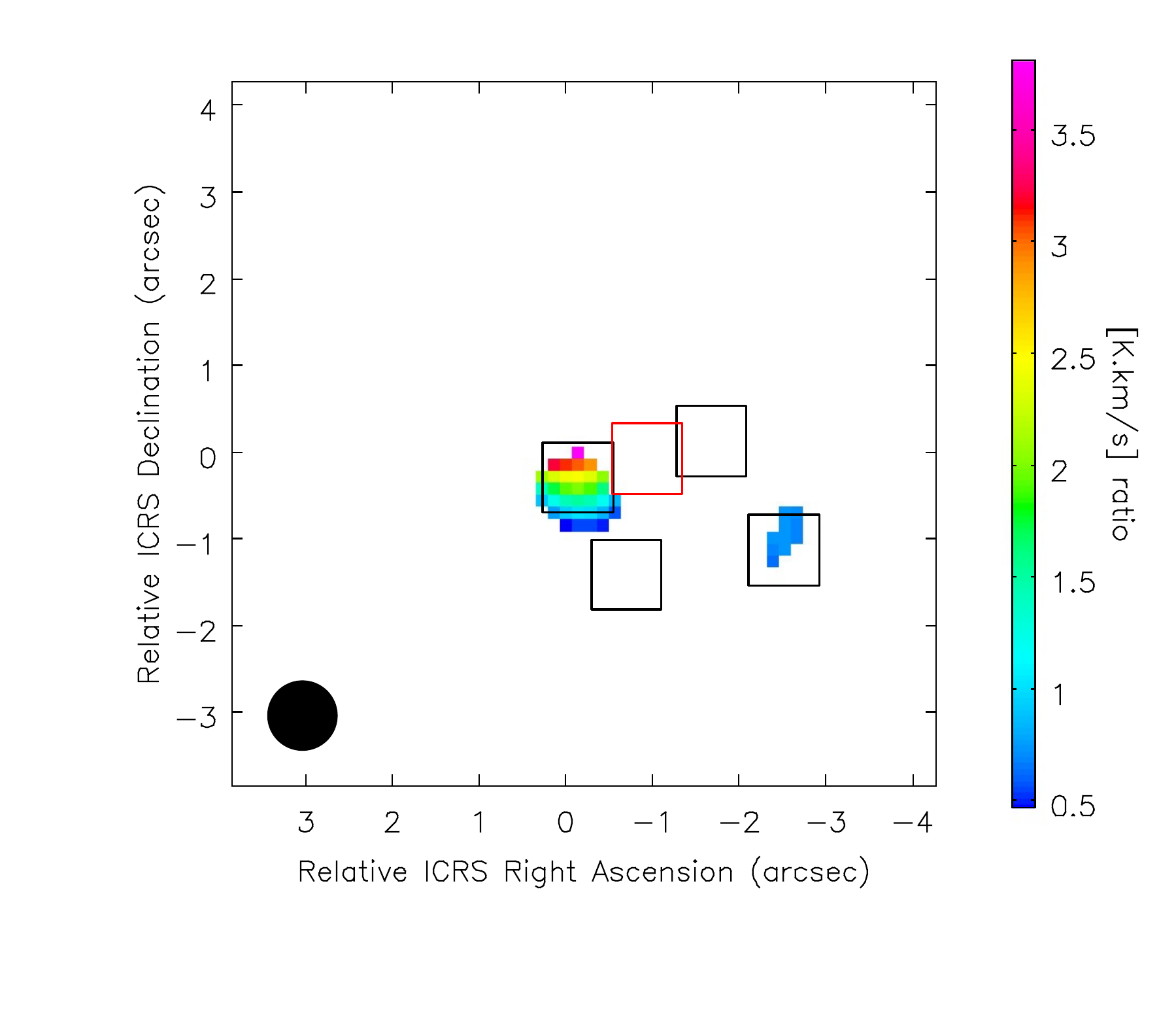} \\
    \centering\small (b) Ratio map of HNCO(5-4) and SiO(3-2) at CND scale. 
  \end{tabular}
  \caption{The comparison maps between  SiO(3-2) and HNCO(5-4) in overlay and ratio maps. In the overlay(a) map the HNCO(5-4) is shown in color map, and SiO(3-2) in contours. The contour starts from {$3.0 \sigma$}, with stpdf of {$3.0 \sigma$}. In (b) the SiO(3-2)/HNCO(5-4) ratio map is displayed. The black and red square boxes mark the $0''.8\times0''.8$ regions listed in Table \ref{tab:table_7regs}, where the AGN is marked in red. The contour starts from {$3.0 \sigma$}, with stpdf of {$3.0 \sigma$}. These maps were pre-smoothed to the common resolution of $0''.8$ and masked by {$3.0 \sigma$} threshold. }
  \label{fig:overlay_n_ratio_h54s32}
\end{figure}

\begin{figure}
  \centering
  \begin{tabular}[b]{@{}p{0.5\textwidth}@{}}
    \centering\includegraphics[width=1.0\linewidth]{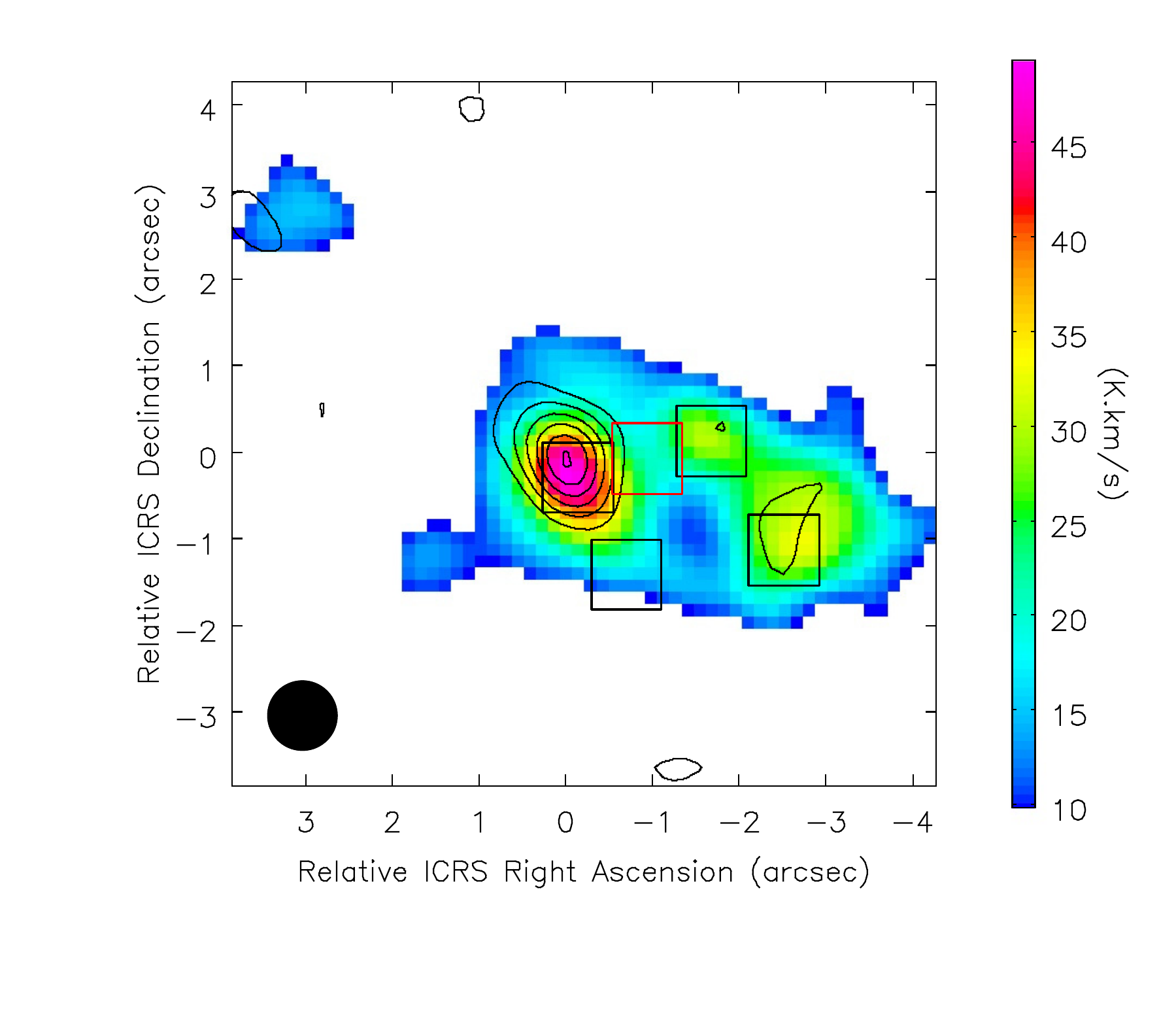} \\
    \centering\small (a) Overlay map of HNCO(6-5) and SiO(3-2) at CND scale. 
  \end{tabular}
  \begin{tabular}[b]{@{}p{0.5\textwidth}@{}}
    \centering\includegraphics[width=1.0\linewidth]{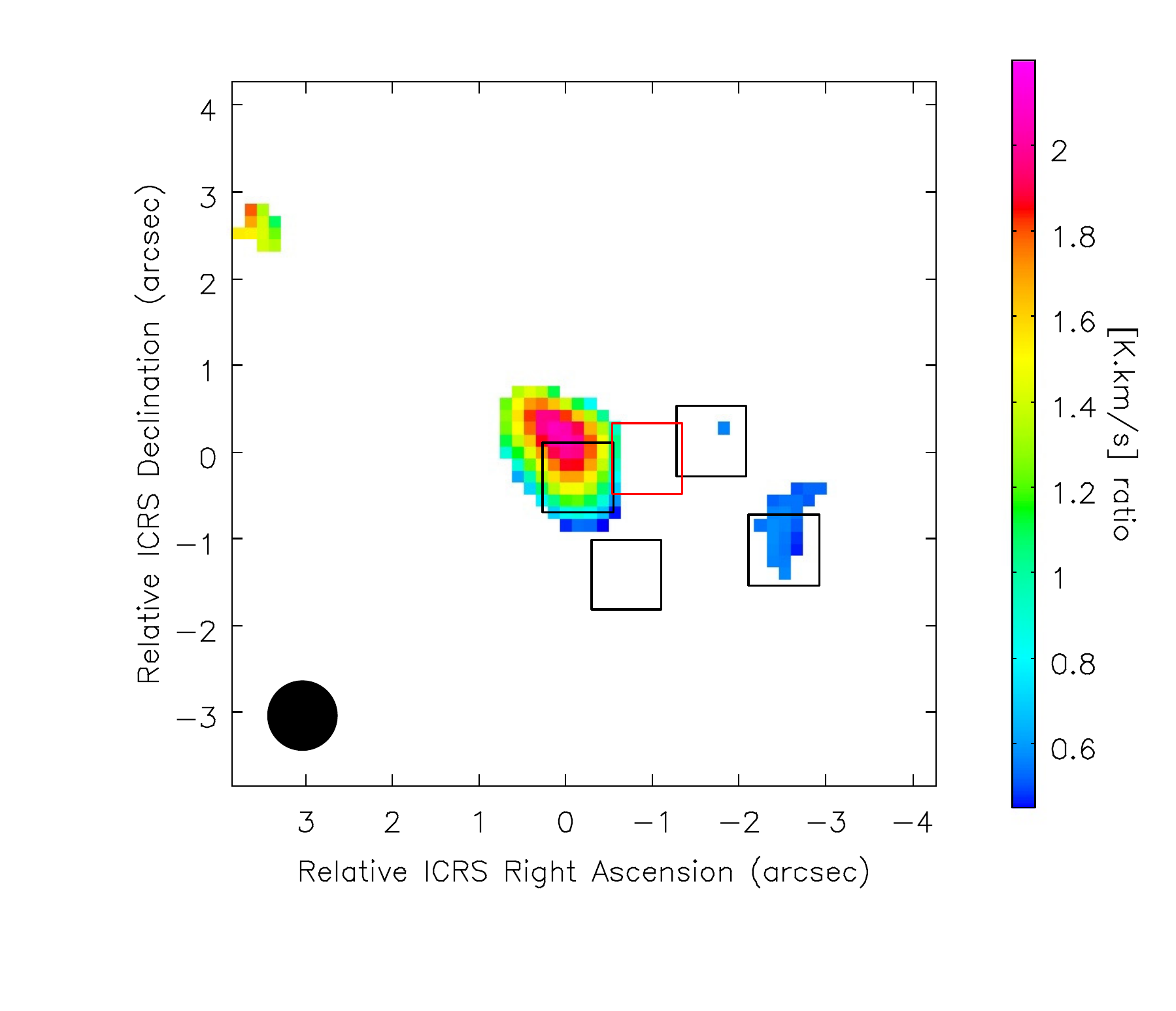} \\
    \centering\small (b) Ratio map of HNCO(6-5) and SiO(3-2) at CND scale. 
  \end{tabular}
  \caption{The comparison maps between SiO(3-2) and HNCO(6-5) in overlay and ratio maps. In the overlay(a) map the HNCO(6-5) is shown in color map, and SiO(3-2) in contours. The contour starts from {$3.0 \sigma$}, with stpdf of {$3.0 \sigma$}. In (b) the SiO(3-2)/HNCO(6-5) ratio map is displayed. The black and red square boxes mark the $0''.8\times0''.8$ regions listed in Table \ref{tab:table_7regs}, where the AGN is marked in red. The contour starts from {$3.0 \sigma$}, with stpdf of {$3.0 \sigma$}. These maps were pre-smoothed to the common resolution of $0''.8$ and masked by {$3.0 \sigma$} threshold. }
  \label{fig:overlay_n_ratio_h65s32}
\end{figure}

In Figures \ref{fig:overlay_n_ratio_h43s32} to \ref{fig:overlay_n_ratio_h65s32} we highlight few transition "pairs": HNCO(4-3)/HNCO(5-4)/HNCO(6-5) versus SiO(3-2) to show both the molecular line ratio and overlay maps of the velocity-integrated intensities in $\int T_{mb} \cdot dv$ (K km s\textsuperscript{-1}). 
{The rest of the SiO/HNCO ratios  show similar trends. 
The selected transition pairs  are few of the most prominent examples demonstrating the trend in line ratios from the east to the west of the CND. 
In particular, we chose to show the ratio HNCO(6-5)/SiO(3-2) as this was the ratio that was modelled in \citet{Kelly+2017} in an attempt to characterize the gas properties traced by these two molecular species. }
In the work by \citet{Kelly+2017}, the PdBI observations of the HNCO(6-5) and SiO(3-2) transitions were compared and it was found that in the overlay map \citep[Figure 4 in][]{Kelly+2017} there is a noticeable spatial offset in the both east and west local peaks between these two transitions. 
As shown in Figure \ref{fig:overlay_n_ratio_h65s32}, we find that for the same transition pair,  the spatial offset between the HNCO(6-5) and SiO(3-2) peak emissions  on both east and west CND is actually not as large as (both $\leq 0''.2$) identified in \citet{Kelly+2017} which reported it to be $\sim0''.4-0''.9$. 

On the other hand, the general SiO(3-2)/HNCO ratios do show a noticeable gradient on the plane of sky across the CND, going from $\geq 1.0$ on the east side to $\leq 1.0$ on the west side. 
This confirms the same trend of chemical differentiation highlighted by \citet{Kelly+2017} across the CND, which may arise from their different chemical "origins" such as the shock chemistry involved in different types of shocks, e.g. fast and slow shocks. 
This may indicate different velocity regimes in the shock fronts of the molecular outflow. 
As mentioned earlier, \citet{Kelly+2017} performed a chemical modeling for the SiO and HNCO emission, and confirmed that fast shocks ($60$ km s\textsuperscript{-1}) are likely to be producing SiO; while HNCO can be associated with slow-shock ($20$ km s\textsuperscript{-1}) chemistry, or simply with the mantle sublimation in the gas that is warm, dense and non-shocked. 
Further characterization of the gas properties will be the key to verifying the perspective provided by the chemical modeling, and to the disentanglement of the gas condition traced by HNCO. 
\section{Characterizing the physical properties of the gas in the CND}
\label{sec:gas_properties}
The relative configuration of the CND to the geometry of the jet-ISM interaction and the outflow provides us with the opportunity to explore shock-driven chemistry related to the molecular outflow, for the large spread of velocities is likely driving different shock chemistry signatures at different locations in the CND \citep{GB+2014,Viti+2014,Kelly+2017,GB+2017}. 
One of our main goals in the current work is to better characterize the gas properties that SiO and HNCO each traces across the CND regions, using a larger number of transitions, observed at higher spatial resolution ($\sim0''.5-0''.8$) than the previous study ($\sim 1''-4''$) by \citet{Kelly+2017}. 
Ultimately, here we explore whether the observed SiO-HNCO properties can be used as a keen probe to the variation of the shocked gas properties across the CND. 

To characterize the SiO and HNCO emissions across the CND region, we performed: (i) an LTE analysis (Section \ref{sec:LTE}), and (ii) a radiative analysis via radiative transfer modeling (Section \ref{sec:radex}). 
The main goal is to characterize the column density of each molecular species, and more importantly the gas temperature and the gas density in the CND. For reference, Table \ref{tab:table_integ_I} also lists all the measured velocity-integrated line intensities (in K km/s) from the observations smoothed to the common resolution $0''.8$. 
\label{sec:ladder}
\begin{table*}[ht!]
  \centering
  \caption{The velocity integrated line intensity of HNCO and SiO transition covered in the current work. These values are extracted from data that have been smoothed to the common $0''.8$ resolution, as indicated in Section \ref{sec:gas_properties}. }
  \label{tab:table_integ_I}
  \begin{tabular}{c|cccccc}
  \hline
    Transition & $\sigma$ & I(AGN) & I(R1) & I(R2)  & I(R3) &  I(R4) \\
    {} & [K km/s] & [K km/s] & [K km/s] & [K km/s] & [K km/s] & [K km/s]\\
    \hline
    \hline
    HNCO(4-3) &   5.29 & 20.12 &   33.07 &   26.73 &   23.29 &   17.78\\
    HNCO(5-4) &   7.46 & $\leq3.0 \sigma$ & 30.56 &   24.68 &   24.27 &   $\leq3.0 \sigma$\\
    HNCO(6-5) &   3.28 & 22.89 &   43.49 &   17.09 &   29.42 &   26.62\\
    \hline
    SiO(2-1) & 9.16 & 47.37 &   108.06 &  38.13 &   36.84 &   30.36\\
    SiO(3-2) & 4.99 & 25.82 &   65.87 &   $\leq3.0 \sigma$ & 16.37 &   15.28\\
    SiO(5-4) & 3.81 & $\leq3.0 \sigma$ & 21.80 &   $\leq3.0 \sigma$ & $\leq3.0 \sigma$ & $\leq3.0 \sigma$\\
    SiO(6-5) & 2.11 & 11.05 &   24.08 &   $\leq3.0 \sigma$ & 6.37 & $\leq3.0 \sigma$\\
    SiO(7-6) & 1.74 & 5.64 & 9.46 & $\leq3.0 \sigma$ & $\leq3.0 \sigma$ & $\leq3.0 \sigma$\\
    \hline
  \end{tabular}
\end{table*}
\subsection{LTE analysis}
\label{sec:LTE}
In quantifying the physical conditions probed by our HNCO and SiO transitions, we first performed a basic LTE analysis. 
We calculate the total column density ($N$) of a given species via the Boltzmann equation in LTE at temperature $T_{k}$: 
\begin{equation}
\label{eq:N_tot}
    N = \frac{N_{u}Z}{g_{u}e^{\frac{-E_{u}}{k_{B}T_{k}}}},
\end{equation}
where $N$ is the total column density of the species, $N_{u}$ is the column density of the upper level $u$, $Z$ is the partition function, $g_{u}$ is the statistical weight of the level $u$, and $E_{u}$ is its energy above the ground state. 
The molecular column density at upper level $u$, $N_{u}$, can be related to the observed integrated line intensity from any given transition, assuming optically thin and a filling factor of unity: 
\begin{equation}
\label{eq:N_u}
    N_{u} = \frac{8\pi k \nu^{2} I}{hc^{3}A_{ul}}, 
\end{equation}
where $I = \int T_{mb} \cdot dv$ is the integrated line intensity (in K km s\textsuperscript{-1}) over spectral axis. 
In general the formalism above provides a lower limit for the true $N$, for in a more realistic setup one also needs to consider an opacity correction factor in Equation \ref{eq:N_u} that in the end underestimates the $N$ in Equation \ref{eq:N_tot} but such correction can be neglected under optically thin condition \citep{Goldsmith_Langer_1999}. 
The HNCO column densities inferred from its three transitions are generally consistent within one order of magnitude, and by using temperature $T_{k}=50-200$ K the HNCO column density is between $4\times10^{14}-9.5\times10^{15}$ cm\textsuperscript{-2} among the CND regions. 
However, the inferred column density of SiO greatly varies  among the available transitions. With temperatures $T_{k}=50-200$ K, the SiO column density is between $6.5\times10^{12}-2.1\times10^{15}$ cm\textsuperscript{-2}. 
The temperature span, $T_{k}=50-200$ K, explored here is based on the previously inferred gas temperature by \citet{Viti+2014} and \citet{Kelly+2017} in the CND. 

In order to constrain the temperature, we can also construct the rotation diagram that relates the quantity $\ln({N_{u}/g_{u}})$ to the $E_{u}$ linearly with a slope of $(-1/T_{rot})$ given that we are mostly in the Rayleigh-Jeans scenario and since we have multiple transitions for each species per selected region. 
The rotational temperature $T_{rot}$ is expected to be equal to kinetic temperature $T_{kin}$ if all levels are thermalized \citep{Goldsmith_Langer_1999}, and in general it can be used as a lower limit estimate for $T_{kin}$. 

\begin{figure*}
  \centering
  \begin{tabular}[b]{@{}p{0.65\textwidth}@{}}
    \centering\includegraphics[width=1.0\linewidth]{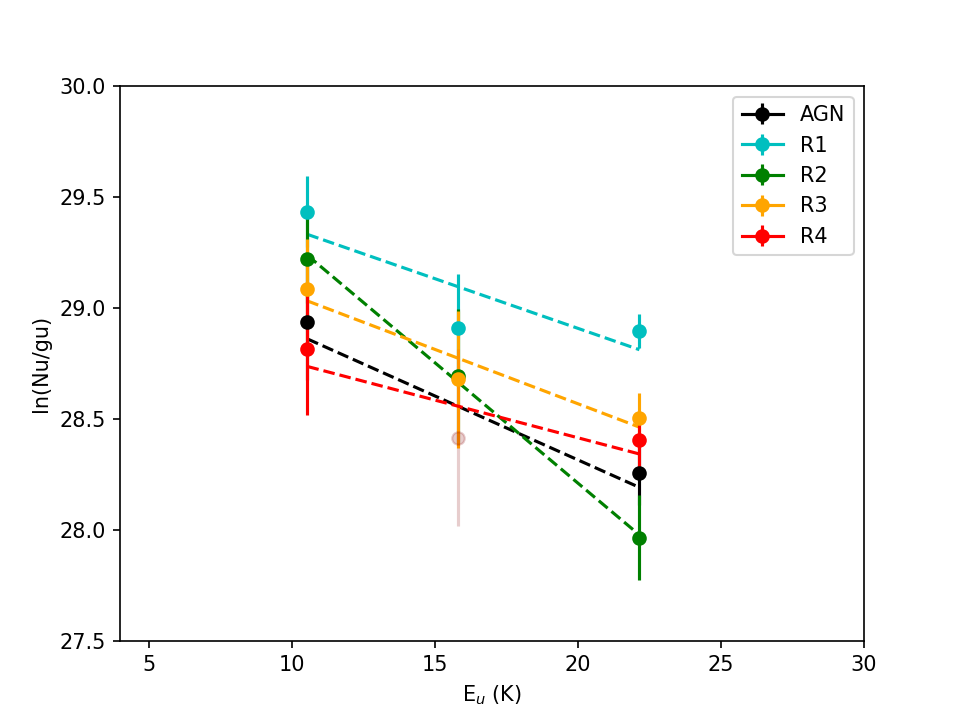} \\
    \centering\small (a) Rotation diagram for all HNCO transitions over all five CND selected regions
  \end{tabular}\\
  \begin{tabular}[b]{@{}p{0.65\textwidth}@{}}
    \centering\includegraphics[width=1.0\linewidth]{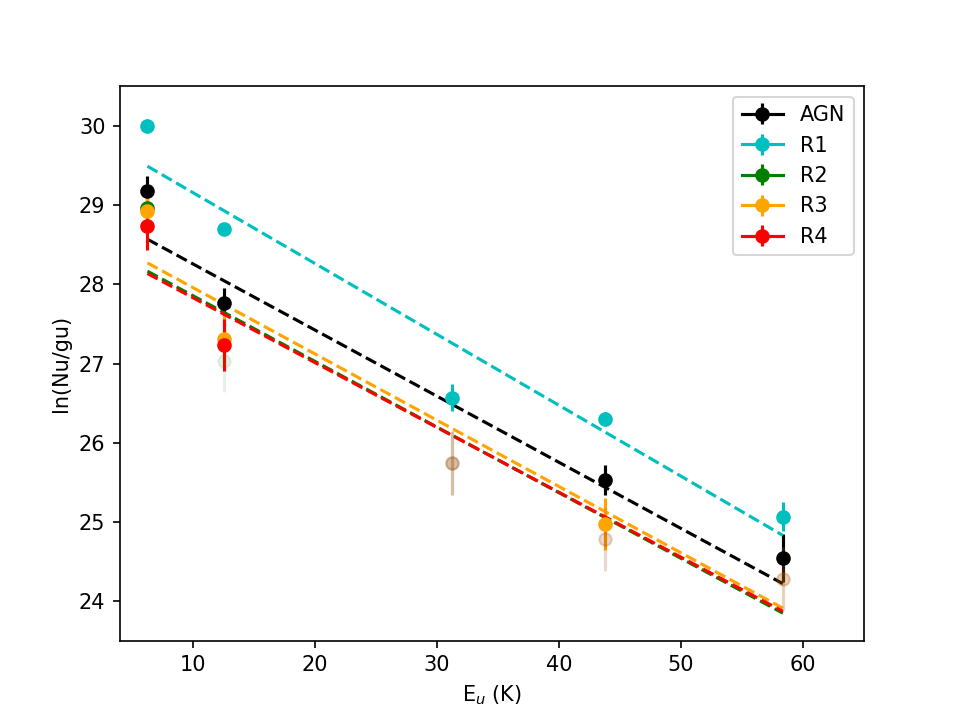} \\
    \centering\small (b) Rotation diagram for all SiO transitions over all five CND selected regions
  \end{tabular}%
  \caption{The rotation diagram derived from (a) HNCO transitions and (b)SiO transitions in all selected regions. The uncertainty is propagated from the measured error from the line intensity. The color coding indicates data from each region selected as shown in the schematic in Figure \ref{fig:mom0_hnco_zoomin}-(d). The faded (gray-ish) marks indicate data points below {$3.0 \sigma$} cut, to which case we will use {$3.0 \sigma$} as upper limit in our further analysis. The fitted rotational temperature are list in Table \ref{tab:table_Trot}. }
  \label{fig:rotdiag_hnco_sio}
\end{figure*}

Figure \ref{fig:rotdiag_hnco_sio} shows the rotation diagrams with the uncertainty constructed across the five CND regions, color coded accordingly, using all available transitions from HNCO and SiO respectively. 
The uncertainties are the propagated error estimate from the measured line intensities. 

The inferred $T_{rot}$ for HNCO is between $9-30$ K, and for SiO it is between $11-12$ K. The individual fitted $T_{rot}$ values are listed in Table \ref{tab:table_Trot}. 
The fitting for HNCO, in general, is within the uncertainties of the line intensities. 
On the other hand, for SiO transitions, the linear fit is generally not well constrained within the errors of the data points, which is consistent with the fact that the column densities derived from different transitions significantly differ at a fixed temperature. 
Taking CND-R1 region as the most prominent example, it is hard to be convinced by a one component fit and it is indeed likely that different transitions may not always trace the same gas components.  
Aside from the potential multiplicity of gas components within our beams, \citet{Goldsmith_Langer_1999} and \citet{Martin+2019_MADCUBA} showed that both the opacity and the gas density that each transition is tracing, can lead to deviations from a linear fit in  the rotation diagram. 
In fact, based on our RADEX analysis (Section \ref{sec:radex}), it is possible for lower-J SiO transition (2-1 and 3-2) to transit into optically thick regime, where the optical depth can be as high as $\sim10$, when the molecular gas density goes below $10^{4-5}$ cm\textsuperscript{-3}, which could contribute to the non-linearity shown in the SiO rotation diagram. 
When the molecular gas density ($n_{H2}$) goes below the critical density ($n_{crit}$) of the given transition, the emission that is not thermalized will also break the linearity. 
This will be discussed further in Section \ref{sec:radex} and Appendix \ref{sec:n_crit}.

In summary,  the main take-away point from the rotation diagrams is that, even within each sub-region of the CND, the gas traced by SiO is not well characterized by a single gas component with LTE condition, which is indeed consistent with such gas being heavily shocked and  a gradient of gas conditions may be present.
This is also consistent with the findings by \citet{Viti+2014} where they concluded that gas within the same selected region of the CND is not well characterized by a single gas phase. 

In order to further test whether the gas is in LTE, in the next Section we perform a non-LTE analysis. 

\subsection{Non-LTE analysis with RADEX}
\label{sec:radex}
For the non-LTE analysis, we use the radiative transfer code RADEX \citep{radex_vandertak_2007} via the Python package SpectralRadex\footnote{https://spectralradex.readthedocs.io} \citep{Holdship+2021_SpectralRadex} using HNCO and SiO molecular data \citep{hnco_moldata_N+1995,hnco_moedata_S+2018,sio_moldata_B+2018} from the LAMDA database \citep{LAMDA_2005}. This allows us to account for how the gas density and temperature affect the excitation of the transitions and to fit three parameters of interest: $n_{H2}$, $T_{kin}$, and $N$ as well as the beam filling factor. 

In order to properly sample this parameter space and obtain reliable uncertainties we couple the RADEX modeling with the Markov Chain Monte Carlo (MCMC) sampler emcee  \citep{FM+2013_emcee} to perform a Bayesian inference of the parameter probability distributions. We assume uniform priors within the ranges given in Table \ref{tab:table_prior} and that the uncertainty on our measured intensities is Gaussian so that our posterior distribution is given by $P(\theta | d) \sim \exp(-\frac{1}{2}\chi^2)$ where $\chi^2$ is the chi-squared statistic between our measured intensities and the RADEX output for a set of parameters $\theta$. We include the intensities of all transitions in our observed frequency range, even those that fall below {$3.0 \sigma$}. These non-detected transitions still contribute information because our modeling approach assumes the integrated intensity contains only molecular emission plus noise. If the molecular emission is weak enough for the noise to dominate, a well fitting model should predict that.

\begin{table}[ht!]
  \centering
  \caption{The prior range adopted for our RADEX parameters as described in Section \ref{sec:radex}. }
  \label{tab:table_prior}
  \begin{tabular}{c|c}
  \hline
    Variable  & Range \\
    \hline
    Gas density $n_{H2}$ [cm\textsuperscript{-3}] & $10^{2}-10^{8}$ \\
    Gas temperature $T_{kin}$ [K] & $10-800$ \\
    N(SiO) [cm\textsuperscript{-2}] & $10^{12}-10^{18}$\\
    N(HNCO) [cm\textsuperscript{-2}] & $10^{12}-10^{18}$\\
    Beam filling factor & $0.0-1.0$\\
    \hline
  \end{tabular}
\end{table}

In general this analysis is confined by the assumption that all of the molecular transitions arise from a single, homogeneous gas component. 
However, this assumption is crude and can only offer an averaged point of view for the gas properties given the limited resolution offered from observations, the different critical densities of the available transitions, and the energetic conditions required for each chemical species. Despite that, if variations in gas temperature and density within each sub-region of the CND are not too steep, a RADEX analysis should still be able to give us an indication of the average gas properties in the non-LTE study. 

\begin{figure*}
  \sidecaption
  \includegraphics[width=18cm]{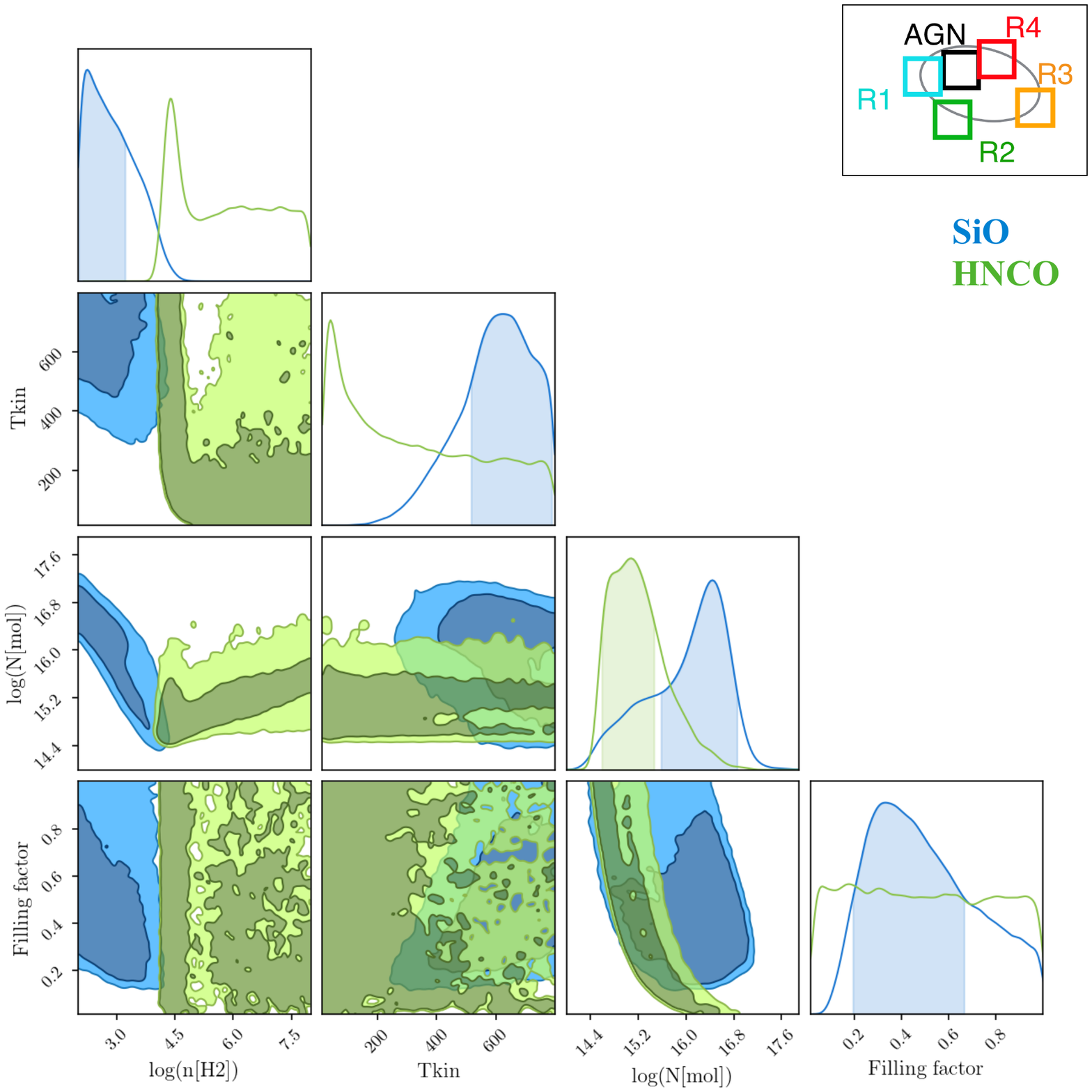}
  \caption{Bayesian inference results for gas properties traced by HNCO (green) and SiO (blue) of CND-R1 region. The corner plots show the sampled distributions for each parameter, as displayed on the x-axis. The 1-D distributions on the diagonal are the posterior distributions for each explored parameter;the reset 2-D distributions are the joint posterior for corresponding parameter pair on the x- and y- axes. In the 1-D distributions, the $1\sigma$ regions are shaded with blue; both $1\sigma$ and $2\sigma$ are shaded in the 2-D distributions. }
  \label{fig:Baye_Overlay_R1}
\end{figure*}

\begin{figure*}
  \sidecaption
    \includegraphics[width=18cm]{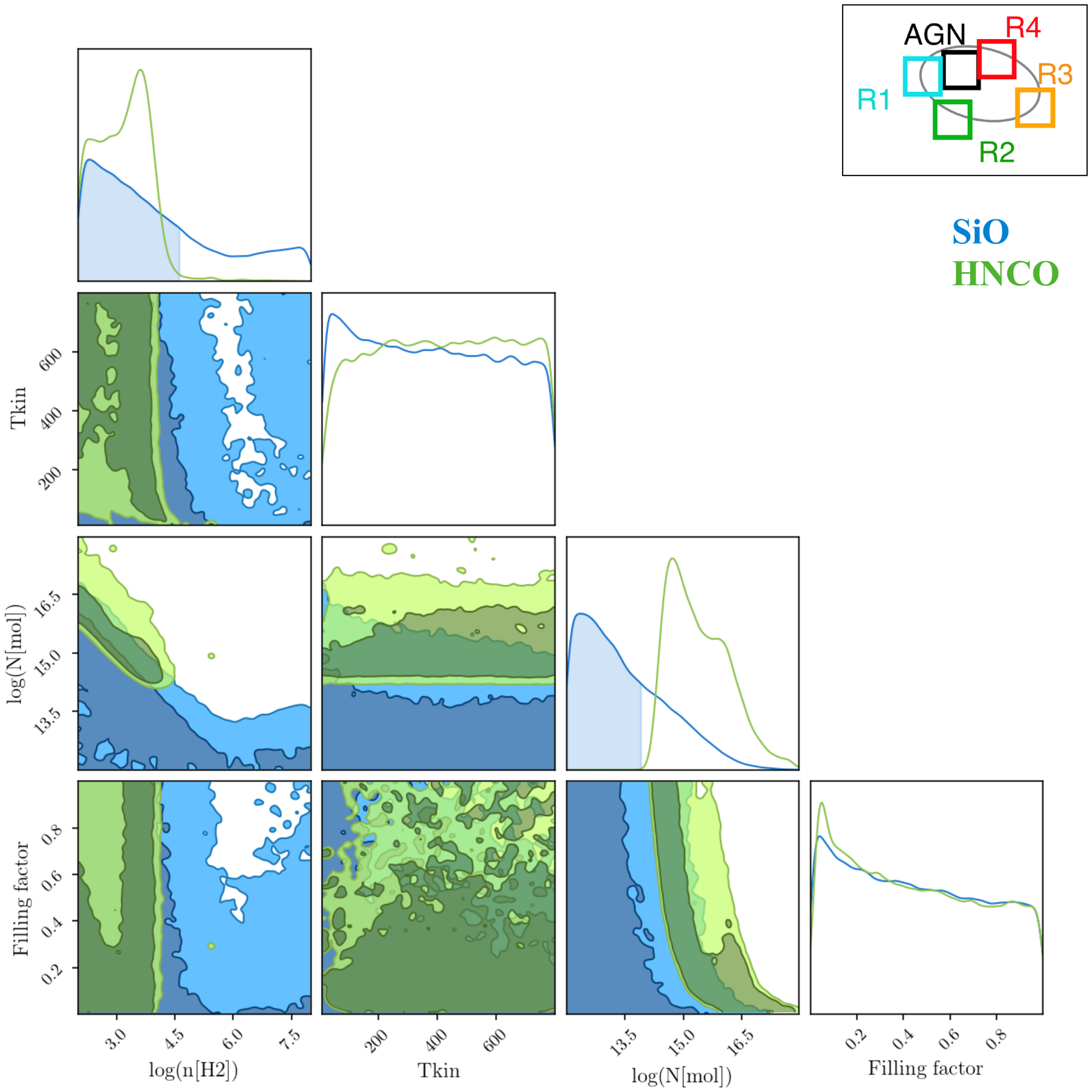} \\
  \caption{As in Figure~\ref{fig:Baye_Overlay_R1} but for the CND-R2 region. }
  \label{fig:Baye_Overlay_R2}
\end{figure*}

\begin{figure*}
  \centering
  \begin{tabular}[b]{@{}p{1.0\textwidth}@{}}
    \centering\includegraphics[width=1.0\linewidth]{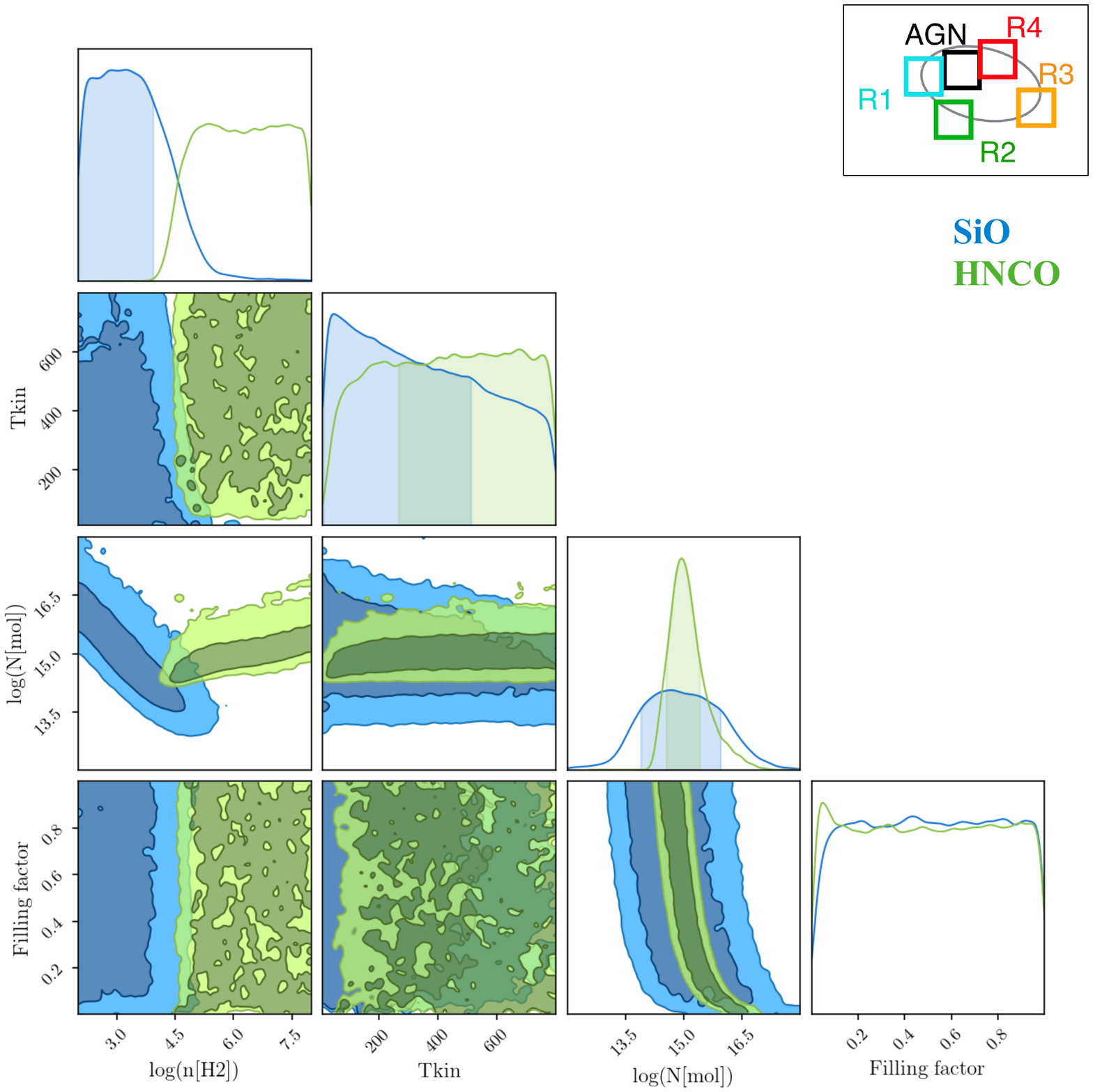} \\
  \end{tabular}%
  \caption{ As in Figure~\ref{fig:Baye_Overlay_R1} but for the CND-R3 region. }
  \label{fig:Baye_Overlay_R3}
\end{figure*}

\begin{figure*}
  \centering
  \begin{tabular}[b]{@{}p{1.0\textwidth}@{}}
    \centering\includegraphics[width=1.0\linewidth]{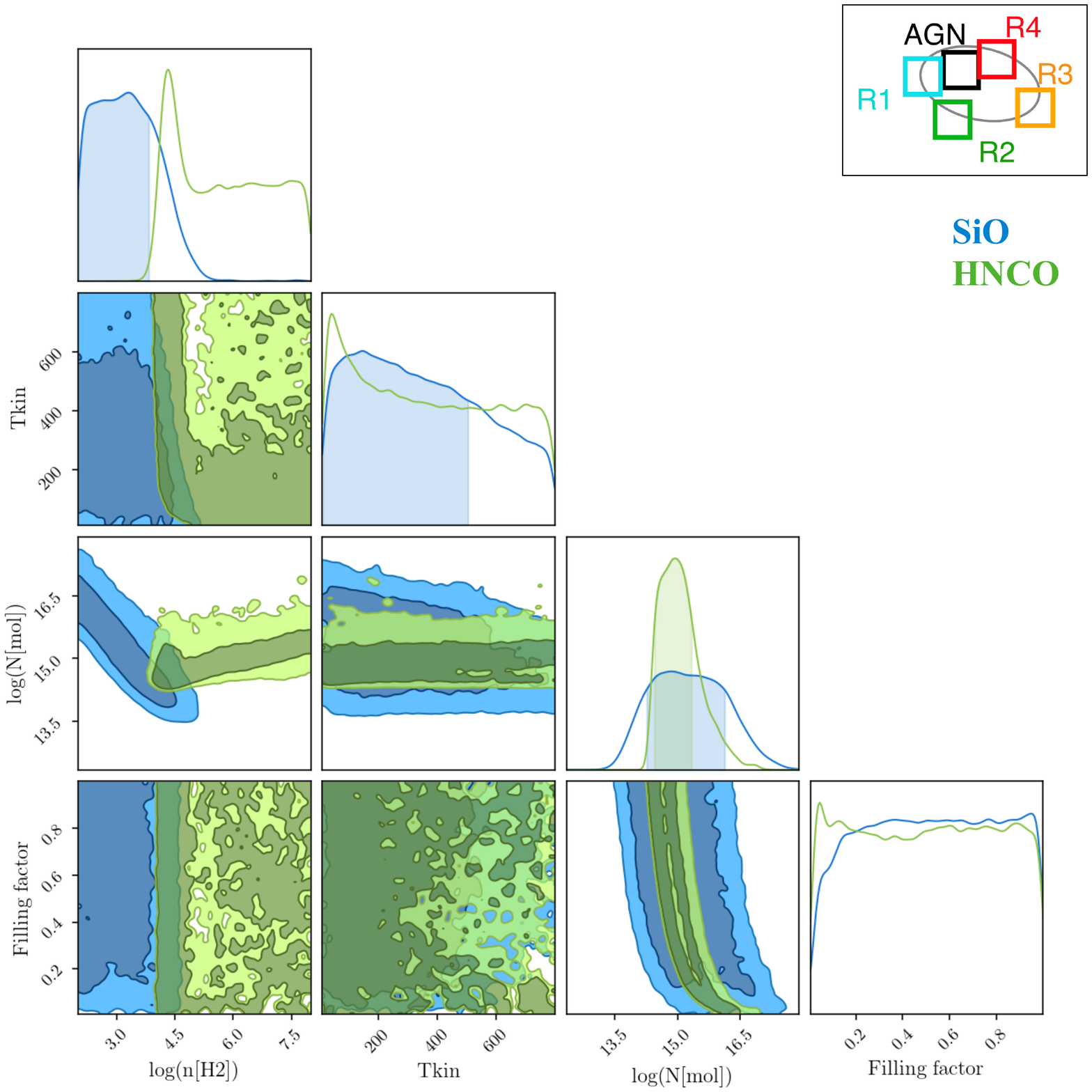} \\
  \end{tabular}%
  \caption{As in Figure~\ref{fig:Baye_Overlay_R1} but for the  CND-R4 region. }
  \label{fig:Baye_Overlay_R4}
\end{figure*}

Based on the chemical modeling performed by \citet{Kelly+2017} which suggested that SiO and HNCO are probing different shock scenarios (fast  versus slow shocks) thus potentially very different gas properties, we choose to separate these two species in our RADEX analysis and infer the gas properties for each species individually. 
In Figure \ref{fig:Baye_Overlay_R1}-\ref{fig:Baye_Overlay_R4} we show the posterior parameter distributions obtained for HNCO (green) and SiO (blue) for each of the four selected CND region (R1-R4). 
The most likely values for these parameters are also listed in Table \ref{tab:table_baye_posterior} with uncertainties calculated by taking an interval around the most likely value that contains 67\% of the probability density, similar to a 1$\sigma$ uncertainty. 
The corner plots for individual species per region can also be found in the Appendix \ref{sec:add_Baye}. Note that we do not include the analysis for AGN, for the signal from AGN is mostly below the {$3.0 \sigma$} threshold, which is probably a result of beam smearing with the current resolution $0.8''$. 

\renewcommand{\arraystretch}{1.5}
\begin{table*}[ht!]
  \centering
  \caption{The inferred gas properties traced by HNCO  and SiO from the Bayesian inference processes  over four selected regions across CND (R1-R4). For poorly constrained cases  we  identify the upper or lower limit of the distribution and for such cases we place the 95 (for upper limit) or 5 (for lower limit) percentile values in  parenthesis. }
  \label{tab:table_baye_posterior}
  \begin{tabular}{c|ccccc}
  \hline
    Species & Parameter &  CND-R1 & CND-R2  & CND-R3 &  CND-R4  \\
    \hline
    \hline
    SiO & {$log_{10}(n_{H_{2}})$} & $\leq3.98$ & $\leq7.48$ & $\leq4.55$ & $\leq 4.86$ \\
    {} & {$T_{kin}$}        & $630_{-110}^{+160}$ & - - & - - & - - \\
    {} & {$log_{10}(N_{SiO})$} & $16.45_{-0.87}^{+0.40}$ & $12.0^{+1.9}_{-0.0}$ & $14.84^{+1.28}_{-0.58}$ & $14.71^{+1.23}_{-0.82}$   \\
    {} & {Beam filling factor} &  $0.33^{+0.34}_{-0.13}$ & - - & - - & - -   \\
    \hline
    HNCO & {$log_{10}(n_{H_{2}})$} & $\geq4.24$  & $\left[2.11,4.10\right]$ & $\geq4.14$ & $\geq4.61$\\
    {} & {$T_{kin}$}        & - - & - - & - - & - - \\
    {} & {$log_{10}(N_{HNCO})$} & $15.08_{+0.39}^{-0.49}$ & $\left[14.41,16.81\right]$  & $14.94_{+0.39}^{-0.49}$ & $14.94_{+0.47}^{-0.40}$ \\
    {} & {Beam filling factor} & - -  & - - & - - & - -\\
    \hline
  \end{tabular}
\end{table*}
The gas conditions derived from the SiO emission present an interesting case. 
In each region, the gas density posterior strongly favours values less than $ 5\times10^4$cm\textsuperscript{-3}. 
This is consistent with the density range found in the chemical modelling of \citet{Kelly+2017} in which a fast shock is needed to enhance the SiO abundance.
However, in R1 and R2 which have detections with higher signal-to-noise ratio (SNR), the posterior actually peaks at the lowest value allowed by the prior. Thus the fit appears to strongly favour extremely low densities. At the same time, the gas temperature is unconstrained in every case except R1 for which we obtain a temperature of T =  $630_{-110}^{+160}$ K. Compared to previous observations, the extremely low density values appear to be not physical, e.g. SiO was found to trace a higher gas density of $\sim10^{5}$ cm\textsuperscript{-3} by \citet{Usero+2004} and \citet{GB+2010} using lower resolution observations toward the CND of NGC 1068. 
As the physical scale we are studying ($0''.8\sim$ 56 pc) is comparable to GMCs scales, we also compare our results to  the  gas density as traced by SiO from the nucleus of our own galaxy: \citet{Huttemeister+1998} surveyed 33 sources in the Galactic center region at pc-scale resolution and invoked shocks as the responsible process for the enhanced SiO abundance. Using the SiO and \textsuperscript{29}SiO line ratios they found that SiO traces hot (T$>$ 100 K) and thin ($n_{H2}\sim$ a few $10^{3}$ cm\textsuperscript{-3}) gas. 
In fact, our measured gas density is quite close to the density described by \citet{Huttemeister+1998}, although we want to stress that our model seems to prefer the lowest possible density (towards $10^{2}$ cm\textsuperscript{-3}), which is even lower than the density inferred by \citet{Huttemeister+1998}. 
On the other hand, the only constrained temperature (for R1) is supported by a multiple-species RADEX analysis of dense gas tracers by \citet{Viti+2014} which found the temperature was $>400$ K. It is also consistent with earlier lower-resolution data from \citet{Krips+2011}.

We therefore suggest caution in interpreting the gas density values inferred from the SiO emission. A variety of gas conditions are expected in a shocked environment which may lead to relative excitation between transitions that cannot be captured with a single RADEX component as discussed in Sect.~\ref{sec:LTE}. In fact, we have estimated the critical densities of the SiO transitions we use in the current work with temperature $T\in (10, 800)$ K as listed in Table \ref{tab:n_crit_sio}, and found that the inferred gas densities from RADEX analysis traced by SiO are actually below the critical density. It may therefore be that the only way to obtain a reasonable fit with a single component is to combine very low densities to produce sub-thermal excitation and then very high temperatures to excite the low E$_u$ transitions as we have observed. These conditions would not therefore be representative of any kind of average conditions in the region.

The best fit for the gas density traced by HNCO is not as well constrained overall but has a lower limit of 10$^{4}$ cm\textsuperscript{-3} in every region except for CND-R2 where lower densities are favoured. No constraint on the gas temperature from the HNCO emission is found in any case.
From these physical parameters, we can consider the enhancement mechanism of HNCO. 
The abundance of HNCO is found to be enhanced in the presence of slow shocks for initial gas densities in the range $10^{3-5}$ cm\textsuperscript{-3} or in a shock-free but warm and dense (\textgreater $10^{4}$ cm\textsuperscript{-3}) environment \citep{Kelly+2017}. 
At low gas density ($10^{3}$ cm\textsuperscript{-3}) the molecular gas and the dust are not coupled, therefore the dust temperature will be lower than the gas temperature, and at this density the dust grain mantles are not sublimated unless shock sputtering occurs. 
At higher densities,  gas and dust are coupled and hence the mantles can be sublimated without the presence of shock. 
For the shock-free scenario, therefore, higher gas density is required for HNCO enhancement. 
Hence we conclude that for all regions, but CND-R2, the enhancement of HNCO is just a consequence of the environment being hotter than the sublimation temperature of HNCO;  however, if we are to believe that the gas density in CND-R2 is, on average, as low as $10^{2}$ cm\textsuperscript{-3}, then the only way to enhance HNCO is via shock sputtering. 

The molecular column densities per species are well constrained in most regions except for CND-R2. 
The SiO column density generally ranges from $10^{14-17}$ cm\textsuperscript{-2} except for CND-R2, where it is much lower  ($\sim10^{12}$ cm\textsuperscript{-2}); this is comparable, but overall larger than the LTE-based estimate of SiO column density presented in Section \ref{sec:LTE}. 
The HNCO column density is approximately $10^{15}$ cm\textsuperscript{-2} in every case, and this is quite consistent with the LTE values derived in Section \ref{sec:LTE}. We may speculate that these results are consistent with the gas traced by SiO being strongly affected by shocks (and hence unlikely to be in LTE), whereas the gas traced by HNCO may be closer to LTE. 

\section{Conclusions}
The molecular gas in the CND of NGC 1068 is outflowing, likely a manifestation of ongoing AGN feedback \citep{GB+2014,GB+2019}. 
As the outflowing gas has a large spread of velocities ($\sim100$ km s\textsuperscript{-1}) which likely drive a range of different shocks  at different locations in the CND, we perform a multi-line molecular study with ALMA of two typical shock tracers in order to determine the chemical signatures of such shocks. In particular we analyse three HNCO lines and five SiO lines of the nearby galaxy NGC 1068 at spatial resolution of $0''.5-0''.8$. 
We briefly summarize below our conclusions: 
   \begin{enumerate}
      \item For both species the strongest peaks all occur on the east side of Circumnuclear Disk (CND), which is not consistent with the lower spatial resolution observations by \citet{Kelly+2017} where they found that HNCO(6-5) peaks in the west of the CND. The cross-species ratio maps of velocity-integrated line intensities of SiO and HNCO, however, show clear spatial differentiation going from large SiO/HNCO ratios in the east to a low SiO/HNCO ratio in the west of the CND;  this is consistent with the trend identified in \citet{Kelly+2017}. 
      \item We performed LTE and non-LTE analyses. For the latter, we coupled a radiative transfer analysis using RADEX with a Bayesian inference procedure, in order to infer the gas properties traced by these two species. The inferred gas densities traced by SiO are generally  constrained to be less than $5\times10^{4}$ cm\textsuperscript{-3}, and are consistent with a scenario where strong shocks ($\ge$ 50 kms$^{-1}$) are present. The inferred gas densities traced by HNCO, however, are not as well constrained overall but tend toward a higher gas density ($\geq10^{4}$ cm\textsuperscript{-3}), except for CND-R2 where the gas density seems to below $10^{4}$ cm\textsuperscript{-3}. The low inferred gas density in CND-R2 would require the presence of slow shocks ($\sim$ 20 km s\textsuperscript{-1}) to produce the observed HNCO. We can not draw the same conclusion for the regions where HNCO yield a higher gas density.
      \item Our work indicates, in agreement with previous studies, that SiO and HNCO trace different gas components within the beam and most importantly different shock conditions and histories. To further disentangle the gas conditions traced by HNCO will require future observations that cover more HNCO transitions at comparable spatial resolution to complete the current investigation. 
   \end{enumerate}

\begin{acknowledgements}
      KYH, SV, and JH are funded by the European Research Council (ERC) Advanced Grant MOPPEX 833460.vii. 
      SGB acknowledges support from the research project PID2019-106027GA-C44 of the Spanish Ministerio de Ciencia e Innovaci{\'o}n. 
      MSG acknowledges support from the Spanish Ministerio de Econom{\'{i}}a y  Competitividad through the grants BES-2016-078922, ESP2017-83197-P and  the research project PID2019-106280GB-100. 
      KYH acknowledges assistance from Allegro, the European ALMA Regional Center node in the Netherlands. 
      This paper makes use of the following ALMA data: ADS/JAO.ALMA\#2013.1.00221.S, ADS/JAO.ALMA\#2015.1.01144.S, and ADS/JAO.ALMA\#2018.1.01506.S. ALMA is a partnership of ESO (represent- ing its member states), NSF (USA) and NINS (Japan), together with NRC (Canada), MOST and ASIAA (Taiwan), and KASI (Republic of Korea), in co- operation with the Republic of Chile. The Joint ALMA Observatory is operated by ESO, AUI/NRAO and NAOJ.
\end{acknowledgements}
\bibliographystyle{aa}
\bibliography{NGC1068,fundamentals,Additional}
\newpage
\appendix
\section{Additional results from LTE analysis}
\label{sec:Trot_table}
Here we attach some additional results from the rotation diagram described in Section \ref{sec:LTE}. 
We performed single-component, linear fitting of the rotation diagram per species per region, in order to constrain the rotational temperature ($T_{rot}$) in each case. 
The results are shown in Table \ref{tab:table_Trot}
\begin{table}[h]
  \centering
  \caption{The single-component fitting of the rotational temperature described in Section \ref{sec:LTE}. }
  \label{tab:table_Trot}
  \begin{tabular}{c|ccccc}
  \hline
    {}  & $T_{rot}$(AGN) & $T_{rot}$(R1) & $T_{rot}$(R2)  & $T_{rot}$(R3) &  $T_{rot}$(R4) \\
    {} &  [K] & [K] & [K] & [K] & [K]\\
    \hline
    HNCO & 17.33 & 22.31 & 9.20 & 20.39 & 29.40 \\
    SiO & 11.98 & 11.18 & 12.07 & 11.95 & 12.21 \\
    \hline
  \end{tabular}
\end{table}
\section{Additional results from non-LTE analysis }
\subsection{Additional plots from RADEX analysis with Bayesian inference processes}
\begin{figure*}
  \centering
  \begin{tabular}[b]{@{}p{0.45\textwidth}@{}}
    \centering\includegraphics[width=1.0\linewidth]{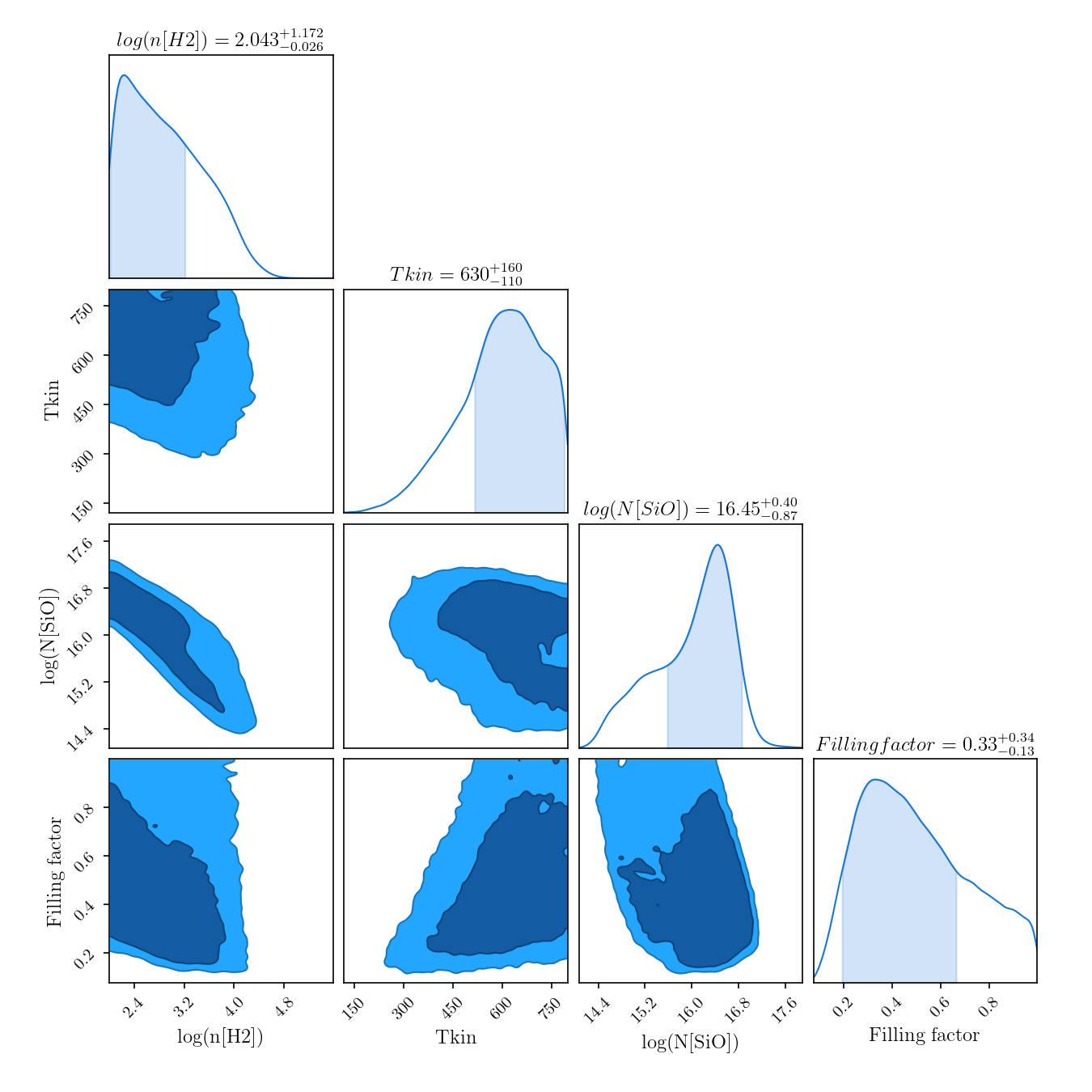} \\
    \centering\small (a) Inferred gas properties traced by SiO, in CND-R1
  \end{tabular}%
  \quad
  \begin{tabular}[b]{@{}p{0.45\textwidth}@{}}
    \centering\includegraphics[width=1.0\linewidth]{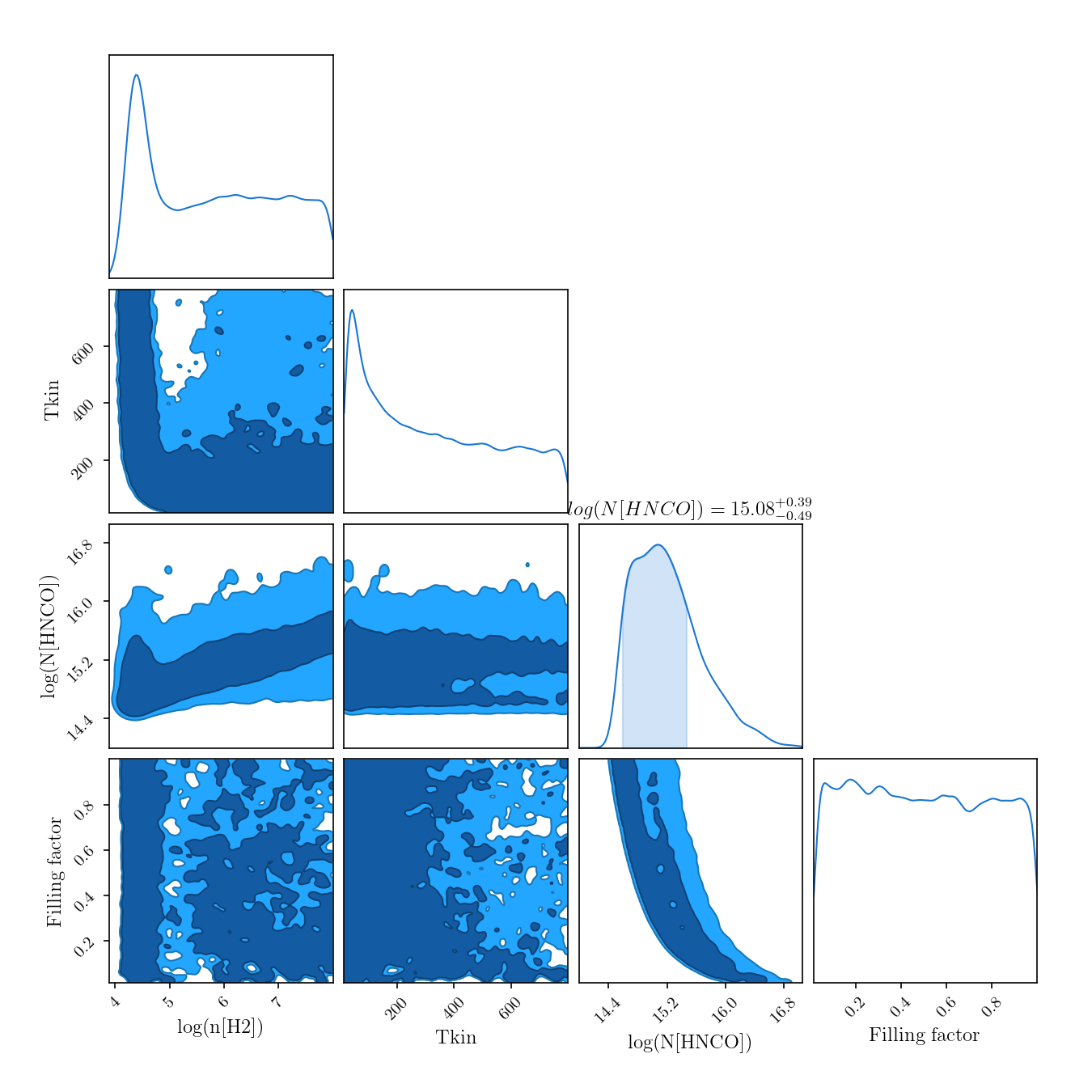} \\
    \centering\small (b) Inferred gas properties traced by HNCO, in CND-R1
  \end{tabular}
  \begin{tabular}[b]{@{}p{0.45\textwidth}@{}}
    \centering\includegraphics[width=1.0\linewidth]{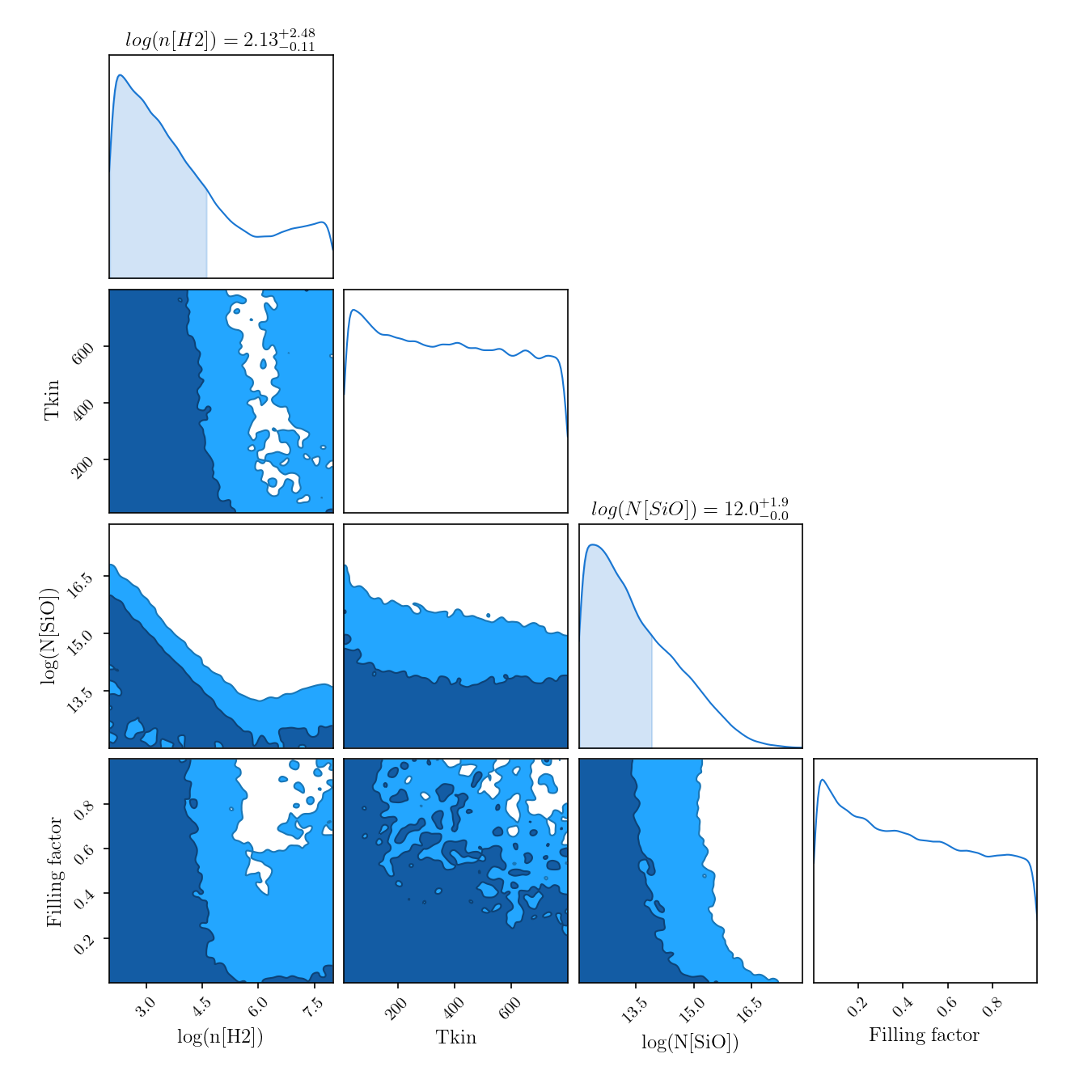} \\
    \centering\small (c) Inferred gas properties traced by SiO, in CND-R2
  \end{tabular}
  \begin{tabular}[b]{@{}p{0.45\textwidth}@{}}
    \centering\includegraphics[width=1.0\linewidth]{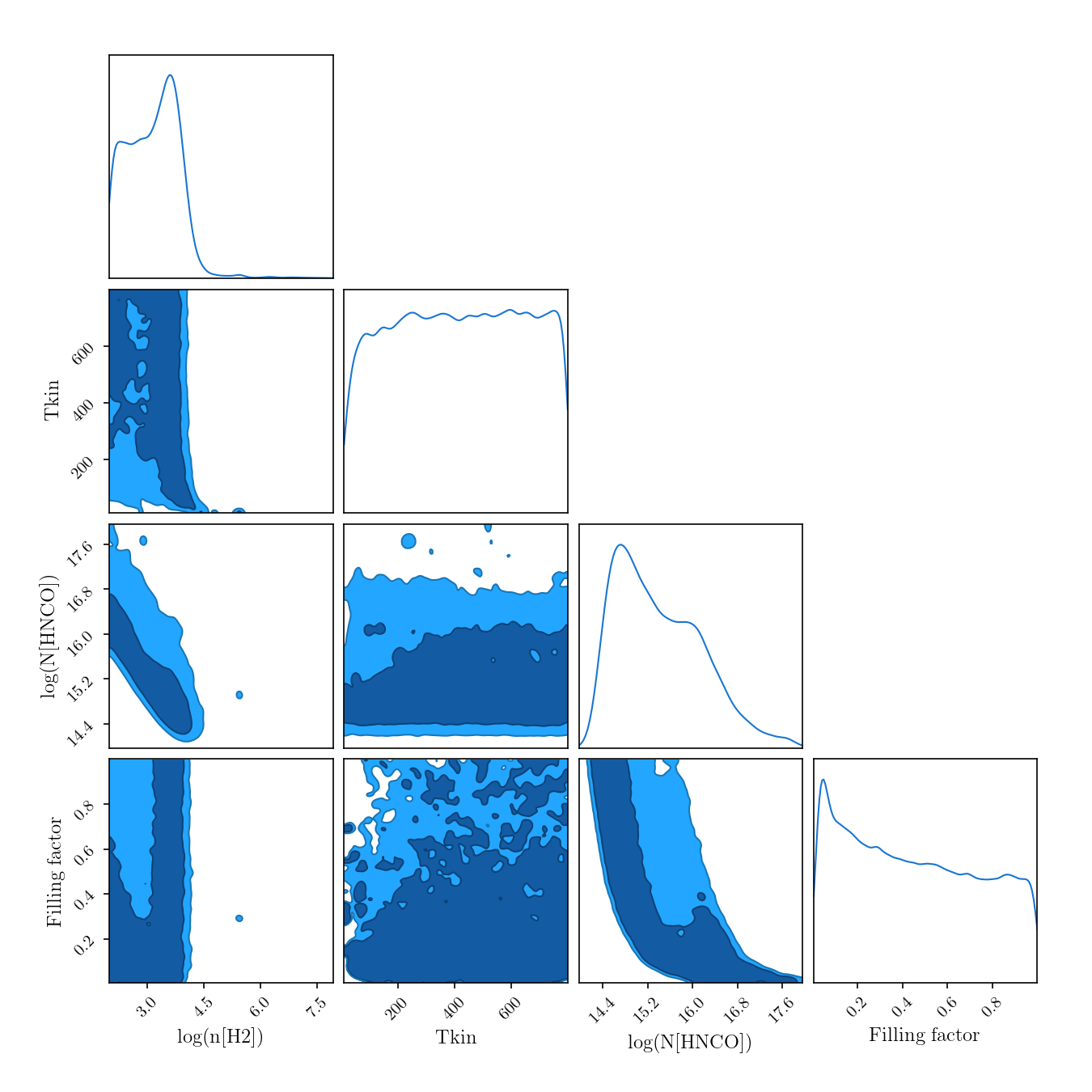} \\
    \centering\small (c) Inferred gas properties traced by HNCO, in CND-R2
  \end{tabular}
  \caption{The corner plots which shows the sampled distributions for each parameter, as displayed on the x-axis. The 1-D distributions on the diagonal are the posterior distributions for each explored parameter, the reset 2-D distributions are the joint posterior for corresponding parameter pair on the x- and y- axes. In the 1-D distributions, the $1\sigma$ regions are shaded with blue; both $1\sigma$ and $2\sigma$ are shaded in the 2-D distributions. On top of each 1-D distribution listed the inferred values if the distribution can be properly constrained. Results from CND-R1 and CND-R2 are presented, with SiO on the left and HNCO on the right panel. }
  \label{fig:Ini_Baye_I}
\end{figure*}

\begin{figure*}
  \centering
  \begin{tabular}[b]{@{}p{0.45\textwidth}@{}}
    \centering\includegraphics[width=1.0\linewidth]{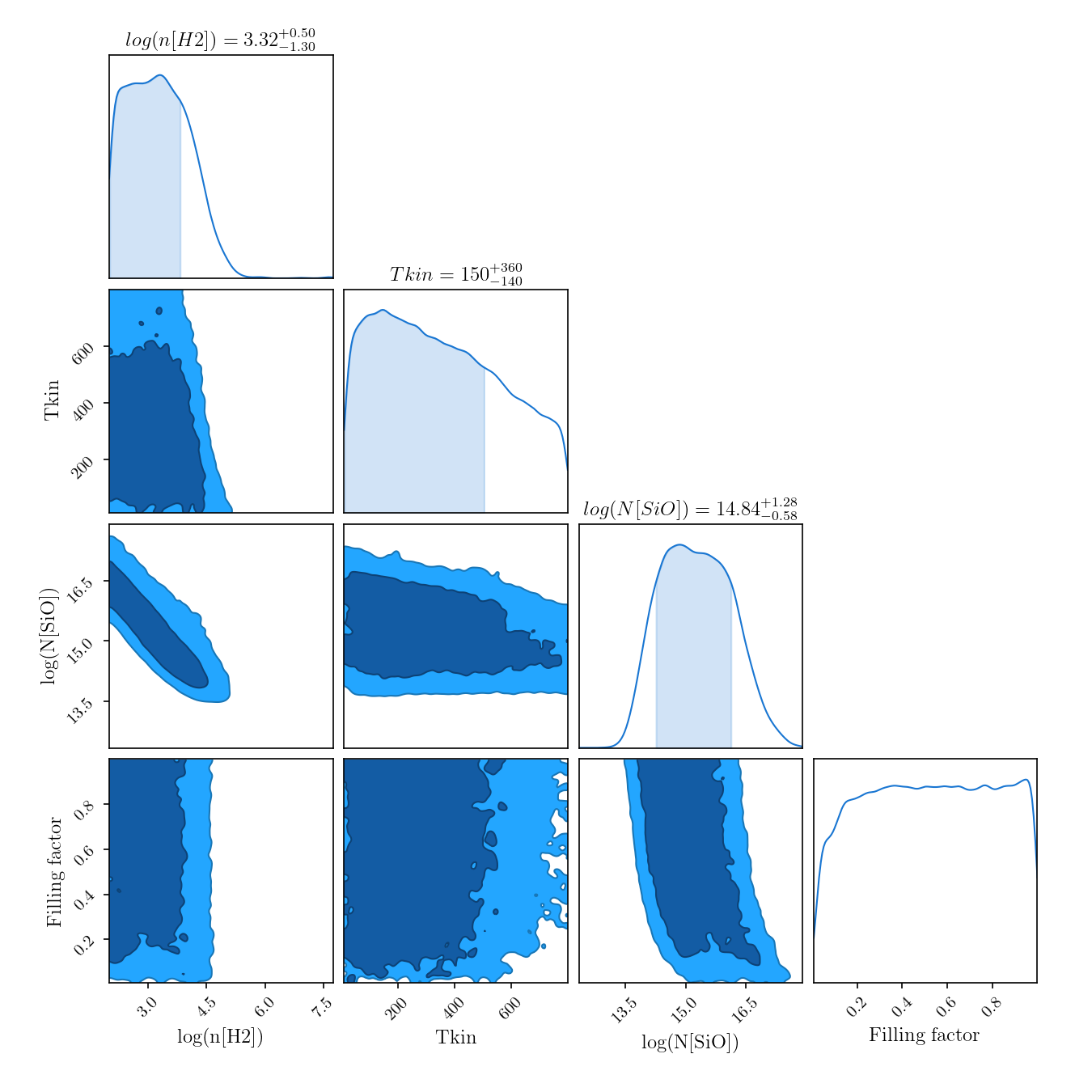} \\
    \centering\small (a) Inferred gas properties traced by SiO, in CND-R3
  \end{tabular}%
  \quad
  \begin{tabular}[b]{@{}p{0.45\textwidth}@{}}
    \centering\includegraphics[width=1.0\linewidth]{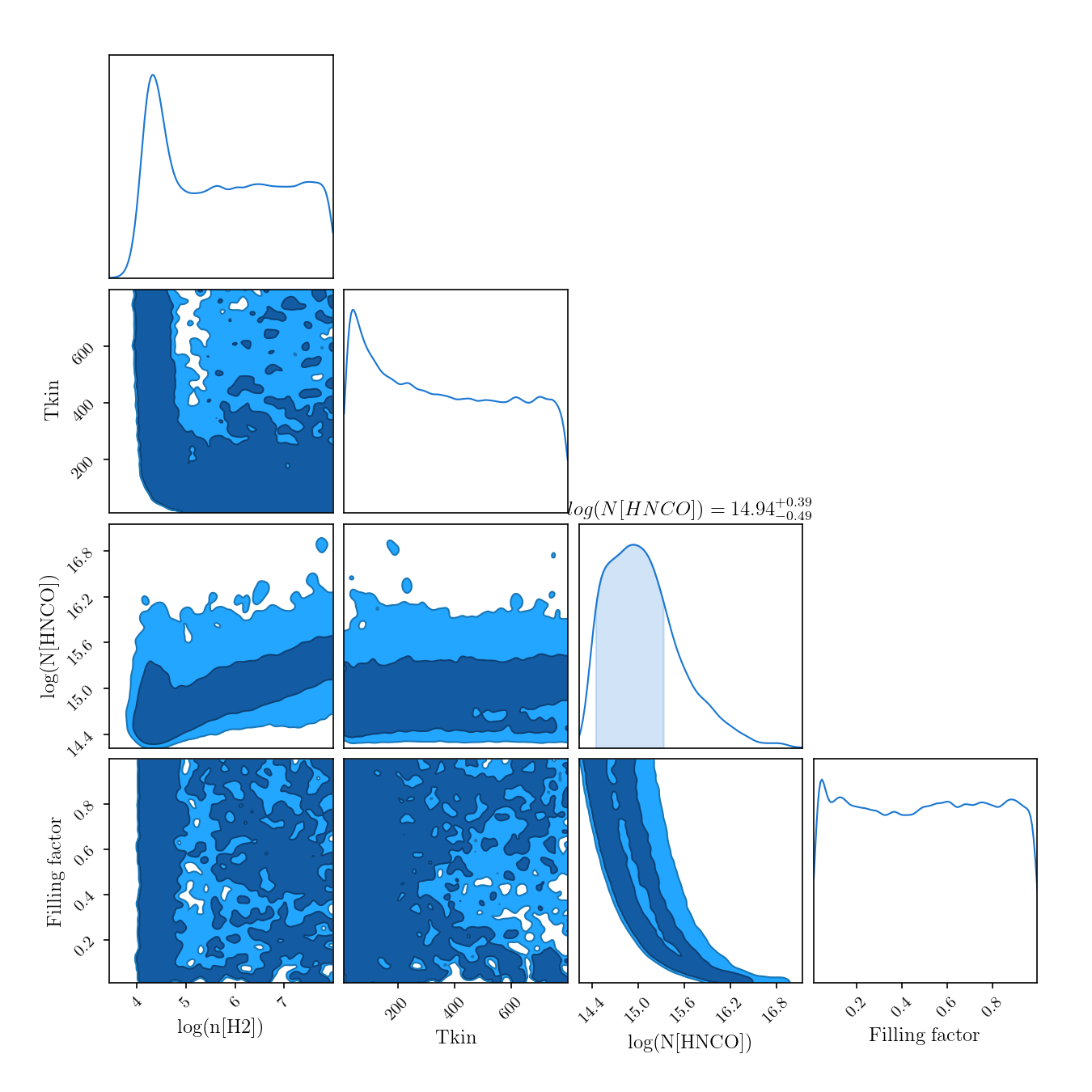} \\
    \centering\small (b) Inferred gas properties traced by HNCO, in CND-R3
  \end{tabular}
  \begin{tabular}[b]{@{}p{0.45\textwidth}@{}}
    \centering\includegraphics[width=1.0\linewidth]{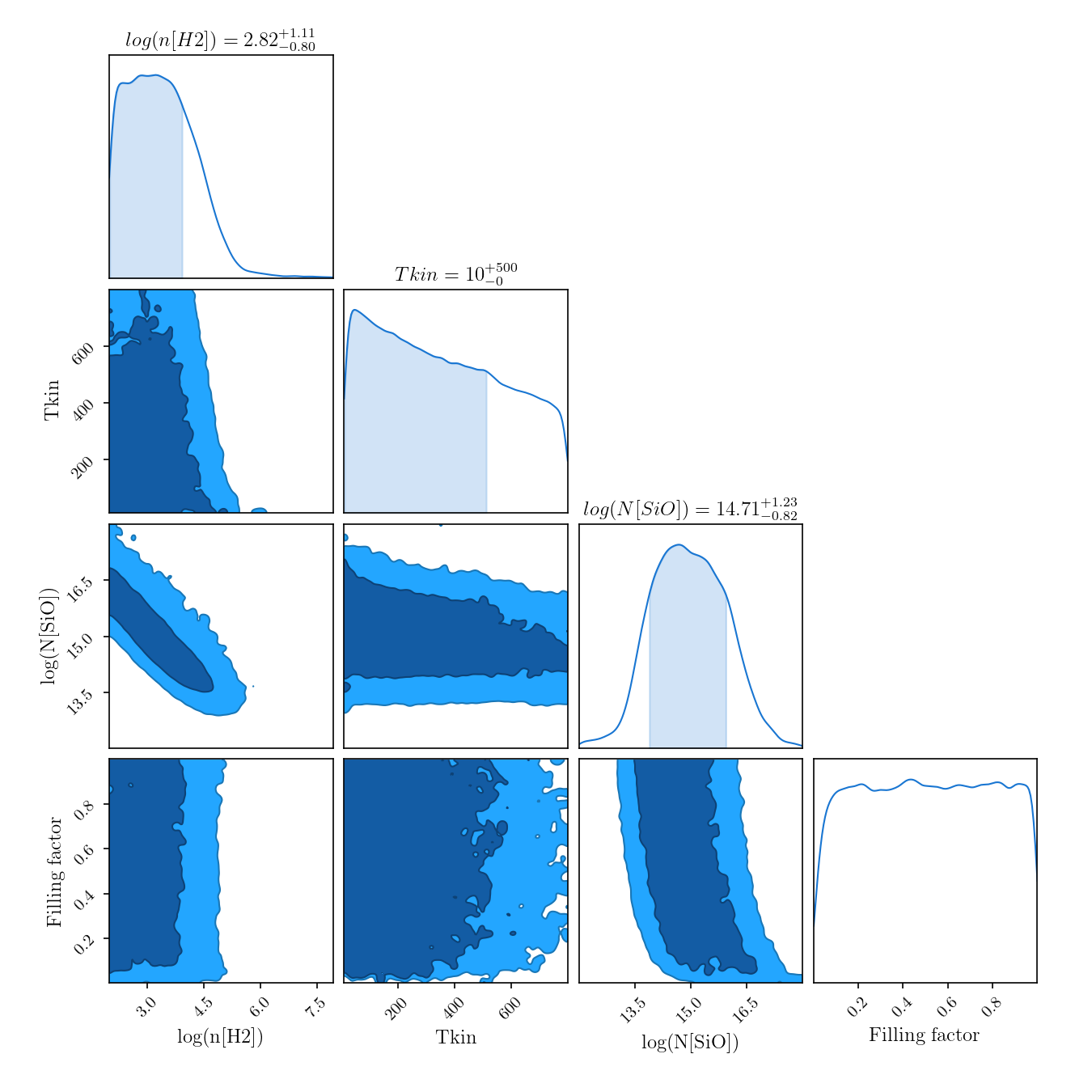} \\
    \centering\small (c) Inferred gas properties traced by SiO, in CND-R4
  \end{tabular}
  \begin{tabular}[b]{@{}p{0.45\textwidth}@{}}
    \centering\includegraphics[width=1.0\linewidth]{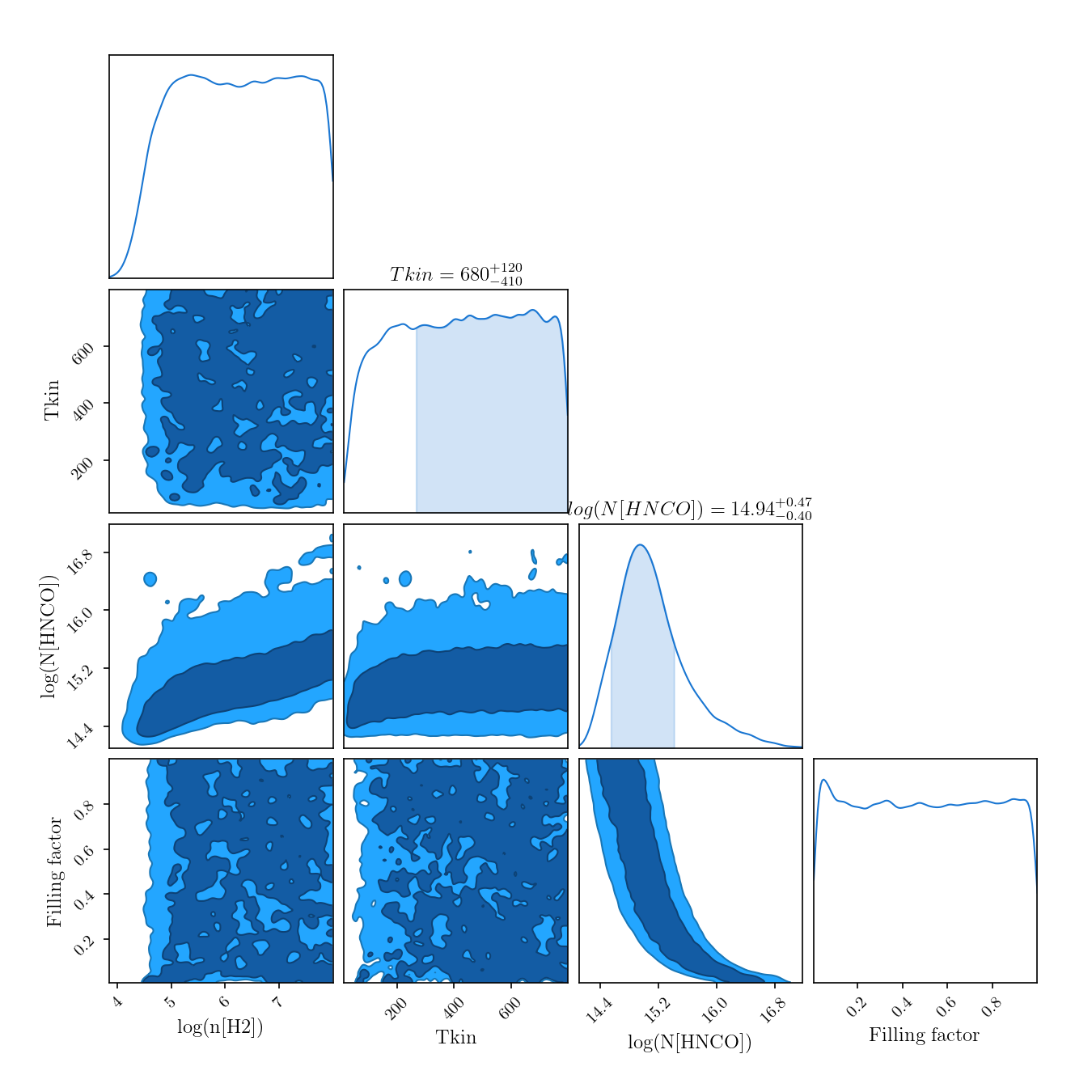} \\
    \centering\small (c) Inferred gas properties traced by HNCO, in CND-R4
  \end{tabular}
  \caption{Same as Figure \ref{fig:Ini_Baye_I}, but for CND-R3 and CND-R4. }
  \label{fig:Ini_Baye_II}
\end{figure*}

In Figure \ref{fig:Ini_Baye_I}-\ref{fig:Ini_Baye_II} we show the corner plots which shows the sampled distributions for each parameter, same as Figure \ref{fig:Baye_Overlay_R1}-\ref{fig:Baye_Overlay_R4}, just we plot each species individually instead of overlay the two species. 
\label{sec:add_Baye}
\subsection{Comparison of the predicted intensity from RADEX with observed values: a posterior predictive check (PPC)}
\begin{figure*}
  \centering
  \begin{tabular}[b]{@{}p{0.45\textwidth}@{}}
    \centering\includegraphics[width=1.0\linewidth]{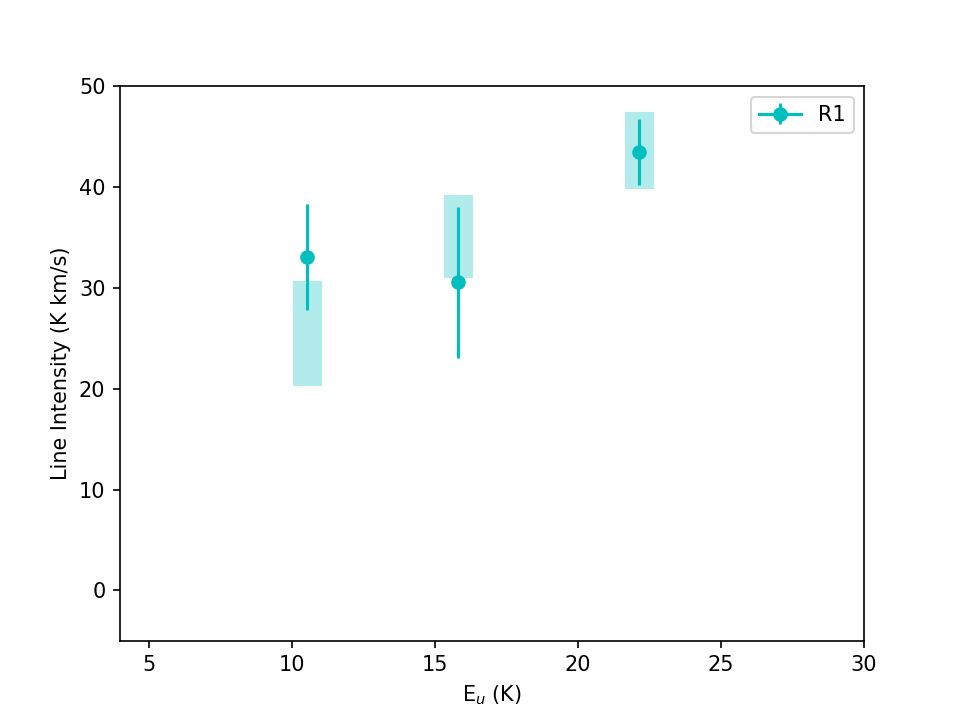} \\
    \centering\small (a) Observed and PPC intensities of HNCO in CND-R1
  \end{tabular}%
  \quad
  \begin{tabular}[b]{@{}p{0.45\textwidth}@{}}
    \centering\includegraphics[width=1.0\linewidth]{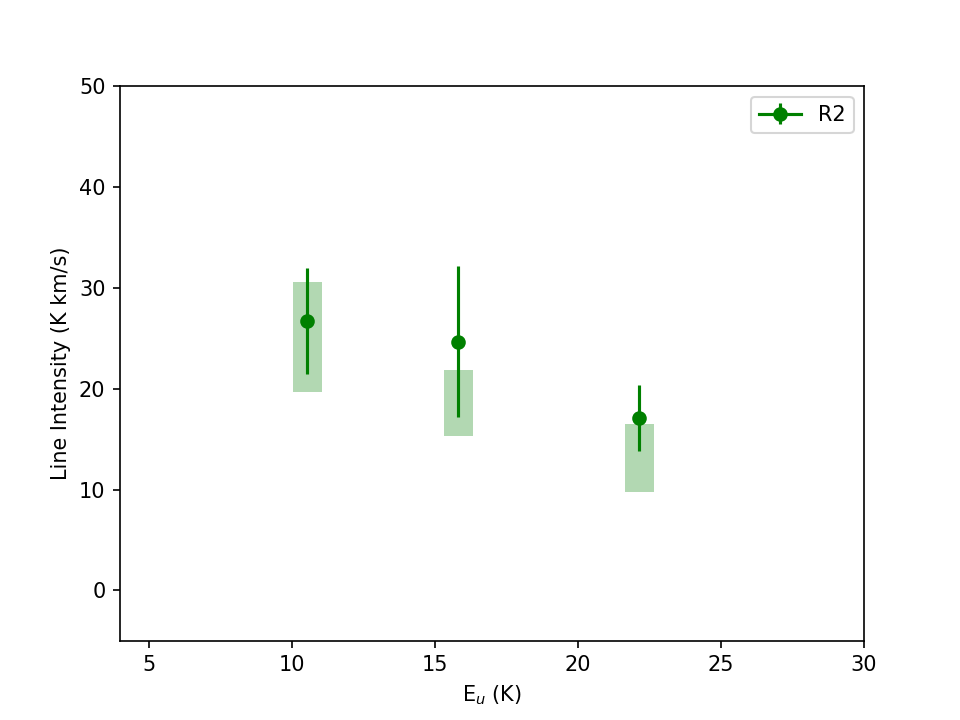} \\
    \centering\small (b) Observed and PPC intensities of HNCO in CND-R2
  \end{tabular}
  \begin{tabular}[b]{@{}p{0.45\textwidth}@{}}
    \centering\includegraphics[width=1.0\linewidth]{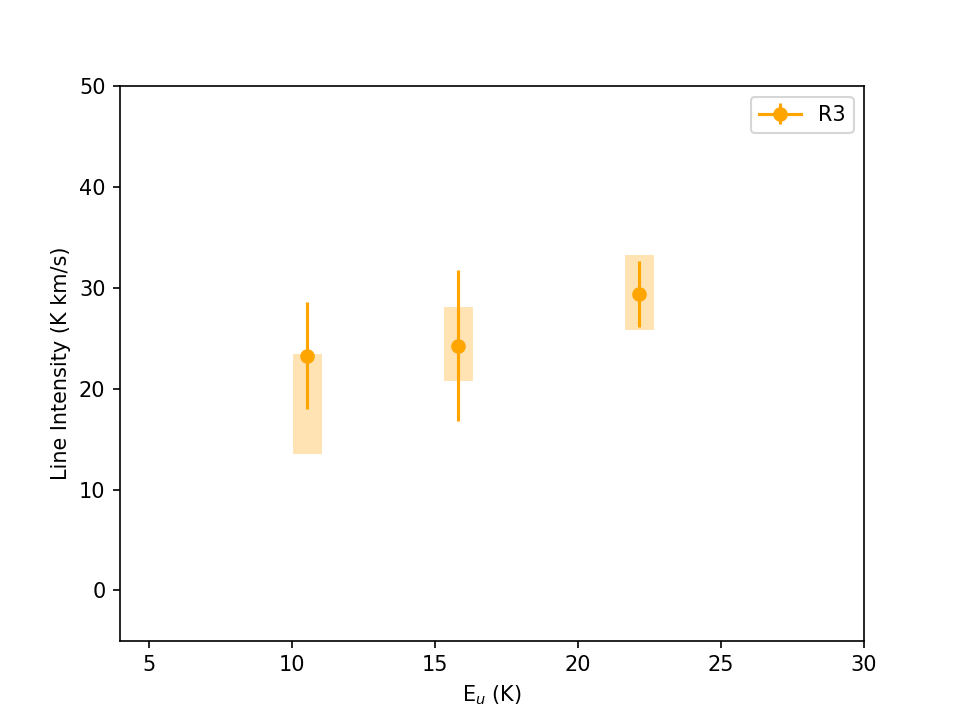} \\
    \centering\small (c) Observed and PPC intensities of HNCO in CND-R3
  \end{tabular}
  \begin{tabular}[b]{@{}p{0.45\textwidth}@{}}
    \centering\includegraphics[width=1.0\linewidth]{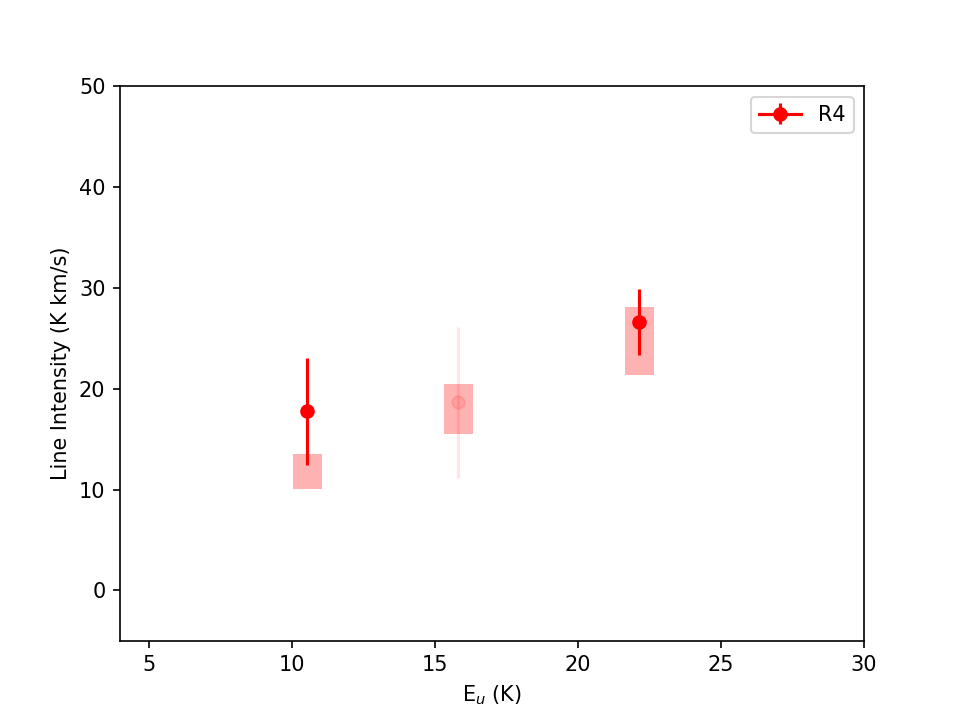} \\
    \centering\small (d) Observed and PPC intensities of HNCO in CND-R4
  \end{tabular}
  \caption{PPC plots of HNCO intensities for four CND regions (R1-R4). }
  \label{fig:PPC_HNCO}
\end{figure*}

\begin{figure*}
  \centering
  \begin{tabular}[b]{@{}p{0.45\textwidth}@{}}
    \centering\includegraphics[width=1.0\linewidth]{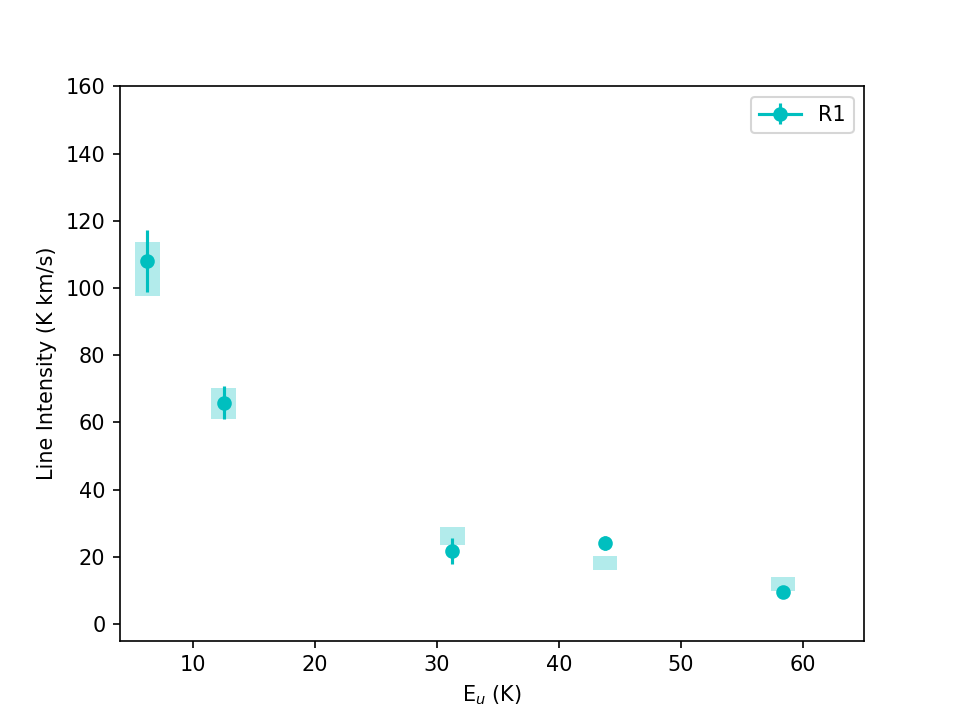} \\
    \centering\small (a) Observed and PPC intensities of SiO in CND-R1
  \end{tabular}%
  \quad
  \begin{tabular}[b]{@{}p{0.45\textwidth}@{}}
    \centering\includegraphics[width=1.0\linewidth]{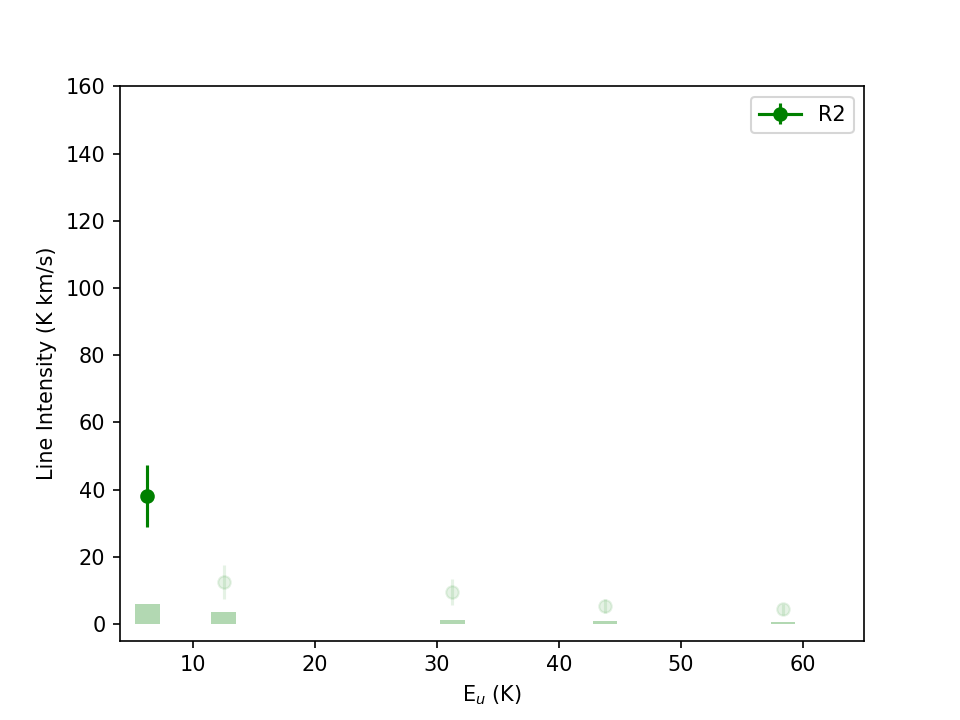} \\
    \centering\small (b) Observed and PPC intensities of SiO in CND-R2
  \end{tabular}
  \begin{tabular}[b]{@{}p{0.45\textwidth}@{}}
    \centering\includegraphics[width=1.0\linewidth]{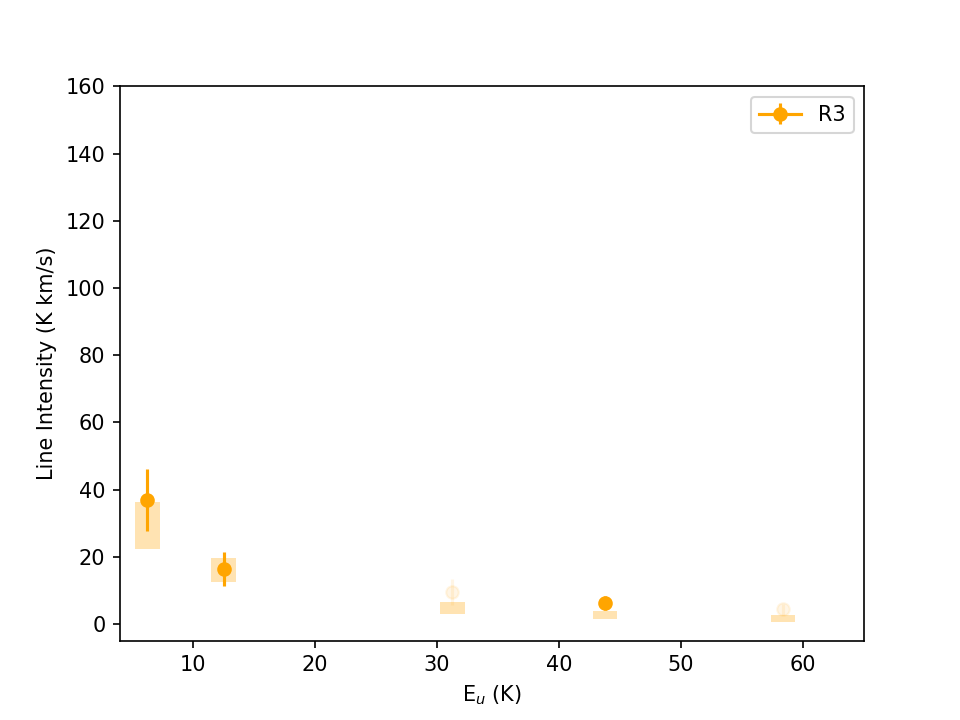} \\
    \centering\small (c) Observed and PPC intensities of SiO in CND-R3
  \end{tabular}
  \begin{tabular}[b]{@{}p{0.45\textwidth}@{}}
    \centering\includegraphics[width=1.0\linewidth]{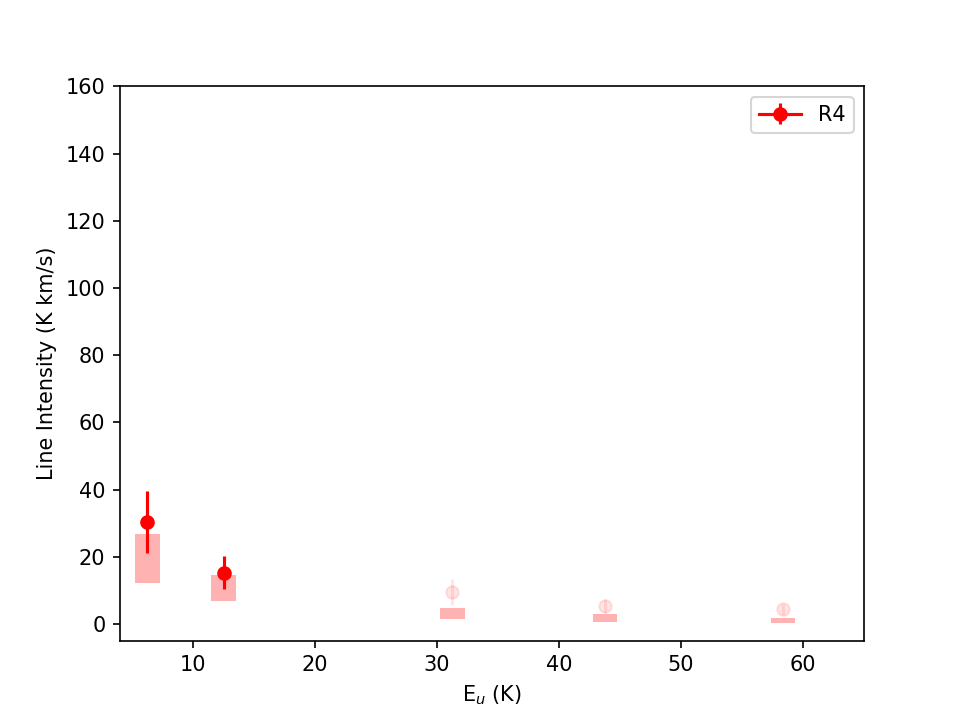} \\
    \centering\small (d) Observed and PPC intensities of SiO in CND-R4
  \end{tabular}
  \caption{PPC plots of SiO intensities for four CND regions (R1-R4). }
  \label{fig:PPC_SiO}
\end{figure*}

In this section, we perform a posterior predictive check for the inferred gas properties in Section \ref{sec:radex} using RADEX and Bayesian inference process. 
This is to verify our posterior distribution produces a distribution of data that is consistent with the actual data, which is the velocity-integrated intensity in our case. 
We sample the predicted line intensities from our posterior between 16-84 percentile, and plot against the observed line intensities. 
The comparisons are shown in Figure \ref{fig:PPC_HNCO}-\ref{fig:PPC_SiO}. 
\renewcommand{\arraystretch}{1.0}
\section{Critical density}
\label{sec:n_crit}
We mention the inferred gas density $n_{H2}$ traced by SiO might be lower than the critical density at given temperature. 
Here we give explicit estimates for the critical densities of SiO with gas temperature range from 10K to 800K. 
The critical density arises from the comparison of collisional deexcitation with radiative deexcitation \citep[e.g.][]{Draine_2011_ISMIGM}, and is written by:
\begin{equation}
    n_{crit} = \frac{A_{u\ell}}{\sum_{j, j<u}k_{uj}}
\end{equation}
, where $k_{uj}$ is the collisional rate coefficient that bears a unit of [cm\textsuperscript{3} s\textsuperscript{-1}], and can be turned into collisional rate [s\textsuperscript{-1}] by multiplying volume density [cm\textsuperscript{-3}]. 
Using SiO molecular data from LAMDA database \citep{LAMDA_2005,sio_moldata_B+2018}, we give estimate of SiO critical densities between T=10K to 800K in Table \ref{tab:n_crit_sio}. 
It is clear that the inferred gas density traced by SiO in all CND regions (R1-R4) in Section \ref{sec:radex} ($10^{2-4}$ cm\textsuperscript{-3}) are all below even the lowest $n_{crit}$, $1.28\times10^{5}$ cm\textsuperscript{-3}, among the temperature range and the transitions we explored. 
\begin{table*}[ht!]
  \centering
  \caption{The critical density of the SiO transitions used in current work at varying gas temperature. }
  \label{tab:n_crit_sio}
  \begin{tabular}{c|ccccccc}
  \hline
    Transition & $n_{crit}$[10K] &  $n_{crit}$[20K] &  $n_{crit}$[50K] &  $n_{crit}$[100K] & $n_{crit}$[200K] & $n_{crit}$[600K] & $n_{crit}$[800K]\\
    {} & [cm\textsuperscript{-3}]& [cm\textsuperscript{-3}]& [cm\textsuperscript{-3}]& [cm\textsuperscript{-3}]& [cm\textsuperscript{-3}]& [cm\textsuperscript{-3}]& [cm\textsuperscript{-3}] \\
    \hline
    SiO(2-1)  &  1.28E+05  &  1.27E+05  &  1.29E+05  &  1.34E+05  &  1.42E+05  &  1.56E+05  &  1.59E+05\\
    SiO(3-2)  &  4.08E+05  &  3.89E+05  &  3.79E+05  &  3.82E+05  &  3.90E+05  &  4.05E+05  &  4.07E+05\\
    SiO(5-4)  &  1.72E+06  &  1.63E+06  &  1.57E+06  &  1.54E+06  &  1.51E+06  &  1.44E+06  &  1.42E+06\\
    SiO(6-5)  &  2.86E+06  &  2.74E+06  &  2.66E+06  &  2.60E+06  &  2.52E+06  &  2.31E+06  &  2.26E+06\\
    SiO(7-6)  &  4.31E+06  &  4.19E+06  &  4.14E+06  &  4.07E+06  &  3.89E+06  &  3.48E+06  &  3.37E+06\\
    \hline
  \end{tabular}
\end{table*}

%
%

\end{document}